\documentclass[11pt]{article}
\pdfoutput=1

\usepackage[mathscr]{euscript}

\usepackage{amsmath,amsfonts,amsbsy,amssymb,array,dsfont}
\usepackage{enumerate,array,latexsym,graphicx,mathrsfs,verbatim,psfrag}
\usepackage{bm} 
\usepackage[normalem]{ulem}
\usepackage{booktabs} 
\usepackage[usenames]{color}
\usepackage[utf8]{inputenc}
\usepackage[english]{babel}
\usepackage{fancybox}

\usepackage{bbm}
\usepackage[small,hang,bf,font={footnotesize,sl}]{caption}
\usepackage{subcaption}
\usepackage{float}

\usepackage{datetime}

\usepackage[nosort]{cite}
\usepackage{chngpage} 

\usepackage{enumitem}
\setlist{itemsep=0pt}

\usepackage[colorlinks=true,      linkcolor=darkblue,      urlcolor=darkblue,      
            filecolor=darkblue,      citecolor=darkblue,       pdfstartview=FitH,     
						pdfpagemode=UseNone,      bookmarksopen=true]{hyperref}  
\usepackage[all]{hypcap}     

\newcommand{\captionfonts}{\small}
\makeatletter  
\long\def\@makecaption#1#2{%
  \vskip\abovecaptionskip
  \sbox\@tempboxa{{\captionfonts #1: #2}}%
 \ifdim \wd\@tempboxa >\hsize
    {\captionfonts #1: #2\par}
  \else
    \hbox to\hsize{\hfil\box\@tempboxa\hfil}%
  \fi
  \vskip\belowcaptionskip}
\makeatother   

\DeclareMathSymbol{\medhatsym}{\mathord}{largesymbols}{"62} 

\DeclareMathSymbol{\medtildesym}{\mathord}{largesymbols}{"65}

\newcommand{\abs}[1]{\ensuremath{\left|#1\right|}}

\newcommand{\comm}[1]{} 

\def\IC{\mathbb{C}}

\def\IR{\mathbb{R}}

\def\IT{\mathbb{T}}

\def\({\left(}
\def\){\right)}
\def\[{\left[}
\def\]{\right]}

\def\coeff#1#2{{\textstyle \frac{#1}{#2}}}

\def\One{{\hbox{ 1\kern-.8mm l}}}

\def\barray{\begin{array}}
\def\earray{\end{array}}
\def\be{\begin{equation}}
\def\ee{\end{equation}}
\def\bea{\begin{eqnarray}}
\def\eea{\end{eqnarray}}
\def\bal{\begin{align}}
\def\eal{\end{align}}

\def\mR{\mathbb{R}}

\newcommand{\xib}{\bar{\xi}}

\numberwithin{equation}{section} 

\makeatletter
\g@addto@macro\bfseries{\boldmath}
\makeatother

\definecolor{cardinal}{rgb}{0.6,0,0}
\definecolor{darkgreen}{rgb}{0,0.4,0}
\definecolor{purple}{rgb}{0.5, 0, 0.5}
\definecolor{golden}{rgb}{0.92, 0.7, 0}
\definecolor{midnight}{rgb}{0, 0, 0.5}
\definecolor{darkblue}{rgb}{0, 0, 0.8}

\def\xib{\bar{\xi}}
\def\Fb{\bar{F}}

\def\omegacomp{{\omega}}

\def\IC{\mathbb{C}}
\def\Neql#1{{\cal N}\!=\!{#1}}
\def\coeff#1#2{\relax{\textstyle {#1 \over #2}}\displaystyle}

\def\IR{\mathds{R}}

\def\cB{{\cal B}}

\def\cD{{\cal D}}
\def\cF{{\cal F}}

\def\cK{{\cal K}}

\def\cN{{\cal N}}
\def\cO{{\cal O}}
\def\cP{{\cal P}}
\def\cQ{{\cal Q}}

\def\nBPS#1{$\frac{1}{#1}$-BPS}

\def\cO{{\cal O}}

\begin{document}


\begin{flushright}
IPHT-T19/135
\end{flushright}

\vspace{6mm}

\begin{center}
\begin{adjustwidth}{-7mm}{-7mm} 
\begin{center}
{\fontsize{20}{20} \bf{Holomorphic Waves of Black Hole Microstructure}} \medskip \\
\end{center}
\end{adjustwidth}
\vspace{10mm}

\centerline{{\bf Pierre Heidmann$^{1,2}$, Daniel R. Mayerson$^1$, Robert Walker$^{1,3}$ and Nicholas P. Warner$^{1,3,4}$}}
\bigskip
\bigskip
\vspace{1mm}

\centerline{$^1$ Institut de Physique Th\'eorique,}
\centerline{Universit\'e Paris Saclay, CEA, CNRS, }
\centerline{Orme des Merisiers,  F-91191 Gif sur Yvette, France}
\vspace{2mm}
\centerline{$^2$ Department of Physics and Astronomy,}
\centerline{Johns Hopkins University,}
\centerline{ 3400 North Charles Street, Baltimore, MD 21218, USA,}
\vspace{2mm}
\centerline{$^3$\,Department of Physics and Astronomy,}
\centerline{University of Southern California,} \centerline{Los
Angeles, CA 90089-0484, USA}
\vspace{2mm}
\centerline{$^4$\,Department of Mathematics,}
\centerline{University of Southern California,} \centerline{Los
Angeles, CA 90089, USA}

\vspace{9mm} 
{\footnotesize\upshape\ttfamily pheidma1 @ jh.edu, ~daniel.mayerson @ ipht.fr,~ walkerra @ usc.edu,~warner @ usc.edu} \\

\vspace{15mm}
 
\textsc{Abstract}

\end{center}
\begin{adjustwidth}{6mm}{6mm} 
 
\vspace{-4mm}
\noindent

\noindent We obtain the largest families constructed to date of $\coeff{1}{8}$-BPS solutions of type IIB supergravity. They have the same charges and mass as  supersymmetric  D1-D5-P  black  holes,  but  they  cap  off  smoothly  with  no  horizon. Their construction relies on the structure of superstratum states, but allows the momentum wave to have an arbitrary shape. Each family is based on an arbitrary holomorphic function of one variable. More broadly, we show that the most general solution is described by two arbitrary holomorphic functions of three variables.  After exhibiting several new families of such ``holomorphic superstrata,'' we reformulate the BPS backgrounds and  equations in holomorphic form and show how this simplifies their structure.  The holomorphic formulation is thus both a fundamental part of superstrata as well as an effective tool for their construction.  In addition, we demonstrate that holomorphy provides a powerful tool in establishing the smoothness of our solutions without constraining  the underlying functions.  We also exhibit new families of solutions in which the momentum waves of the superstrata are highly localized at infinity but diffuse and spread into the infra-red limit in the core of the superstratum.  Our work also leads to some results that can be tested within the dual CFT.

\bigskip

\end{adjustwidth}

\vspace{8mm}
 

\thispagestyle{empty}

\newpage


\baselineskip=11.5pt
\parskip=1pt
\setcounter{tocdepth}{2}

\tableofcontents


\baselineskip=15pt
\parskip=3pt

\section{Introduction}
\label{sect:introduction}

Perhaps the most important, and remarkable, property of microstate geometries is that they provide the only known viable mechanism for supporting microstate structure at the horizon scale of a black hole \cite{Gibbons:2013tqa}. Indeed, such geometries are smooth and horizonless and yet very closely approximate a black-hole geometry while capping off just above where the horizon of the corresponding black hole would form.    As a result, such geometries can provide invaluable laboratories for studying putative descriptions of microstate structure at the horizon scale.  

The construction of microstate geometries has largely focussed on supersymmetric solutions in six dimensions but, even for these, the computations can be rather daunting.  Nevertheless, quite a number of physically interesting microstate geometries are ultimately rather simple.  Some can be reduced to   (2+1)-dimensional backgrounds that are best described as  smoothly capped BTZ geometries \cite{Bena:2016ypk,Bena:2017upb,Tyukov:2017uig,Bena:2017xbt,Bena:2018mpb,Bena:2019azk}.  That is, the geometry is asymptotic, at infinity, to AdS$_3$.  It then transitions to an  AdS$_2 \times S^1$ BTZ throat, and then, at some large depth (high red-shift), the $S^1$ starts to shrink again creating the smooth cap.  The cap region closely approximates a global AdS$_3$ geometry that has the same curvature radius as, but is highly redshifted relative to, the AdS$_3$ at infinity.  The throat between the two AdS$_3$ regions can be adjusted to create an arbitrarily high redshift. In some classes of these geometries, the massless scalar wave equation is even separable  \cite{Tyukov:2017uig,Bena:2019azk,Heidmann:2019zws,Walker:2019ntz}. This has enabled a great deal of analysis of such geometries and an investigation of their possible effects on microstructure  \cite{Tyukov:2017uig,Raju:2018xue,Bena:2018bbd,Bena:2018mpb,Bena:2019azk}. 

One of the purposes of this paper is to provide explicit, and far more extensive, families of microstate geometries.  We will lay out a range of relatively simple microstate geometries of the D1-D5-P black hole in Type IIB supergravity, each of which depends on a freely choosable holomorphic function of one variable.   This holomorphic function encodes the shape of momentum-wave excitation of the D1-D5 system and the different examples depend on how the wave moves on the angular directions of the AdS$_3 \times S^3$.

As one might expect, our new families of solutions arise from the superstratum construction.  Originally, the existence of superstrata was conjectured  \cite{Bena:2011uw} as the result of a double supertube transition.  It was argued that this transition would produce families of BPS solutions that depend upon arbitrary functions of (at least) two variables.  This construction was ultimately realized in \cite{Bena:2015bea} but the technical complexity limited the explicit computations to relatively simple families of Fourier modes.  Moreover, the construction in  \cite{Bena:2015bea}  focussed to ``angular momentum waves,'' or, more precisely, fluctuations whose momentum in the AdS$_3$ is exactly matched by a corresponding angular momentum on $S^3$.  As a result, these superstrata did not enter deeply into  the black-hole regime because the corresponding black hole would not have a large horizon. 

The breakthrough came in \cite{Bena:2016ypk,Bena:2017xbt} in which superstrata with pure momentum waves were constructed. These led to microstate geometries  that lie deep within the black hole regime and the six-dimensional geometry was that of a deformed $S^3$  fibered over a capped BTZ geometry.  

The full ten-dimensional solution lies within IIB supergravity and involves a compactification on $\IT^4$ or K3.  In this paper, as is standard for superstrata,  we will work with the six-dimensional $(1,0)$ supergravity, coupled to two tensor multiplets, that naturally results from such a compactification and is suggested by string scattering calculations\footnote{The universal sector of such a compactification is to consider fields that ``have no legs'' on the $\IT^4$ or K3, (apart from tensors that involve the volume form of $\IT^4$ or K3).  This results in precisely six-dimensional $(1,0)$ supergravity, coupled to two tensor multiplets.}\cite{Giusto:2011fy,Giusto:2012gt,Bena:2015bea}.  

When the pure momentum  waves are combined with the original superstratum modes on $S^3$, it strongly suggested that there should be superstrata that are arbitrary functions of {\it three variables}: basically built from the three sets of Fourier modes associated with the three compact $U(1)$ isometries of AdS$_3 \times S^3$.  There was, however, a potential obstruction to such superstrata in that the interactions of generic modes  might not lead to smooth solutions.  It was recently shown \cite{Heidmann:2019zws}  that a new,  broader classes of ``supercharged superstrata'' \cite{Ceplak:2018pws} can be used to obviate this obstruction and this opened the way to the construction of fully general superstrata as functions of three variables.

The holographically dual D1-D5 CFT also implies that there must be such generic superstrata.  The superstrata are dual to coherent superpositions of CFT states, $$ \bigotimes_{k,m,n,q} |k,m,n,q\rangle^{\rm NS}\,,  $$obtained by acting with the generators of the ``small,'' anomaly-free $\Neql{4}$ superconformal algebra on the ``length-$k$'' strands in the  NS-NS ground state of the D1-D5 CFT:
\begin{equation}
\begin{aligned}
 |k,m,n,q\rangle^{\rm NS}  ~=~ (J^+_0)^{m-1} (L_{-1})^{n-1}  \bigg(q \, G_{-\frac12}^{+1}G_{-\frac12}^{+2} + \Big(1 + \frac{q(1-k)}{k}\Big)\, J^+_0 L_{-1}\bigg)\, |O^{--}\rangle_{k} \,,
\end{aligned}
\label{SSCFTstates}
\end{equation} 
where $q=0,1$; $n-q \ge 0$, $k-q >0$ and $k \ge m  \ge q$.   For more details, see \cite{Ceplak:2018pws}. 

Encoding these states in  six-dimensional supergravity solutions starts with a very particular ansatz for the fundamental tensor gauge fields that requires these fields to  depend upon the phases: 
\begin{equation}
v_{k,m,n} ~\equiv~ (m+n) \frac{\sqrt{2}\,v}{R_y} + (k-m)\,\varphi_1 - m\,\varphi_2 \,.
 \label{phases}
\end{equation} 
The coordinates $(v,\varphi_1,\varphi_2)$, refer to the background  AdS$_3  \times S^3$ ``supertube'' geometry:
\begin{equation}
\begin{aligned}
d s^2_{6} ~=~  \sqrt{Q_1 Q_5}\, \bigg[  \, &   -\frac{ (r^2+ a^2)}{a^2 \, R_y }\,dt^2 ~+~  \frac{dr^2}{(r^2+ a^2)} ~+~  \frac{ r^2}{a^2 \, R_y }\,dy^2\\
&+ ~ d\theta^2 ~+~ \sin^2 \theta \, \bigg(d\varphi_1 - \frac{1}{R_y }\,dt\bigg)^2~+~ \cos^2 \theta \,\bigg(d\varphi_2 - \frac{1}{R_y }\,dy\bigg)^2   \bigg] \,.
\end{aligned}
\label{STmet}
\end{equation}
where 
\begin{equation}
  y ~\equiv~  y ~+~ 2\pi  R_y \,, \label{yperiod}
\end{equation}
and  $u$ and $v$, are the standard null coordinates:
\begin{equation}
  u ~=~  \coeff{1}{\sqrt{2}} (t-y)\,, \qquad v ~=~  \coeff{1}{\sqrt{2}}(t+y) \,. \label{tyuv}
\end{equation}
BPS solutions must be independent of  $u$, because it is the ``null time'' generated by the commutator of the supersymmetries. One then uses the ansatz for the tensor gauge fields as the starting point for solving the linear system of BPS equations.  As one solves the BPS equations, one then finds that the complete solution will generically depend non-trivially on combinations of phases of the form  (\ref{phases}).

The important point is that  a generic superposition of modes $(k,m,n,q)$  must encode a generic function of three variables.  The fact that one can take any superposition of the states (\ref{SSCFTstates}) strongly suggests\footnote{There may be some limit on the modes arising from the fact that the total length of the excited strands is limited by $N_1 N_5$.  However we expect that such issues will only arise when one quantizes the classical system.} that holography requires there to be no restriction on the function of three variables. The two choices of supercharge sector, $q=0,1$, actually means that there must be two independent arbitrary functions of three variables.  

Much of the study of superstrata to date has gone in the opposite direction in that it has focussed on ``single-mode superstrata,'' in which only one of the Fourier modes is non-trivial.  This produces an extremely simple class of solutions because the energy-momentum tensor only depends on the RMS value of the Fourier mode and so the geometry is independent of $(u,v,\varphi_1,\varphi_2)$.   Some of these geometries also have conformal Killing tensors \cite{Bena:2017upb,Heidmann:2019zws,Walker:2019ntz}.  

The single mode superstrata have the virtue of being simple and yet capture some of the very interesting physical properties of capped BTZ geometries.  On the other hand, such geometries are extremely coherent, highly atypical states and this leads to some behaviours, like sharp echoes \cite{Bena:2019azk}, that are very atypical for microstate structure.  While the states of the ``supergraviton gas,'' described via (\ref{SSCFTstates}) are still rather specialized\footnote{These states do not include twisted sector states of the CFT, and, in particular,  are not sufficiently numerous to account for the black hole entropy \cite{deBoer:1998us,Maldacena:1999bp,Shigemori:2019orj}.}, one might hope that  the corresponding superstrata may well start to exhibit somewhat more typical behaviour when it comes to the microstructure.  It is therefore important to make a systematic construction and analysis of more generic superstrata. 

In this paper, we will show that there is a natural ``toric'' description of superstrata in which one should think of the superstratum as being described by two {\it holomorphic functions} of three {\it complex} variables.  Two of these complex variables may be viewed as describing the $S^3$ in $\IC^2$, and thus encode the angle, $\theta$, while the third lies entirely on the AdS$_3$:
\begin{equation}
\chi ~\equiv~\frac{a}{\sqrt{r^2+ a^2}} \, \sin \theta \, e^{i \varphi_1} \,, \qquad \eta ~\equiv~\frac{a}{\sqrt{r^2+ a^2}} \, \cos \theta \, e^{i \big (\frac{\sqrt{2} v}{R_y} - \varphi_2\big)}\,, \qquad \xi ~\equiv~\frac{r}{\sqrt{r^2+ a^2}} \, e^{i \frac{\sqrt{2} v}{R_y} }\,. \label{cplxcoorddefn}
\end{equation}
This holomorphic formulation leads to huge simplifications in the form of the solutions and in the actual process of solving of the BPS equations.  

While we will not be able to solve the generic superstratum, we will show, by explicit construction that the holomorphic formulation makes it possible to construct families of solutions that depend on arbitrary holomorphic functions  of one complex variable.  These families are generated by picking specific powers for two of the complex variables and then allowing everything else to be an arbitrary function of the third variable.   While we will not give an explicit proof, our analysis makes it very plausible that the generic superstratum does indeed depend on two holomorphic functions of three variables.

In practice, solving the BPS equations involves introducing several sets of Fourier coefficients into the tensor gauge fields.  {\it A priori,} it seems that there might be several functions of three variables. However, one finds that smoothness of the solution requires  one to impose ``coiffuring constraints'' that reduce the solution to precisely two arbitrary functions of three variables.  These coiffuring constraints have always seemed a little mysterious, but in the holomorphic formulation they are simple and ``natural:'' they amount to requiring that certain fields only depend on the modulus-squared of the underlying holomorphic function (or its derivatives).   We will discuss this more in Section \ref{sect:superstrata}.

One of the other features of single-mode superstrata is that the effect of the momentum wave is smeared to its RMS value in the geometry and it is impossible to localize such a momentum wave.   Now that we have holomorphic waves, we can start with the \nBPS{4} AdS$_3 \times S^3$ geometry dual to the ground state of the D1-D5 CFT and study how a momentum wave that is much more localized at infinity  flows to the IR as it creates  the \nBPS{8}  geometry dual to the corresponding D1-D5-P state. We will give some explicit  examples of this  in Section \ref{sect: 10nExample}.

Finally, one of the classes of momentum waves that we consider here induces a variation in the D1-D5 charge density along the original \nBPS{4} supertube.  These  charge density variations follow, in supergravity, from smoothness of the solution.    Thus we find a new set of gravity predictions that can be used to test, once again, the precision holography of the D1-D5 CFT.  

Since one of our primary goals here is to exhibit families of new superstrata that prove broadly useful, we will defer some of the more detailed discussion of superstratum construction and start with the metrics of several interesting new families of black hole microstate geometries that are based on some form of holomorphic wave.  We will also extract some of the physically interesting details from them and highlight some features that can be compared with correlators in the holographically dual CFT.

In Section \ref{sect:metrics}, we give an overview of the metrics and our construction of the new superstrata.  Our goal is to highlight the important features of the process without getting into any of the more technical details.  In Section \ref{sect:Examples},  we catalog five families of superstrata, each of which involves a single holomorphic function of one variable.   A summary of these solutions may be found in Appendix \ref{app:compendium}.  In Section \ref{sect: 10nExample}, we use one of the simpler superstratum families to obtain some examples in which the momentum waves are localized around the $y$-circle.  

In Sections \ref{sect:superstrata}  and  \ref{sect:superstrata:complex}, we go into the  details of superstrata construction, first showing how the holomorphic structure arises and then recasting the BPS equations in terms of the new  holomorphic variables.  Section \ref{sect:conclusions} contains our conclusions. A more technically complicated ``hybrid superstratum'' that involves two independent holomorphic functions may be found in Appendix \ref{app:MoreEx}; the remaining Appendices (\ref{app:SphericalCoordinates} and \ref{app:Proof}) contain some important mathematical details that play a role in our work.

\section{The structure of  superstrata}
\label{sect:metrics}

Our purpose here is to give a broad overview of the construction of superstrata and exhibit some of the key physical and mathematical steps.  We have kept the technical details to a minimum so as to make it more accessible. A more complete exposition can be found in Sections \ref{sect:superstrata} and \ref{sect:superstrata:complex}.

The starting point for the construction  is the reduction of  IIB supergravity on $\IT^4$ or K3.  The universal sector of such a compactification consists of restricting to fields that ``have no legs'' on the $\IT^4$ or K3 (apart from tensors that involve the volume form of $\IT^4$ or K3).  This results in six-dimensional $(1,0)$ supergravity, coupled to two tensor multiplets.   It is this theory that is the ``work horse'' of superstratum construction.

\subsection{BPS solutions in six dimensions}
\label{sect:BPS6D}

The  \nBPS{8} solutions of this particular supergravity theory have been extensively studied. First, the six-dimensional metric must take the form  of a fibration over a four-dimensional base, $\cB$ \cite{Gutowski:2003rg,Cariglia:2004kk}: 
\begin{equation}
\label{sixmet}
d s^2_{6} ~=~-\frac{2}{\sqrt{\cP}}\,(d v+\beta)\,\Big[d u+\omega + \frac{\mathcal{F}}{2}(d v+\beta)\Big]+\sqrt{\cP}\,d s^2_4\,,
\end{equation}
where the  metric, $d s^2_4$, on $\cB$ is ``almost hyper-K\"ahler,'' $\beta$ and $\omega$ are one-forms on $\cB$, while $\cF$ and $\cP$ are functions.  The vector field $\frac{\partial}{\partial u}$ is the Killing vector required by supersymmetry and, in principle, everything can depend on the other five coordinates.

Supersymmetry requires that the three tensor gauge fields (one lies in the gravity multiplet) are determined by three pairs
\begin{equation}
\label{EMpieces}
(Z_1, \Theta_2) \,, \qquad (Z_2, \Theta_1)  \,, \qquad (Z_4, \Theta_4)  \,,
\end{equation}
where the $Z_I$ are functions and the $\Theta_J$ are two-forms whose components lie only on $\cB$ \footnote{The labelling may seem a little anomalous but this reflects the underlying $SO(2,1)$ structure.  The absence of $(Z_3, \Theta_3)$ is a historical artifact: In five dimensions one gets another gauge field, whose pieces, $(Z_3, \Theta_3)$, come from $\cF$ and $d \beta$}.  These may be viewed as the electrostatic potentials and magnetic components of the tensor gauge fields.  We will not need the details of the tensor fields here and we refer the reader to \cite{Bena:2011dd,Bena:2015bea,Bena:2017geu,Bena:2017xbt} and \cite[Appendix E.7]{Giusto:2013rxa}  for  details. It is also convenient to define
 \begin{equation}
\Theta_3 ~\equiv~  d \beta    \,.
 \label{Theta3}
\end{equation}

The quantities, $Z_I$ and $\Theta_J$, are  extremely important to solving the BPS conditions and will thus be essential elements to our discussion.  We will describe the BPS equations in detail in Sections \ref{sect:superstrata} and \ref{sect:superstrata:complex}, where we will show how the  $Z_I$ and $\Theta_J$ are determined by a ``first layer'' of  equations and how these solutions then provide sources for the ``second layer'' of BPS equations that determine $\cP$, $\cF$ and $\omega$.  In particular, the BPS equations require the warp factor, $\cP$, to be related to the electrostatic potentials via:
\begin{equation}
\cP   ~=~     Z_1 \, Z_2  -  Z_4^2 \,.
\label{Pform}
\end{equation}
%

\subsection{The supertube and the superstratum}
\label{sect:stss}

To construct superstrata one makes several assumptions that greatly simplify the BPS equations.  First, one takes 
 the four-dimensional base to be flat  $\IR^4$ written in spherical bipolar coordinates:
 \begin{equation}
 d s_4^2 ~=~ \Sigma \, \left(\frac{d r^2}{r^2+a^2}+ d\theta^2\right)+(r^2+a^2)\sin^2\theta\,d\varphi_1^2+r^2 \cos^2\theta\,d\varphi_2^2\,,
 \label{ds4flat}
\end{equation}
where $\Sigma$ is the distance from the supertube locus
 \begin{equation}
\Sigma~\equiv~  r^2+a^2 \cos^2\theta     \,.
 \label{Sigdefn}
\end{equation}
One also takes the connection, $\beta$, to be:
\begin{equation}
\label{betadefn}
\beta ~\equiv~  \frac{R_y\, a^2}{\sqrt{2}\,\Sigma}\,(\sin^2\theta\, d\varphi_1 - \cos^2\theta\,d \varphi_2) \,.
\end{equation}

The pure D1-D5 system  forms the canvas upon which the superstrata will be constructed.  This is given by setting
\begin{equation}
Z_1 ~=~ \frac{Q_1}{\Sigma}  \,, \qquad Z_2 ~=~ \frac{Q_5}{\Sigma} \,, \qquad   Z_4 ~=~ 0  \,, \qquad \Theta_I ~=~ 0  \,, \qquad  \cF ~=~ 0 \,, \qquad    \omega ~=~ \omega_0  \,,
\label{STcomps}
 \end{equation}
where
\begin{equation}
 \omega_0 ~\equiv~  \frac{a^2 \, R_y \, }{ \sqrt{2}\,\Sigma}\,  (\sin^2 \theta  d \varphi_1 + \cos^2 \theta \,  d \varphi_2 ) \,.
\label{angmom0}
\end{equation}
One therefore has
\begin{equation}
\cP ~=~ \frac{Q_1 Q_5}{\Sigma^2}   \,.
\label{Pst}
 \end{equation}
Using this, the metric (\ref{sixmet}) is smooth, and reduces to (\ref{STmet}), provided that one imposes the supertube regularity condition:
\begin{equation} 
Q_1Q_5 ~=~R_y^2 \,  a^2  \,.
\label{STreg}
\end{equation} 
This defines the holographic dual of the maximally-spinning, \nBPS{4}, RR-vacuum of D1 and D5 branes wrapped on the common $S^1$ along the $y$-direction. 
 
Superstrata are the holographic duals of the  \nBPS{8} states obtained by exciting the left-moving states (\ref{SSCFTstates}) on top of the present RR vacuum solution.  In supergravity, this means turning on non-trivial excitations in almost all of the fields and allowing these excitations to depend on $(r,\theta,v,\varphi_1,\varphi_2)$.

\subsection{The complex formulation of superstrata}
\label{sect:cplx}

As we will show in much of the rest of this paper, the way in which superstrata depend on $(r,\theta,v,\varphi_1,\varphi_2)$ is most simply encoded in 
the complex coordinates:
 \begin{equation}
\xi ~\equiv~\frac{r}{\sqrt{r^2+ a^2}} \, e^{i \frac{\sqrt{2} v}{R_y} } \,, \quad \chi ~\equiv~\frac{a}{\sqrt{r^2+ a^2}} \, \sin \theta \, e^{i \varphi_1} \,, \quad \eta ~\equiv~\frac{a}{\sqrt{r^2+ a^2}} \, \cos \theta \, e^{i \big (\frac{\sqrt{2} v}{R_y} - \varphi_2\big)} \,.
\label{cplxcoords} 
\end{equation}
It is also  useful  to define
\begin{equation}
\mu ~\equiv~ \eta/\chi ~=~\cot \theta \, e^{i \big (\frac{\sqrt{2} v}{R_y} - \varphi_1- \varphi_2\big)}  \,. \label{cplxcoords2}
\end{equation}
Note that in these coordinates: 
 \begin{equation}
   \chi^{k-m} \, \eta^m \, \xi^n ~=~\chi^{k}\mu^{m}\xi^{n}~=~   \Delta_{k,m,n} \,e^{i v_{k,m,n}}       \,.
 \label{coordpowers}
\end{equation}
where $v_{k,m,n}$ is defined in (\ref{phases}) and
 \begin{equation}
 \Delta_{k,m,n} ~\equiv~  \bigg(\frac{r}{\sqrt{r^2+a^2}}\bigg)^n  \, \bigg(\frac{a}{\sqrt{r^2+a^2}}\bigg)^k  \, \sin^{k-m}\theta \,  \cos^{m}\theta\,.
  \label{Deltadefn}
 \end{equation}
The coordinates $(\chi,\eta,\xi)$ are not all independent in that they satisfy the constraint:
 \begin{equation}
  |\chi|^2  ~+~ |\eta|^2 ~+~ |\xi|^2 ~=~   1  \,,
 \label{constraint}
\end{equation}
and so they sweep out a hemisphere  $S^5$.  Indeed, the limit $r \to \infty$ corresponds to circle defined by $\chi =  \eta = 0$.  This is the $S^1$ swept out by $y$ while the $S^3$ at infinity in  $ds^2_4$ corresponds to the sphere $ |\chi|^2  + |\eta|^2 = \epsilon^2$ as $\epsilon \to 0$.

The most general hybrid superstratum involves two holomorphic functions, $G_1$ and $G_2$, of three variables \cite{Heidmann:2019zws}. It is convenient to think of them in terms of their Taylor expansions with Fourier coefficients $b_{k,m,n}$ and $c_{k,m,n}$:

\begin{equation}
G_1(\xi,\chi,\eta) ~\equiv~   \sum_{k,m,n} \,b_{k,m,n} \, \xi^n \,  \chi^{k-m} \, \eta^m \,, \qquad G_2(\xi,\chi,\eta)  ~\equiv~   \sum_{k,m,n} \,c_{k,m,n} \, \xi^n \,  \chi^{k-m} \, \eta^m \,,
\label{holfns}
\end{equation}
The function $G_1$ encodes superposition of states with $q=0$ in (\ref{SSCFTstates}), the original superstratum modes \cite{Bena:2016ypk,Bena:2017xbt}, while $G_2$ encodes superposition of states with $q=1$, the supercharged modes \cite{Ceplak:2018pws}. Taking both functions to be 0 corresponds to the non-excited supertube solution detailed in the previous section.

As we will see below,  the warp factor, $\cP$, in the metric only depends on the charges and $|G_1|$:
\begin{equation}
\cP   ~=~     \frac{1}{\Sigma^2}\,\bigg(Q_1 Q_5 ~-~  \frac{R_y^2}{2}\, |G_1|^2\,\bigg)   \,.
 \label{cPhol} 
\end{equation}
%

\subsection{Regularity of superstrata}
\label{sect:regss}

Fundamentally, the superstratum is required to be regular and have the correct asymptotics at infinity.  In particular, the metric (\ref{sixmet}) must be smooth, must not have any closed timelike curves (CTC's)  and be asymptotic to AdS$_3 \times S^3$.  Since the construction of superstrata involves solving, on $\cB$, systems of linear differential equations with sources, one must first find inhomogeneous solutions to solve for the sources and then allow the possible addition of homogeneous solutions.   As one would expect, non-trivial homogeneous solutions either have singular sources on $\cB$, or diverge at infinity.   This means that regularity requires the careful choice of homogeneous solutions combined with constraints on the parameters, akin to  (\ref{STreg}). 

If one starts from the differential equations that determine  (\ref{EMpieces}), one expects that each of the pairs $(Z_I, \Theta_J)$ to have independent solutions.  However, very early in the superstratum program, it was found that if these flux components remain  independent then the  second layer of BPS equations that determine the metric constituents, $\cP$, $\cF$ and $\omega$, would necessarily have singular solutions.  It was shown in \cite{Bena:2015bea, Bena:2016agb} that if these fluxes were all related via ``coiffuring''  constraints then these particular singularities would be removed. Physically, these coiffuring constraints  were motivated by the structure of the dual CFT and are loosely motivated by horizon smoothness issues in black rings \cite{Bena:2014rea}, but mathematically they seemed rather {\it ad hoc}.

We will describe in Section \ref{sect:superstrata}, how  ``coiffuring''  amounts to requiring that the metric constituents, $\cP$, $\cF$ and $\omega$, are always constructed from sesquilinear pairings of the $G_j$ and ${\overline G}_k$.  That is, $\cP$, $\cF$ and $\omega$ can all be expressed as quadratics in  $G_j, {\overline G}_k$, and their primitives or derivatives, but these quadratics are linear in each of the $G_j$ and ${\overline G}_k$  (and their primitives or derivatives) considered separately.   Indeed, the fact that $\cP$ only depends on $|G_1|^2$ is the simplest possible such form.  In Section  \ref{sect:Examples}, Appendix \ref{app:compendium}  and Appendix \ref{app:MoreEx} we will see more complicated sesquilinear pairings of the $G_j$ and ${\overline G}_k$.

While the homogeneous solutions depend on the particular superstratum, there is a universal and important family that reflects the residual ability to re-define the coordinate $u$:
\begin{equation} 
u~\to~  u ~+~ f(x^i,v) \qquad \Leftrightarrow \qquad  \omega ~\to~  \omega ~-~ d_4f + (\partial_v f) \,\beta \,,\quad \mathcal{F} ~\to~ \mathcal{F} ~-~ 2\, (\partial_v f)\,,
\label{gaugetrf}
\end{equation}
where $d_4$ is the exterior derivative on the $\IR^4$ base with metric  (\ref{ds4flat}).  It is obvious from (\ref{sixmet})  that such a transformation of $u$  is equivalent to the shifts in $\mathcal{F}$ and $\omega$.   We will use this ``gauge transformation'' frequently.  In particular, we note that such transformations can be used to cancel oscillatory parts of $\mathcal{F}$.  We also note that such gauge transformations are not necessarily holomorphic.

In practice, checking the regularity of a superstratum solution boils down to removing obvious singularities and focussing on ``dangerous places'' at which CTC's might occur.  These dangerous places are where the $\varphi_1$ and $\varphi_2$  circles pinch off ($\theta =0$ and $\theta = \frac{\pi}{2}$ respectively) in  (\ref{ds4flat}), where the whole S$^3$ pinches off ($r=0,$ $\theta=0$) and yet $\omega$ may not vanish in these directions, thereby creating a CTC in  (\ref{sixmet}). It  usually suffices to look at disk defined by $r=0$ in order to check these issues.  

The locus   $r=0$ and $\theta =\frac{\pi}{2}$ is the original supertube locus, on which one has $\Sigma=0$.  This is where we are now creating the momentum waves and so it is the locus that generates the most significant physical constraints.  Indeed, regularity on this locus usually leads to a constraint between the charges that generalizes  (\ref{STreg}). 

 Finally, one must carefully examine the $y$-circle for CTC's and smoothness at the center of the cap.   
 
 Usually the best and most efficient way to examine the regularity of a superstratum metric  is to write it as an $S^3$ fibration over a three-dimensional base, $\cK$.  That is, one collects all the terms involving $d \theta$, $d \varphi_1$ and $d \varphi_2$ and completes the squares for all the mixed terms between these differentials and $dt, dy$ and $dr$.  This defines the metric on the $S^3$ fiber.   One then collects the residual parts of the metric, which only involve $dt, dy$ and $dr$, and this defines the metric on the base, $\cK$. One then completes the square in $dy$ on $\cK$.  This makes it easier to verify that the compact directions have a regular metric without CTC's\footnote{This does not necessarily mean that the metric in the compact directions needs to be positive-definite everywhere: see the discussion in section \ref{10n-y-circle}.} and that the entire metric has the correct asymptotics at infinity and is well-behaved in the cap.

To summarize, we typically go  through the check-list
\begin{itemize}
\item[(i)] Adjust the gauge and homogeneous solutions so that the metric is asymptotic to AdS$_3\times$S$^3$.
\item[(ii)] Use homogeneous solutions to remove any singularities in $\cF$ and $\omega$ at $r=0$.
\item[(iii)] Make sure that the warp factor, $\cP$, in (\ref{cPhol}) is always positive.
\item[(iv)] Examine the potential singularities and CTC's at the center of $\IR^4$:   $r=0$, $\theta = 0$.
\item[(v)] Look at the supertube locus: $r = 0$ and $\theta = \frac{\pi}{2}$ where circles pinch off in (\ref{ds4flat})  and eliminate singularities and CTC's in (\ref{sixmet}).  
\item[(vi)]  Make sure that the $y$-circle has no CTC's and caps off smoothly.
\item[(vii)] Make sure there are no  CTC's elsewhere.
\end{itemize}
Item (iii) is generically not a problem:  in the explicit examples we study in Section \ref{sect:Examples}, $\cP>0$ follows from judicious application of the Cauchy-Schwarz inequality.   The  CTC constraints are usually more stringent than the positivity of $\cP$.   The last item, while essential, is usually very hard to prove rigorously and is often done by looking at numerical sections of the metric.

Finally, there is an important issue that can arise in making gauge transformations: There are two canonical  choices of the $u$-coordinate.  First, one would like the metric to be asymptotic to AdS$_3 \times S^3$ in its standard global form (\ref{STmet}) in which $u$ and $v$ are related to $t$ and $y$ via   (\ref{tyuv}).  Making such a choice of gauge for $u$ enables us to read off all the conserved charges with ease. Thus we will often have to make a gauge transformation of our solutions to achieve this end. On the other hand, there is another canonical meaning for $y$:  If the complete metric is written as an $S^3$ fibration over $\cK$, then the ``y-circle'' is canonically thought of as the circle that pinches off at $r =0$ in this three-dimensional space time.   Sometimes the re-definitions of $u$ that yield an asymptotically AdS$_3 \times S^3$ geometry can upset the identification of the $y$ circle as the one that pinches off at $r=0$.  

One can address this issue by simply using two coordinate charts: one around $r=0$ and the other at large values of $r$.  One then has a non-trivial  re-definition of $u$ in going between these charts.  Alternatively, one can make a gauge choice that depends upon $r$ in such a way that it has the desired effect at infinity while vanishing at $r=0$.  The price here is a non-zero radial component in $\omega$.  We will see an example of this in Section \ref{sect:10ni}.

\subsection{Conserved charges}

Superstrata have five conserved quantized charges: the number of each type of brane, $N_1$, $N_5$, the momentum, $N_P$, and the two angular momenta on $\cB$, $j_L$ and $j_R$. Reading off the supergravity charges is then an essential part of the analysis of the solution. It is a fairly standard procedure but we summarize it here. 

The D1 and D5 supergravity charges, $Q_1$ and $Q_5$, are given by the $r^{-2}$ terms in the large $r$ expansion of the gauge fields $Z_1$ and $Z_2$.

The angular momenta are obtained from the asymptotic form of $\omega$ and $\beta$\footnote{The purely oscillatory terms do not contribute since they average out in the Komar integrals for the conserved quantities.}:
\begin{equation}
\beta_1+\beta_2 +\omega_1 + \omega_2~=~ \frac{ \sqrt{2}}{r^2}\, \big[ \, (J_R - J_L \cos 2\theta )  ~+~ {\it oscillating \ terms}\, \big] ~+~ \cO(r^{-4}) 
\label{asmpmoms}
\end{equation} 
where $ \beta_1, \omega_1$ and $ \beta_2, \omega_2$ are the components along $d\varphi_1$ and $d\varphi_2$ respectively. 

\noindent The momentum charge is given by the RMS value of the expansion of $\cF$ at infinity: 
\begin{equation}
\cF  ~=~ - \frac{1}{r^2} \,\big(2\, Q_P  ~+~{\it oscillating \ terms}   \big) ~+~ \cO(r^{-4})   \,.
\label{Fexp}
\end{equation}

Finally, the supergravity  charges are related to the quantized charges via \cite{Bena:2015bea}:
\begin{equation}
Q_1 ~=~  \frac{(2\pi)^4\,N_1\,g_s\,\alpha'^3}{V_4}\,,\quad Q_5 = N_5\,g_s\,\alpha'\,, \quad Q_P   ~=~    \frac{ (2\pi)^4 \,g_s^2 \,\alpha'^4}{V_4\, R_y^2}\, N_P \,,\quad J_{L, R} ~=~  \frac{(2\pi)^4 g_s^2 \alpha'^4}{V_4\,R_y}\, j_{L, R}\,,
\label{Q1Q5_N1N5a}
\end{equation}
where $V_4$ is the volume of $\IT^4$ in the IIB compactification to six dimensions.  It  is convenient to define $\cN$ via: 
\begin{equation}
\cN ~\equiv~ \frac{N_1 \, N_5\, R_y^2}{Q_1 \, Q_5} ~=~\frac{V_4\, R_y^2}{ (2\pi)^4 \,g_s^2 \,\alpha'^4}~=~\frac{V_4\, R_y^2}{(2\pi)^4 \, \ell_{10}^8} ~=~\frac{{\rm Vol} (\IT^4) \, R_y^2}{ \ell_{10}^8} \,,
\label{Q1Q5_N1N5b}
\end{equation}
where $\ell_{10}$ is the ten-dimensional Planck length and  $(2 \pi)^7 g_s^2 \alpha'^4  = 16 \pi G_{10} ~\equiv~ (2 \pi)^7 \ell_{10}^8$. The quantity, ${\rm Vol} (\IT^4)  \equiv (2\pi)^{-4} \, V_4$, is sometimes introduced \cite{Peet:2000hn} as a ``normalized volume'' that is equal to $1$ when the radii of the circles in the $\mathbb{T}^4$ are equal to $1$. The quantized charges are then more succinctly related to the supergravity charges via:
\begin{equation}
N_1\, N_5 ~=~ \cN \, \frac{Q_1 \, Q_5}{R_y^2}\,,  \qquad   N_P ~=~ \cN \, Q_P  \,,  \qquad   j_{L,R} ~=~ \cN \,  \frac{J_{L,R}}{R_y}    \,.
\label{quantch}
\end{equation}
%

\section{Some examples}
\label{sect:Examples}

Here we focus on the metric structure of several families of superstrata, that is, we focus on the functions $\cF$, $\cP$ and on the one-forms $\beta$ and $\omega$. We will not derive the expressions for the gauge fields $(Z_I,\Theta_J)$ here; they can be easily derived from the formula \eqref{1stLayerGenHolo} detailed in Section \ref{sect:superstrata:complex}. Moreover, a complete, self-contained summary of these solutions is given in Appendix \ref{app:compendium}; additional families are discussed in Appendix  \ref{app:MoreEx}.   

 It  turns out that the supercharged solutions ($q=1$ in (\ref{SSCFTstates})) are much more complicated than their non-supercharged analogues ($q=0$).  Therefore, for the sake of readability,  we will only consider non-supercharged excitations in this section.  That is, we will set the holomorphic function, $G_2$, in  (\ref{holfns}) to zero.  An example  with both supercharged and non-supercharged excitations is given in Appendix \ref{app:MoreEx}.  We will examine our first example in great detail, and be  somewhat more cursory in the study of the other examples. 

We construct solutions with arbitrary functions of one of the three complex variables \eqref{cplxcoords} and give examples in each class. Functions of $\xi$, $\chi$ or $\eta$  encode very different features. Indeed, $\xi$ does not depend on the S$^3$ angles and lies entirely on the AdS$_3$ whereas $\chi$ and $\eta$ mix the two parts but $\chi$ is more centered around the supertube locus ($\theta=\frac{\pi}{2}$) and $\eta$ is more centered around the center of the $\mR^4$ ($\theta=0$). For solutions with an arbitrary function of $\xi$, we construct the superposition of ($1,0,n$), ($1,1,n$) and ($2,1,n$) superstratum modes in the three next sections. For solutions with an arbitrary function of $\chi$, we construct the superposition of ($k,0,1$) modes in Section \ref{sect:ki01} and for solutions with an arbitrary function of $\eta$, we construct the superposition of ($k,k,0$) superstrata in Section \ref{sect:kiki0}.

The two last examples have an interesting feature in that  their regularity requires a non-trivial adjustment of the electrostatic potential $Z_{1}$ and magnetic two-form flux $\Theta_2$.  This is the supergravity dual of the development of a particular CFT vev that has been described in \cite{Giusto:2015dfa}.

\subsection{The $(1,0,n)$ superstrata}
\label{sect:10ni}

The single-mode $(1,0,n)$ superstratum was first constructed in \cite{Bena:2016ypk} and has been well-studied since \cite{Bena:2017upb,Tyukov:2017uig,Bena:2017xbt,Raju:2018xue,Bena:2018mpb,Heidmann:2019zws,Bena:2019azk,Walker:2019ntz}.  For $k=1$ and $m=0$ there are no supercharged modes in the CFT and the corresponding supergravity solutions are singular if one tries to incorporate them \cite{Ceplak:2018pws}.  Thus $G_2 $ must necessarily be set to zero.
The underlying holomorphic function, $G_1$, is given by the most general superposition of the $(1,0,n)$ modes determined by an arbitrary function $F$ fluctuating with the complex variable $\xi$ \eqref{cplxcoords}:
\begin{equation}
G_1(\xi,\chi,\eta)   ~=~    \chi \, F(\xi) ~=~\frac{a\,F(\xi)}{\sqrt{r^2+ a^2}} \, \sin \theta \, e^{i \varphi_1} \,,  \qquad\quad F(\xi) ~=~ \sum_{n=1}^\infty \, b_n \, \xi^n \,.
 \label{G10n} 
\end{equation}
For simplicity, we will assume that the $b_n$ are real. (One can easily generalize to allow for complex $b_n$.) 

\noindent The connection one-form $\beta$ is identical to the one of the seed supertube solution \eqref{betadefn}. The warp factor of the six-dimensional metric \eqref{sixmet} is now given by:
\begin{equation}
\sqrt{\cP}   ~=~    \frac{1}{\Sigma}\,\sqrt{Q_1 Q_5 ~-~  \frac{a^2 R_y^2}{2\,(r^2+a^2) }\, |F|^2\, \sin^2 \theta}   \,,
 \label{cP10n} 
\end{equation}
where $\Sigma$ is given in \eqref{Sigdefn}. The BPS equations then lead to the solutions:
\begin{equation}
\cF  ~=~   \frac{1}{a^2} (  |F|^2 - c)  \,, \qquad \omega  ~=~ \omega_0 ~+~ \frac{R_y}{\sqrt{2} \, \Sigma}\,   (c-|F|^2 )\, \sin^2 \theta \,d\varphi_1  \,,
 \label{Fom10n} 
\end{equation}
where $c$ is a constant and $\omega_0$ is given by (\ref{angmom0}). (The constant, $c$, and the one-form, $\omega_0$, are, in fact, homogeneous solutions of the BPS equations.)

\subsubsection{Gauge transformations}
\label{sect:hom-gauge}

To get an asymptotically AdS metric we must arrange that $\cF$ vanish at infinity.  The easiest way to achieve this is to define 
\begin{equation}
F_\infty (v)  ~\equiv~    \lim_{|\xi| \to 1}  \, F(\xi) ~=~    \lim_{r  \to \infty}  \, F(\xi)    \,,
 \label{Finfty} 
\end{equation}
and set 
\begin{equation}
c  ~=~   \frac{1}{{\sqrt{2}\,\pi R_y} }\, \int_0^{\sqrt{2}\,\pi R_y}  \, | F_\infty (v)|^2  \, dv\,.
 \label{cval10n} 
\end{equation}
This removes the mean-square value of $|F_\infty (v)|$ from $\cF$ at infinity.  The difference, $| F_\infty (v)|^2 - c$, is then purely oscillatory, and so we may define a periodic function, $f$, via:
\begin{equation}
f  ~\equiv~ \frac{1}{2\,a^2} \,  \int^v \Big[ \, | F_\infty (v)|^2 ~-~ c \,\Big]    \,  dv\,, \qquad \partial_v f ~=~ \frac{1}{2\,a^2} \, \Big[\, | F_\infty (v)|^2 ~-~ c \,\Big]  \,.
 \label{gtrf10n} 
\end{equation}
Using  $f$ to make  a gauge transformation, as in (\ref{gaugetrf}), one arrives at:
\begin{align}
 \label{F-gauged-10n} \mathcal{F} &= \frac{1}{a^2} (|F|^2 - |F_\infty|^2),\\
  \omega &= \left(1 - \frac{1}{2a^2}(|F_\infty|^2-c)\right)\omega_0 + \frac{R_y}{\sqrt{2}\Sigma} (|F_\infty|^2 - |F|^2)\sin^2\theta d\varphi_1 . \label{om-gauged-10n} 
\end{align}
These functions now fall off as $\frac{1}{r^2}$ at infinity and the non-oscillating terms yield the  asymptotic charges.   One can remove the oscillating terms in $\frac{1}{r^2}$ in $\cF$ using another gauge transformation at the cost of adding some oscillating $\frac{dr}{r^3}$ terms to $\omega$.

It may seem somewhat surprising that this solution has oscillating terms that fall off rather weakly at infinity.  The root cause of this is that this solution contains some magnetic fluxes for the $B$-field of IIB supergravity that fall off very slowly at infinity. This is a peculiarity of taking $k=1$.  Higher values of $k$ result in stronger fall-off in the field through the stronger fall off in $\Delta_{k,m,n}$ in (\ref{Deltadefn}).

We also note that the gauge transformation  (\ref{gtrf10n}) makes a non-trivial re-definition of $u$ that persists all the way to $r=0$.  In particular, $\cF$, in  (\ref{F-gauged-10n}) limits to a very non-trivial oscillating function, $- \frac{1}{a^2}\,| F_\infty (v)|^2$, at $r=0$.   To get a simple smooth cap, one must undo this re-definition of the $u$ coordinate.   Alternatively, one can use a modified gauge transformation, like the one defined by:
\begin{equation}
 \partial_v \hat f ~\equiv~ \frac{1}{2\,a^2} \, |\xi|^p \Big[\, c ~-~ | F_\infty (v)|^2 \,\Big]  \,,
 \label{gtrf10n-b} 
\end{equation}
for any suitably large $p$. Since $|\xi| \to 1$ as $r \to \infty$ and $|\xi| \sim r$ as $r \to 0$, this reduces to the gauge transformation   (\ref{gtrf10n}) at infinity and yet vanishes in the cap. Thus we get the simple, global AdS$_3 \times S^3$ metric at infinity and we preserve the standard form of the metric in the cap.  The only other consequence is that we also get a radial component of the angular momentum vector, $\omega_r$, proportional to  $\partial_r \hat f$.  This makes only sub-leading corrections to the metric in both limits.

\subsubsection{Regularity}
\label{sect:10nreg}

The metric (\ref{sixmet}) potentially has singularities at $\Sigma =0$ and may also have closed time-like curves (CTC's).  The function $\sqrt{\mathcal{P}}$ has a factor of $\Sigma^{-1}$ and this means that the  singular terms 
can only arise in the metric in the $(\varphi_1, \varphi_2)$ directions.   Indeed, we find that the only dangerous component is $d\varphi_1^2$ whose coefficient is   
 \begin{equation}
-  \frac{\sqrt{2}\, R_y\, a^2}{\sqrt{\mathcal{P}} \,\Sigma}\, \sin^2\theta\, \bigg(\omega_1 ~+~  \frac{R_y\, a^2\, \cF}{2\,\sqrt{2}\,\Sigma}\, \sin^2\theta \bigg)~+~ \sqrt{\mathcal{P}} \, (r^2+a^2)\sin^2\theta\,\,,
 \label{danger1}
\end{equation}
In the solution (\ref{cP10n}), (\ref{F-gauged-10n})-(\ref{om-gauged-10n}) above, this reduces to
 \begin{equation}
 \frac{1}{ \Sigma}  \, \bigg(Q_1 Q_5  ~-~  \frac{a^2 R_y^2}{2\,(r^2+a^2) }\, |F|^2\, \sin^2 \theta \bigg)^{-
\frac{1}{2}} \, 
\Big[  Q_1 Q_5\, (r^2+a^2) ~-~    a^2 R_y^2 \,(a^2 +  \coeff{1}{2}\,c )\, \sin^2\theta \, \Big] \,  \sin^2\theta\,.
 \label{danger10n}
\end{equation}
To remove the singularity at $r=0, \theta =\frac{\pi}{2}$ one must impose the condition:
 \begin{equation}
\frac{Q_1 Q_5}{R_y^2}  ~=~  a^2 +  \frac{1}{2}\,c ~=~ a^2 +\frac{1}{{2\sqrt{2}\,\pi R_y} }\, \int_0^{\sqrt{2}\,\pi R_y}  \, | F_\infty (v)|^2 ~=~a^2 ~+~  \frac{1}{2}\,  \sum_{n=1}^\infty \, b_n^2\,,
 \label{reg10n}
\end{equation}
where we have used the Taylor expansion  (\ref{G10n}).  This is the generalization of (\ref{STreg}) to the $(1,0,n)$ family of  superstrata. 

The coefficient of $d\varphi_1^2$ in the metric then reduces to:  
 \begin{equation}
\bigg(Q_1 Q_5  ~-~  \frac{a^2 R_y^2}{2\,(r^2+a^2) }\, |F|^2\, \sin^2 \theta \bigg)^{-
\frac{1}{2}} \, Q_1 Q_5\, \sin^2\theta = (\mathcal{P}\Sigma^2)^{-1/2} \sin^2\theta  \,,
 \label{metcoeff1}
\end{equation}
which is smooth and positive provided that $\mathcal{P}\Sigma^2>0$ everywhere. Using the holomorphicity of $F$, one can prove rigorously that this is always true for any $F$. We refer the interested reader to the Appendix \ref{app:ProofP10n} for all the details.

\subsubsection{CTC's and the $y$-circle}
\label{10n-y-circle}
 
To check for closed time-like curves along all the circle directions, it is simplest to set $dt=dr=d\theta=0$ and extract the residual metric.  It is also simplest to work in the coordinate gauge where the solution is given by  (\ref{Fom10n}).   One then completes the squares  in the metric so as to reveal the potentially dangerous curves.  The result may be written as:
\begin{align}\nonumber
 \mathcal{P}^{-1/2} & \left(  \frac{R_y^2|F|^2}{2\Sigma}\frac{r^2}{r^2+a^2} \cos^2\theta d\varphi_2^2  ~+~  \frac{Q_1Q_5}{\Sigma}\sin^2\theta d\varphi_1^2 \right.\\
 & \left.~+~  \frac{R_y^2}{2a^2\Sigma} \left( \frac{2\,Q_1Q_5}{R_y^2} - |F|^2\right)\left[ \frac{r^2}{R_y^2} dy^2   + a^2\cos^2\theta\left(d\varphi_2-\frac{dy}{R_y}\right)^2\right]    \,\right) \,,
  \label{redmet1}
 \end{align}
%
where we have used  (\ref{reg10n}) to eliminate the constant, $c$.  This is positive definite if and only if 
\begin{equation}
 |F(\xi)|^2 ~<~  \frac{2\,Q_1Q_5}{R_y^2} \,.
 \label{CTCcond1}
  \end{equation}
We will return to this condition momentarily.

It is also instructive to examine the $y$-circle by writing (\ref{sixmet}) in terms of an $S^3$ fibration over a base defined by $(t,y,r)$. For the reduced metric,   (\ref{redmet1}), this means  completing the squares in the $d\varphi_i$ and leaving a residual pure $dy^2$ term.  If one also restores the $dr^2$ term, the result is:
\begin{equation}
  \frac{\Sigma \sqrt{\mathcal{P}} } {(r^2+a^2)} \, dr^2 ~+~   \frac{1} {2a^2 } \, \bigg(Q_1Q_5 - \frac{a^2 \, R_y ^2}{2\,(r^2 +a^2)} \,|F(\xi)|^2 \bigg)^{-1} \,\bigg( \frac{2\,Q_1Q_5}{R_y^2} - |F(\xi)|^2\bigg) \, \Sigma \, \sqrt{\cP}\,r^2 \, dy^2 \,.
  \end{equation}
 Since $F$ vanishes as $r \to 0$ and $\Sigma^2 \cP \to Q_1Q_5 $, this expression behaves as:
\begin{equation}
\frac{\sqrt{Q_1Q_5}}{a^2 } \,\bigg( dr^2 ~+~  \frac{dy^2}{R_y^2} \bigg) \,,
 \label{dysq}
  \end{equation}
which means the $(r,y)$-metric caps off smoothly.

Returning to the condition  (\ref{CTCcond1}), we first note that this is satisfied if
\begin{equation}
 |F(\xi)|^2 ~\leq~  c ~=~ \frac{2\,Q_1Q_5}{R_y^2} - 2 a^2 ~=~   \frac{1}{{\sqrt{2}\,\pi R_y} }\, \int_0^{\sqrt{2}\,\pi R_y}  \, | F_\infty (v)|^2  \, dv ~=~  \sum_{n=1}^\infty\, b_n^2 
 \label{CTCcond2}
  \end{equation}
If this condition is not satisfied then there is a danger that $y$ circles might lead to CTC's.  However, we do not need to require this inequality to be satisfied everywhere;  rather, we only need to demand  that it is satisfied on \emph{some interval} of \emph{every} $y$-circle.   If this happens then part of every $y$-circle is space-like and the $y$-circles are not CTC's (though they may contain some timelike arcs).

It is easy to see that this condition is met. For any circular loop with constant $|\xi|= \rho$, observe that 
\begin{equation}
 \frac{1}{\sqrt{2}\pi R_y } \oint_{|\xi|=\rho} \,| F (\xi)|^2 \,  dv  ~=~  \sum_{n=1}^\infty\, b_n^2  \, \rho^{2n} ~\leq~  c.
\end{equation}
Since the average value of $| F (\xi)|^2$ around the $y$-circle satisfies (\ref{CTCcond2}), then if  (\ref{CTCcond2}) is violated somewhere on the circle then there must be another arc on which  (\ref{CTCcond2}) is satisfied. 

This does not yet completely prove the absence of CTCs; one could imagine that there are other closed loops in the $\xi$ (unit) disc, more general than the circular, constant $|\xi|$ loops that we just discussed.  However, we can show that for a non-constant holomorphic function $F$, one always has
\begin{equation}
c ~-~ |F(\xi)|^2 ~>~ 0 
 \label{CTCcond3}
  \end{equation}
along some interval of any simple, closed curve in the $\xi$-plane that winds around $r=0$. The mathematical proof of this claim is given in Appendix \ref{app:Proof}. This rigorously proves the absence of CTC's along along the circle directions.

\subsubsection{Conserved charges}
\label{sect:10ncharg}

The second term in $\omega$ (\ref{om-gauged-10n}) falls off at infinity as $r^{-4}$ and so does not contribute to the angular momentum.  The first term in  (\ref{om-gauged-10n}) has only a purely oscillatory term that falls off at infinity as $r^{-2}$. This means that the angular momentum is exactly the same as the round supertube with $F = 0$:
\begin{equation}
J_L ~=~ J_R ~=~  \frac{R_y}{2} \,a^2  \, ,
\label{JandP}
\end{equation}
while using (\ref{Fexp}) one finds the momentum charge: 
\begin{equation}
Q_P=~ \frac{1}{4\sqrt{2}\pi R_y}\int_0^{\sqrt{2}\pi R_y}dv\ (\xi_{\infty} F'_{\infty}(v)\bar{F}_\infty(v) + c.c. )~=~     \frac{1}{2}\,  \sum_{n=1}^\infty \, n\, b_n^2   \,.
\label{QP10n}
\end{equation}
The quantized charges are then obtained by using (\ref{quantch}).  
%
\subsection{The $(1,1,n)$ superstrata}
\label{sect:11ni}
In \cite{Walker:2019ntz} it was shown that the $(k,m,n)$ and $(k,k-m,n)$ single-mode superstrata are rather similar solutions. In six dimensions they are almost related by sending $\theta\to \frac{\pi}{2}-\theta$ and making coordinate redefinitions in $(\varphi_{1},\varphi_{2})$. If one performs the requisite spectral transformations followed by dimensional reduction to five dimensions, then they indeed become identical \cite{Walker:2019ntz}. Thus it is no surprise that the analysis of the $(1,1,n)$ multi-mode family built from
\begin{equation}
G_1(\xi,\chi,\eta)   ~=~    \eta \, F(\xi) ~=~\frac{a\,F(\xi)}{\sqrt{r^2+ a^2}} \, \cos \theta \, e^{i\left(\frac{\sqrt{2}v}{R_{y}}-\varphi_{2} \right)} \,,  \qquad\quad F(\xi) ~=~ \sum_{n=1}^\infty \, b_n \, \xi^n \, ,
 \label{G11n} 
\end{equation}
is almost identical to that of the $(1,0,n)$ multi-mode family presented above. To avoid duplication of the previous sections analysis we will be brief in discussing its construction.

\noindent The connection one-form $\beta$ is the same as for the seed supertube \eqref{betadefn}. The metric warp factor is now given by:
\begin{equation}
\sqrt{\cP}   ~=~    \frac{1}{\Sigma}\,\sqrt{Q_1 Q_5 ~-~  \frac{a^2 R_y^2}{2\,(r^2+a^2) }\, |F|^2\, \cos^2 \theta}   \, .
 \label{cP11n} 
\end{equation}
The BPS equations are then solved by:
\begin{equation}
\cF  ~=~   \frac{1}{a^2} \left(   |\xi|^2\,  |F|^2 - c \right)  \,, \qquad \omega  ~=~ \omega_0 ~+~ \frac{R_y}{\sqrt{2} \, \Sigma}\, \left(  c\, \sin^2 \theta \,d\varphi_1 + |\xi|^2|F|^2 \cos^{2}\theta\, d\varphi_{2} \right)\,.
 \label{Fom11n} 
\end{equation}
Ensuring the metric is asymptotically AdS now fixes $c$ in the same manner as the $(1,0,n)$ multi-mode
\begin{equation}
c  ~=~   \frac{1}{{\sqrt{2}\,\pi R_y} }\, \int_0^{\sqrt{2}\,\pi R_y}  \, | F_\infty (v)|^2  \, .
 \label{cval11n} 
\end{equation}
This is identical to the $(1,0,n)$ family since $\mathcal{F}$ has the same limit as $r\to \infty~.$

Performing the same gauge transformation (\ref{gtrf10n}) as in the $(1,0,n)$ case, one arrives at the final expressions:
\begin{align}
  \mathcal{F} &= \frac{1}{a^2} (|\xi|^2| F|^2 - |F_\infty|^2),\\
   \omega &= \left(1 - \frac{1}{2a^2}(|F_\infty|^2-c)\right)\omega_0 + \frac{R_y}{\sqrt{2}\Sigma}\left(|F_\infty|^2\sin^2\theta d\varphi_1+ |\xi|^2|F|^2\cos^2\theta d\varphi_2\right) ,
\end{align}
which have the correct fall off at infinity.

The analysis of the regularity of the metric proceeds precisely as for the $(1,0,n)$ superstrata in Section \ref{sect:10nreg}; in particular, we again have the condition:
\begin{equation}
\frac{Q_1 Q_5}{R_y^2}  ~=~ a^2 +  \frac{1}{2}\,c \,.
\end{equation}
One can similarly check that $\mathcal{P}>0$ everywhere. The analogy of (\ref{redmet1}) for the $(1,1,n)$ metric with $dt=dr=d\theta=0$ in the gauge (\ref{Fom11n}) is:
\begin{align}\nonumber
 \mathcal{P}^{-1/2} & \left( \frac{R_y^2|F|^2}{2\Sigma}\frac{r^2}{r^2+a^2} \sin^2\theta \left(d\varphi_1-\frac{dy}{R_y}\right)^2 ~+~   \frac{Q_1Q_5}{\Sigma}\cos^2\theta \left(d\varphi_2-\frac{dy}{R_y}\right)^2\right.\\
 & \left.~+~  \frac{R_y^2}{2a^2} \left( \frac{2\,Q_1Q_5}{R_y^2} - |F|^2\right)\left[ \frac{r^2}{R_y^2} dy^2   + a^2\sin^2\theta d\varphi_1^2\right]    \,\right) \,,
 \end{align}
so we again see that this is positive-definite if and only if (\ref{CTCcond1}) is satisfied, which we expect to hold (see Section \ref{10n-y-circle}).

Finally, the conserved supergravity charges are given by:
\begin{equation}
\begin{split}
Q_P \,=\frac{1}{2}\sum_{n=1}^{\infty}(n+1)b_{n}^{2}~ \,,\qquad  J_L\,=\, \frac{R_y}{2} \,a^2 \, , \qquad J_R\,=\, \frac{R_y}{2} \left(a^2\,+\, c \right) \,,
\end{split}
\end{equation}
from which one obtains the quantized charges by using (\ref{quantch}).

\subsection{The $(2,1,n)$ superstrata}
\label{sect:21ni}

The single-mode $(2,1,n)$ superstratum has been studied in \cite{Bena:2017upb,Ceplak:2018pws,Heidmann:2019zws,Walker:2019ntz}.    For $k\ge 2$ there are supercharged modes in the CFT \cite{Ceplak:2018pws} and so this is one of the simplest  examples in which such modes arise, and for which there are two independent holomorphic functions.  We give the full supercharged solution in   Appendix \ref{sect:21niApp};  here we will restrict to the significantly simpler solutions constructed from the original $(2,1,n)$ modes \cite{Bena:2017upb}. Thus we set $G_2 \equiv 0$.  The remaining holomorphic function, $G_1$, is then given by the most general superposition of the original $(2,1,n)$ modes determined by an arbitrary function $F$ fluctuating with the complex variable $\xi$  \eqref{cplxcoords}:
\begin{equation}
G_1(\xi,\chi,\eta)   ~=~    \chi \, \eta \, F(\xi) ~=~ \frac{a^2 \, F(\xi)}{a^2+r^2} \, \sin \theta\,\cos\theta\, e^{i \big (\frac{\sqrt{2} v}{R_y} + \varphi_1- \varphi_2\big)} \,,  \qquad\quad F(\xi) ~=~ \sum_{n=1}^\infty \, b_n \, \xi^n \,, 
 \label{G21n} 
\end{equation}
where $b_n$ are real Fourier coefficients. Once again, the connection one-form $\beta$ is still given by its seed value \eqref{betadefn}. The metric warp factor is  given by:
\begin{equation}
\sqrt{\cP}   ~=~    \frac{1}{\Sigma}\,\sqrt{Q_1 Q_5 ~-~  \frac{a^4 R_y^2}{2\,(r^2+a^2)^2 }\, |F|^2\, \sin^2 \theta\, \cos^2 \theta}   \,.
 \label{cP21n} 
\end{equation}
We decompose $\omega$ as follows
\begin{equation}
\qquad \omega  ~=~ \omega_0 \,+\, \varpi_r \, dr\,+\, \varpi_\theta \, d\theta\,+\, \varpi_1 \, d\varphi_1\,+\, \varpi_2 \, d\varphi_2 \,,
 \label{om21n} 
\end{equation}
where $\omega_0$ is defined in (\ref{angmom0}). The quantities $\mathcal{F}$ and $ \omegacomp$ are determined by the BPS equations sourced by $G_1$:
\be 
\begin{split}
\mathcal{F} &\,=\, \frac{1}{4\,a^2}(1 - |\xi|^2)(1 - |\chi|^2)\,\left| F \right|^2 \,+\, \frac{1}{a^2} K(\xi,\xib)  \,, \qquad \varpi_r  \,=\,  \varpi_\theta\,=\, 0\, ,\\
\varpi_1 &\,=\, - \frac{R_y}{2\sqrt{2}\Sigma}\sin^2\theta \,\left[\,|\eta|^2\,\left| F \right|^2 \,+\,  K(\xi, \xib) \, \right],\\
\varpi_{2} &\,= \, \frac{R_y}{2\sqrt{2}\Sigma}\cos^2\theta \,\left[\,|\xi|^2 |\chi|^2 \,\left| F \right|^2 \,+\,  K(\xi, \xib) \,  \right], 
\end{split}
\label{21nMultiGen}
\ee
where we have defined
\be 
K(\xi, \xib) \equiv \int_0^{|\xi|} x \left| F\Big(x\,e^{i\frac{\sqrt{2}}{R_y}v}\Big) \right|^2 dx\,.
\ee
Once again, note that the solutions are sesquilinear in $F$.

One can also add non-trivial solutions of the homogeneous BPS equations.   The ones needed here are a combination of a trivial harmonic solution with a constant coefficient, $c^{(0)}$, and a gauge transformation \eqref{gaugetrf} involving a holomorphic function, $c^{(1)}(\xi)$:
\begin{equation}
\begin{split}
\mathcal{F}  ~\to~  & \mathcal{F}\,-\, \frac{c^{(1)}(\xi)+c^{(1)}(\bar{\xi})}{a^2}\,,\\
\omega ~\to~
&\omega \,+\, \frac{c^{(0)}}{2\,a^2}\,\omega_0  \,+\, \frac{c^{(1)}(\xi)+c^{(1)}(\bar{\xi})}{2\,a^2} \,\beta \,+\,  \frac{i R_y}{2\sqrt{2}\,r(r^2+a^2)} \,\left(c^{(1)}(\xi)-c^{(1)}(\bar{\xi}) \right)\,dr,\\
\end{split}
\label{lastHomSol21n}
\end{equation}
where $\beta$ and $\omega_0$ are defined in \eqref{betadefn} and \eqref{angmom0}, $c^{(0)}$ is an arbitrary constant.  The function, $c^{(1)}$, is {\it a priori} arbitrary, but it will be fixed in the next section. This leads to the final expressions:
\begin{equation}
\begin{split}
\mathcal{F} &\,=\, \frac{1}{4\,a^2}(1 - |\xi|^2)(1 - |\chi|^2)\,\left| F \right|^2 \,+\, \frac{1}{a^2}\left[ K(\xi,\xib) \,-\, \left(c^{(1)} + \bar{c}^{(1)}\right)\right]\,,\\
\omega &\,=\, \left(1 + \frac{c^{(0)}}{2\,a^2}\right) \omega_0 + \frac{i\,R_y}{2\sqrt{2}\, r\, (r^2 + a^2)}\,\left( c^{(1)}-\bar{c}^{(1)}\right) \,dr \\
&\phantom{\,=\,}- \frac{R_y}{2\sqrt{2}\Sigma}\left( |\eta|^2\, \left| F \right|^2 + \left[ K(\xi,\xib) \,-\, \left(c^{(1)} + \bar{c}^{(1)}\right)\right] \right)     \sin^2\theta\, d\varphi_1\\
&\phantom{\,=\,} + \frac{R_y}{2\sqrt{2}\Sigma}\left( |\xi|^2 |\chi|^2 \,\left| F \right|^2 + \left[ K(\xi,\xib) \,-\, \left(c^{(1)} + \bar{c}^{(1)}\right)\right] \right)     \cos^2\theta \,d\varphi_2\,,
\end{split}
\end{equation}

\subsubsection{Regularity}
\label{sec:21nreg}

First we address the behavior as $r \to \infty$, where we want the  to be asymptotic to AdS$_3\times$S$^3$. This requires $\cF$ and $\omega$ to decay as $r^{-2}$. As before, $\cF$ goes to a constant and a $v$-dependent oscillating term at the boundary. To make $\cF$ vanish at infinity we fix $c^{(1)}$ via:
\begin{equation}
c^{(1)}(\xi) \,\equiv \, \sum^{\infty}_{n=1} \frac{b_n^2}{4(n+1)} \,+\, \sum_{n,\,\ell=1}^{\infty}  \frac{b_n b_{n+\ell}}{2(1+n)+\ell} \,\, \xi^\ell \,,\\
\end{equation}
With this choice, one then  that $\mathcal{F}$ and $\omega$ decay at least as $r^{-2}$ and the solutions are asymptotic to AdS$_3\times$S$^3$.

At the supertube locus, $r=0,\,\theta = \frac{\pi}{2}$, there are divergences in the solution.  The only dangerous term comes from the metric component along the supertube, that is the coefficient of $d\varphi_1^2$ given by \eqref{danger1}.  To remove the singularity, one must impose
\begin{equation}
\frac{Q_1\,Q_5}{R_y^2}\,=\, a^2\,+\, \frac{1}{2}\,c^{(0)}\,.
\label{reg21n}
\end{equation}

To analyze the behaviour at the ``center of space,'' $r=0,\,\theta = 0$, one has to switch to new polar coordinates ($\widetilde{r},\,\widetilde{\theta}$) and take the limit $\widetilde{r}\rightarrow0$:
\be
r \,\sim\, \widetilde{r}\, \cos \widetilde{\theta}\,, \qquad \sin \theta \,\sim\, \frac{\widetilde{r}}{a}\, \sin \widetilde{\theta} \qquad \Rightarrow \qquad \xi \sim  \frac{\widetilde{r}}{a} \, \cos \widetilde{\theta}\, e^{i\frac{\sqrt{2}}{R_y}v}\,, \qquad \chi \sim \frac{\widetilde{r}}{a}\, \sin \widetilde{\theta}  \, e^{i\phi_{1}}\,.
\label{polarcoordtheta=0}
\ee
The $\theta$ and $\varphi_1$ components of the angular-momentum one-form vanish when $\widetilde{r}\rightarrow 0$. However, $\omegacomp_2$ goes to a constant, which means there are CTC's.  We therefore must cancel this constant by fixing $c^{(0)}$ according to:
\begin{equation}
c^{(0)} \,\equiv \, \sum^{\infty}_{n=1} \frac{b_n^2}{2(n+1)} \,=\,  \frac{1}{{\sqrt{2}\,\pi R_y} }\, \int_0^{\sqrt{2}\,\pi R_y}  \, K_\infty(v)\, dv  \,,
\end{equation}
where we have defined
\begin{equation}
 K_\infty(v) \,\equiv \,  \lim_{|\xi|\rightarrow 1} K(\xi,\xib) =  \lim_{r\rightarrow \infty} K(\xi,\xib)  \,.
\end{equation}
Moreover, in order to have a well-defined solution, one must have $\sqrt{\cP}$ real and so $\cP>0$ everywhere. This can be proven rigorously for generic $F$ using the holomorphicity of the function. We refer the interested reader to the Appendix \ref{app:ProofP10n} for the details of the proof.

Investigating the absence of CTC's in this geometry is a daunting task, which we will not complete. Indeed, checking that there are no closed time-like curves for generic $F(\xi)$ along all the circle directions, as it has been done in Section \ref{10n-y-circle} for the ($1,0,n$) solutions, is very non-trivial with the solutions above. This issue should take a much simpler form for particular choices of $F(\xi)$. From the dual CFT, we expect there to be no constraints on the possible holomorphic functions that can show up as $(2,1,n)$ superstrata, so we also do not expect any choice of holomorphic function $F$ to lead to the presence of CTC's. (See, also, the discussion in Section \ref{sect:conclusions}.)

\subsubsection{Conserved charges}

The angular momenta $J_R$ and $J_L$ and the momentum charge $Q_P$ can be read off from the asymptotic behaviour of $\omega$, $\beta$ and $\cF$ through the expressions \eqref{asmpmoms} and \eqref{Fexp}. For the solutions constructed above, we obtain:
\begin{equation}
\begin{split}
Q_P \,= \, \frac{1}{8} \,\sum^{\infty}_{n=1} b_n^2 \,,\qquad  J_L\,=\, \frac{R_y}{2} \,a^2 \, , \qquad J_R\,=\, \frac{R_y}{2} \left(a^2\,+\, \frac{1}{2}\,c^{(0)} \right) ~=~\frac{R_y}{2} \left(a^2\,+\, \frac{1}{2}\,\sum^{\infty}_{n=1} \frac{b_n^2}{2(n+1)}\right) \,.
\end{split}
\end{equation}
Again,  the quantized charges are obtained from  these supergravity charges by using (\ref{quantch}).      

\subsection{The $(k,0,1)$ superstrata}
\label{sect:ki01}
The single-mode $(k,0,1)$ superstratum has been constructed and studied in \cite{Bena:2018mpb}.  For $m=0$ there are no supercharged modes in the CFT and the corresponding supergravity solutions are singular if one tries to incorporate them. Thus we once again set $G_2 \equiv 0$.
The underlying holomorphic function, $G_1$, is given by the most general superposition of the $(k,0,1)$ modes determined by an arbitrary function $F$ fluctuating with the complex variable $\chi$ \eqref{cplxcoords}:
\begin{equation} 
G_1(\xi,\chi,\mu) \,=\, \xi \,\chi \, \partial_\chi F(\chi)\,,\qquad G_2(\xi,\chi,\mu) \,=\, 0\,.
\label{G1k01}
\end{equation}
We define the real Fourier modes of $F$ as before:
\begin{equation} 
F(\chi) \,=\, \sum_{k=1}^{\infty} \, b_k \, \chi^k\,.
\label{Fourierk01}
\end{equation}
We could also take $G_1(\xi,\chi,\mu) \,=\, \xi \,F(\chi)$ in analogy with the previous examples, but it simplifies the results if we use (\ref{G1k01}) as we will not have to introduce the primitive of $F$. The connection $\beta$ is given by \eqref{betadefn}.  We decompose $\omega$ as in \eqref{om21n}. The quantities determining the metric are given by:
\begin{equation} 
\begin{split}
\sqrt{\cP} &\,=\, \frac{1}{\Sigma} \,\sqrt{Q_1 Q_5 \,-\, \frac{R_y^2\,|\xi|^2|\chi|^2 }{2}\,\partial_\chi \partial_{\bar{\chi}} |F|^2}\,,\qquad \cF \,=\,\frac{|F|^2}{\Sigma}\,,\\
 \varpi_r & \,=\, \frac{i\, R_y\, }{2\sqrt{2}}\frac{|\xi|^2}{r \Sigma}\,\left[ \chi\, \partial_\chi |F|^2 - \bar{\chi}\, \partial_{\bar{\chi}} |F|^2 \right]\, ,\qquad  \varpi_\theta  \,=\, 0\,,\\
 \varpi_1 &\,=\, \frac{R_y}{\sqrt{2}\,\Sigma}\left( \frac12 |\xi|^2 \left[ \chi\, \partial_\chi \left(\frac{|F|^2}{1-|\chi|^2} \right)+\bar{\chi}\, \partial_{\bar{\chi}} \left(\frac{|F|^2}{1-|\chi|^2} \right) \right]  \,-\,  \frac{1-|\xi|^2}{1-|\chi|^2} \, |F|^2\right)\sin^2\theta, \\
 \varpi_{2} &\,= \frac{R_y}{\sqrt{2}\,\Sigma}\left(  \, \frac12 |\xi|^2\left[ \chi\, \partial_\chi \left(\frac{|F|^2}{1-|\chi|^2} \right)+\bar{\chi}\, \partial_{\bar{\chi}} \left(\frac{|F|^2}{1-|\chi|^2} \right) \right] \right) \cos^2\theta\,.
\end{split}
\label{k01MultiGen}
\end{equation}
We will also need non-trivial solutions of the homogeneous BPS equations.  {\it A priori} the calculation of such solutions is non-trivial.  However, the holomorphic structure of the BPS solutions renders this relatively easy.  Indeed, the homogeneous solution that we will need to add to $\cF$ and $\omega$
can be generated trivially from (\ref{k01MultiGen}) by 
the mapping:
\begin{equation} 
|F|^2 ~\to~  |F|^2 ~-~ c(\chi) ~-~c(\bar{\chi}) \,, 
\label{homsolk01}
\end{equation}
where $c$ is an arbitrary holomorphic function. It is not, {\it a priori}, obvious that this replacement results in homogeneous solutions to the BPS equations:  It is simply an empirical fact based on the direct computations and is a consequence of the holomorphic structure of the BPS system.  We will use this technique for generating homogeneous solutions in other superstrata and we will  comment on its possible significance in Section \ref{sect:conclusions}.

\subsubsection{Regularity}
\label{sec:k01reg}\label{sec:k01CTC}

The investigation of regularity proceeds much as before, but there is a new and interesting issue near the supertube locus and this leads to a non-trivial result that is also apparent in the correlators of the holographically dual CFT \cite{Giusto:2015dfa}.

First we handle the straightforward issues.   As $r \to \infty$, one has $|\xi|\sim 1- \frac{a^2}{r^2},$ $|\chi|\sim \frac{a^2 \sin^2\theta}{r^2}$ and  one can easily verify that $\cF$  and $\omega$ given by \eqref{k01MultiGen} decay as $r^{-2}$.   Thus, unlike the previous examples, the solutions determined by \eqref{k01MultiGen} do not need any modification to get a metric that is  asymptotic to AdS$_3\times$S$^3$. Moreover, the plane where the $\varphi_1$-cycle (resp. $\varphi_2$-cycle) shrinks at $\theta=0$ (resp. $\theta=\frac{\pi}{2}$) do not lead to any singularity when $r\neq0$. Indeed, it is straightforward to check that the scalars are indeed independent of $\varphi_1$ on this plane (resp. $\varphi_2$) and that the forms have no legs along the shrinking direction.  As before, to examine the neighborhood of $r=0,\,\theta = 0$ ($\xi = 0, \, \chi =0$) one needs to change to the polar coordinates \eqref{polarcoordtheta=0}.  The components angular-momentum one-form all vanish and so do not produce CTC's.

It is at $r=0,\,\theta = \frac{\pi}{2}$ that things get interesting:  there are divergences. To reveal them more clearly, it is useful to use  polar coordinates ($\rho,\vartheta$) around the supertube locus  for small $\rho$:
\begin{equation} 
r \, =\, \rho\, \cos \vartheta\,, \quad \sin \theta \,=\, 1-\frac{\rho^2}{2a^2}\, \sin^2 \vartheta \quad \Rightarrow \quad \xi \sim  \frac{\rho}{a} \, \cos \vartheta\, e^{i\frac{\sqrt{2}}{R_y}v}\,, \quad \chi \sim \left( 1-\frac{\rho^2}{2a^2}\, \sin^2 \vartheta\right)  \, e^{i\varphi_{1}}\,.
\end{equation}
The divergence in $\cF$ has the following form:
\begin{equation} 
\cF \,=\, \frac{F\left(e^{i\phi_{1}} \right)F\left(e^{-i\phi_{1}} \right)-c\left(e^{i\phi_{1}} \right)-c\left(e^{-i\phi_{1}} \right)}{\rho^2 \, \sin^2 \vartheta}\,+\, \underset{\rho\rightarrow 0}{\cO}(1)\,.
\label{eq:SingularF}
\end{equation}
We can cancel this singularity by fixing $c$ via:
\begin{equation} 
c(\chi) \,=\, \frac{1}{2}\,\sum_{k=1}^{\infty} \, b_k^2 \,+\, \sum_{k,\,\ell =1}^{\infty} b_k\,b_{k+\ell}\,\, \chi^\ell\,,
\end{equation}
where $b_k$ are the Fourier coefficients of $F$ \eqref{Fourierk01}. This choice of $c(\chi)$  does not change the asymptotics at infinity and does not introduce singularities elsewhere. It also cancels  the $\rho^{-1}$ divergence in $\varpi_r$, the $\rho^{-4}$ divergence in $\varpi_{1}$ and the $\rho^{-2}$ divergence in $\varpi_2$. 

Next, one needs to examine \eqref{danger1} and cancel the $\rho^{-2}$ divergence in the metric components along $d\varphi_1^2$.   As $\rho \rightarrow 0$, one has
\begin{equation} 
\frac{\sqrt{Q_1 Q_5}}{R_y^2}\,\,g_{\varphi_1 \varphi_1}\, \sim \, \frac{1}{\rho^2}\, \left[\frac{Q_1 Q_5}{R_y^2} \,-\, a^2 \,-\, \frac{1}{2} \sum_{k=1}^{\infty} \, k\, b_k^2 \,\,-\,\, \sum_{k,\,\ell =1}^{\infty} k\,b_k\,b_{k+\ell} \, \cos(\ell \,\varphi_1) \right]\,.
\label{k01div}
\end{equation}
This contains a fluctuating component that cannot be cancelled by a simple constraint like  (\ref{STreg}), (\ref{reg10n}) or (\ref{reg21n}).  However this divergence is easily fixed.  The expression (\ref{k01div}) represents a charge density fluctuation that is being induced along the supertube  by this class of momentum waves.  We therefore have to address this by allowing such a variation in the original supertube charge density.  This means that the {\it harmonic} part of $Z_1$ needs to be allowed to vary according to 
\begin{equation} 
Z_1 ~=~  \frac{Q_{1}}{\Sigma} ~+~  \frac{R_y^2}{2\,  Q_5\, \Sigma} \, \big(\lambda (\chi) ~+~ \lambda (\bar{\chi})\big) ~+~ ...  \,,
\label{Z1mod}
\end{equation}
where  $\lambda$ is an arbitrary holomorphic function and the ``$...$" denotes the fluctuating modes associated with the momentum wave\footnote{Note that we must require $\lambda (0)   ~=~ 0$ to avoid modifying the $Q_1$ overall charge since the boundary $r\rightarrow \infty$ corresponds to $\chi\rightarrow 0$.}.   (See Section \ref{sect:basesols} for more details of the momentum wave terms.). This modifies $\cP$ in (\ref{k01MultiGen}):
\begin{equation} 
\sqrt{\cP} ~=~  \frac{1}{\Sigma} \,\sqrt{Q_1 Q_5 ~+~  \frac{R_y^2}{2}\,  \big(\lambda (\chi) ~+~ \lambda (\bar{\chi}) -  \,|\xi|^2|\chi|^2\, \partial_\chi \partial_{\bar{\chi}} |F|^2 \big)}\,,  
\label{Pmod}
\end{equation}
but the additional term in $Z_1$ makes no other changes in the superstratum solution. (This is because of the near-trivial form, (\ref{Z2Theta1}), of some of the other fields in the superstratum Ansatz.) Thus this addition to $Z_1$ still solves the   BPS equations with only minor modifications to the metric.

Returning to  (\ref{danger1}), the modification of $Z_1$, or $\cP$, amounts to replacing  
\begin{equation} 
Q_1 Q_5 ~\to~ Q_1 Q_5~+~    \frac{R_y^2}{2} \big( \lambda (e^{i \varphi_1}) \,+\,  \lambda (e^{-i \varphi_1})\big) \,,  
\label{QQrepl}
\end{equation}
in (\ref{k01div}).  One can then cancel the divergence in $g_{\varphi_1\varphi_1}$ by taking 
\begin{equation} 
\frac{Q_1 Q_5}{R_y^2} ~=~ a^2 \,+\, \frac{1}{2} \sum_{k=1}^{\infty} \, k\, b_k^2 \,=\, a^2\,+\,  \frac{1}{4\pi i}\oint_{|\chi|=1} \bar{F}\partial_\chi F \,d\chi \,, \qquad 
\lambda(\chi) ~=~   \sum_{k,\,\ell =1}^{\infty}  k\,b_k\,b_{k+\ell}\,\, \chi^\ell \,.
\label{lambdaDef}
\end{equation}
The zero modes thus gives us the usual regularity constraint, akin to  (\ref{STreg}), (\ref{reg10n}) and (\ref{reg21n}), while the non-trivial modes are cancelled by the supertube charge density fluctuations, $\lambda (\chi)$.   Having cancelled the divergence, the metric is regular in the neighborhood of $r=0$.

The charge density fluctuation might however induce CTC's.  Specifically, the charge density fluctuations must not cause $Z_1$ to go negative or, equivalently, $\cP$ must remain positive everywhere. We have proven that this is indeed true for generic holomorphic function $F$ in the Appendix \ref{app:ProofPk01}.

It is thus very gratifying that the required charge density variations (\ref{QQrepl})  in $Z_1$, preserve the smoothness of the metric and place no restrictions on the Fourier coefficients, and hence on $F(\chi)$. 

Finally, as for the $(2,1,n)$ family above, we will not complete the daunting analysis of CTC's in this geometry for generic $F(\chi)$. Holography suggests we should not expect any choice of holomorphic function $F$ to lead to the presence of CTC's. (See, also, the discussion in Section \ref{sect:conclusions}.)

\subsubsection{Conserved charges}

The angular momenta $J_R$ and $J_L$ and the momentum charge $Q_P$ can be read off from the asymptotic form of $\omega$, $\beta$ and $\cF$ through the expressions \eqref{asmpmoms} and \eqref{Fexp}. For the solutions constructed above, we obtain the following supergravity charges:
\be 
\begin{split}
Q_P \,= \, \frac{1}{2} \,\sum^{\infty}_{k=1} b_k^2\,=\, \frac{1}{4\pi i} \oint_{|\chi|=1}\frac{|F|^2}{\chi}\, d\chi\,,\qquad  J_L\,=\, J_R\,=\, \frac{R_y}{2} \,a^2 \,,
\end{split}
\end{equation}
from which the quantized charges are obtained by using  (\ref{quantch}).        

\subsection{The ($k,k,0$) superstrata}
\label{sect:kiki0}

In this final section, we construct the solutions obtained from superposition of ($k,k,0$) superstrata. They form the simplest example of family of solutions that fluctuate according to the third complex variable, $\eta$ \eqref{cplxcoords}. Unlike the functions of the variable $\chi$, which correspond to fluctuations around the supertube locus, the function of variable $\eta$ correspond to fluctuations around the center of $\mR^4$. For this purpose, it is convenient to introduce the ``distance'' to the center of $\mR^4$,
\be 
\Lambda \,\equiv\, r^2\,+\, a^2 \, \sin^2 \theta \,.
\label{Lambda}
\ee
For $n=0$, there are no supercharged modes in the CFT and we once again set $G_2 \equiv 0$.
The underlying holomorphic function, $G_1$, is given by the most general superposition of the $(k,k,0)$ modes determined by an arbitrary function $F$ fluctuating with the complex variable $\eta$ \eqref{cplxcoords}:
\be 
G(\xi,\chi,\mu) \,=\, F(\eta)\,,\qquad F(\eta) \,=\, \sum_{k=1}^{\infty} b_k \,\eta^k\,,
\ee
where $b_k$ are taken to be real. The connection $\beta$ is given by \eqref{betadefn}. The metric warp factor is  given by:
\begin{equation}
\sqrt{\cP}   ~=~    \frac{1}{\Sigma}\,\sqrt{Q_1 Q_5 ~-~  \frac{R_y^2}{2}\, |F|^2}   \,.
 \label{cPkk0} 
\end{equation}
As for $\cF$ and $\omega$, the BPS equations lead to the solutions:
\begin{equation}
\cF  ~=~   \frac{|F|^2 - c}{\Lambda} \,, \qquad \omega  ~=~ \omega_0 ~+~ \frac{R_y}{\sqrt{2} \,\Lambda \Sigma}\,   \left(c-|F|^2 \right)\, \left[ (a^2+r^2) \sin^2 \theta\,d\varphi_1 \,+\, r^2 \cos^2 \theta \,d\varphi_2 \right] \,,
 \label{Fomkk0} 
\end{equation}
where $c$ is a constant, corresponding to a homogeneous solution, and $\omega_0$ is given by (\ref{angmom0}). One can compare the expression of $\cF$ with the supersposition of ($k,0,1$) superstrata \eqref{k01MultiGen} and note that the fluctuations along $\eta$ are indeed centred around the center of $\mR^4$, $\Lambda \sim 0$, whereas the fluctuations along $\chi$ are around the supertube locus, $\Sigma \sim 0$ as expected.

\subsubsection{Regularity}
\label{sec:kk0reg}

The investigation of regularity proceeds much as before, but once again there  is an interesting issue near the center of $\mR^4$ that is related to the development of vevs in the dual  CFT \cite{Giusto:2015dfa}.

We first handle the straightforward regularity analysis.   As $r \to \infty$, one can easily verify that $\cF$  and $\omega$ given by \eqref{Fomkk0} decay as $r^{-2}$. Moreover, the plane where the $\varphi_1$-cycle (resp. $\varphi_2$-cycle) shrinks at $\theta=0$ (resp. $\theta=\frac{\pi}{2}$) do not lead to any singularity when $r\neq0$. Indeed, it is straightforward to check that the scalars are indeed independent of $\varphi_1$ on this plane (resp. $\varphi_2$) and that the forms have no legs along the shrinking direction. 

At the center of $\mR^4$, $r=0,\,\theta = 0$, there are divergences that need to be regularized. To reveal them more clearly, one needs to change to the polar coordinates \eqref{polarcoordtheta=0}. 
The divergence in $\cF$ has form:
\begin{equation} 
\cF \,=\, -\frac{\sum_{k=1}^\infty {b_k}^2 \,+\, 2\,\sum_{\ell=1}^{\infty} \left( \sum_{k=1}^{\infty} b_k\,b_{k+\ell}\right) \, \cos \left[\ell \left(\sqrt{2}\frac{v}{R_y} -\varphi_{2} \right)\right]\,-\,c}{\widetilde{r}^2}\,+\, \underset{\widetilde{r}\rightarrow 0}{\cO}(1)\,.
\label{kk0div}
\end{equation}
This contains a fluctuating component that cannot be cancelled by the constant term $c$.  However, the expression (\ref{kk0div}) represents a charge density fluctuation that is being induced along the center of $\mR^4$  by the class of $(k,k,0)$ momentum waves.  We therefore have to address this issue by modifying the tensor gauge fields as for the superposition of ($k,0,1$) superstrata. That is, we add a solution of the BPS equations parametrized by an arbitrary function $\lambda(\eta)$ given by
\be 
\begin{split}
Z_1 &\,\rightarrow\, Z_1 \,+\, \frac{R_y^2}{2\,Q_5\,\Sigma}\left(\lambda(\eta)+\lambda(\bar{\eta}) \right)\,,\qquad \Theta_2 \,\rightarrow\, \Theta_2 \,+\, \frac{R_y}{Q_5}\,\left[\eta \,\partial_\eta \lambda(\eta)\, \Omega_{y} \,+\,\bar{\eta}\,\partial_{\bar{\eta}} \lambda(\bar{\eta})\, \bar{\Omega}_{y} \right]\,,\\
\cF&\,\rightarrow\, \cF \,-\, \frac{\lambda(\eta)+\lambda(\bar{\eta})}{\Lambda}\,,\qquad \omega \,\rightarrow\,\omega\,+\, \frac{R_y \,\left(\lambda(\eta)+\lambda(\bar{\eta}) \right)}{\sqrt{2}\,\Lambda \Sigma}\left((a^2+r^2)\sin^2\theta \,d\varphi_1 +r^2\cos^2\theta d\varphi_2 \right)\,,
\end{split}
\ee
where $\Omega_y$ is a self-dual form that will be detailed later \eqref{SDyz}. Then, by fixing
\be 
\begin{split}
c\,=\, \sum_{k=1}^{\infty} {b_k}^2\,=\,\frac{1}{2\pi i} \oint_{|\eta|=1} \frac{|F(\eta)|^2}{\eta} \,d\eta \,,\qquad \quad \lambda(\eta) \,=\, \sum_{k,\ell=1}^{\infty}  b_k \,b_{k+\ell} \,\eta^\ell\,,
\label{lambdaDefkk0}
\end{split}
\ee
all the divergences vanish. Moreover, this specific choice makes also all the non-vanishing scalars, as $\cP$, to be independent of the $S^3$ angles and all $\theta$-components, $\varphi_1$-components and $\varphi_2$-components of the forms to vanish. We have then regularized the solutions at the center of $\mathbb{R}^4$.

At the supertube locus, $r=0$, $\theta =\frac{\pi}{2}$, we have the same type of divergences as for the solutions that are fluctuating with $\xi$. They are simply cancelled by investigating the metric component along $d\varphi_1^2$ and by requiring that
\begin{equation} 
\frac{Q_1 Q_5}{R_y^2} \,=\, a^2 \,+\, \frac{1}{2} \sum_{k=1}^{\infty} \,b_k^2\,=\, a^2 \,+\, \frac{1}{4\pi i} \oint_{|\eta|=1} \frac{|F(\eta)|^2}{\eta} \,d\eta  \,. \\
\label{RegCond}
\end{equation}

Moreover, to have regular six-dimensional solutions, one needs first to require that $\sqrt{\cP}$ is well-defined everywhere and that there are no closed time-like curves along all the circle directions. The first condition, $\cP\geq 0$ can be proven for generic holomorphic function $F$ (see Appendix \ref{app:ProofPkk0}). As for the second condition, this is more subtle and we leave the question for future works. However, we strongly believe that the solutions are indeed free of CTC's for generic holomorphic function $F$ and that it could be shown in a similar fashion as for the ($1,0,n$) superstrata detailed in Section \ref{10n-y-circle} and Appendix \ref{app:Proof}.

\subsubsection{Conserved charges}

The angular momenta $J_R$ and $J_L$ and the momentum charge $Q_P$ can be read off from the asymptotic form of $\omega$, $\beta$ and $\cF$ through the expressions \eqref{asmpmoms} and \eqref{Fexp}. For the solutions constructed above, we obtain the supergravity charges:
\begin{equation}
\begin{split}
Q_P& \,= \, \frac{1}{2}\, \sum_{k=1}^{\infty} \,b_k^2\,=\,\frac{1}{4\pi i} \oint_{|\eta|=1} \frac{|F(\eta)|^2}{\eta} \,d\eta\,,\qquad \quad  J_L\,=\, \frac{R_y}{2} \,a^2 \, ,\\
 \qquad J_R&\,=\, \frac{R_y}{2} \left(a^2\,+\, \frac{1}{2}\,\sum_{k=1}^{\infty} \,b_k^2 \right) ~=~\frac{R_y}{2} \left(a^2\,+\, \frac{1}{4\pi i} \oint_{|\eta|=1} \frac{|F(\eta)|^2}{\eta} \,d\eta\right) \,.
\end{split}
\label{ConservedCharges}
\end{equation}
%

\subsection{Yet more superstrata}
\label{sect:genss}

We have computed other examples of superstrata and wish to remind the reader that these examples may be found in Appendix \ref{app:MoreEx}. We have extended the family of ($2,1,n$) superstrata above to the same solutions with supercharged modes in Appendix \ref{sect:21niApp}. This essentially results in solutions fluctuating with two arbitrary functions of one variable. Moreover, we have also obtained the solutions for  $(k,0,n)$ superstrata for fixed $n$ and arbitrary $k$.  This may be found in Appendix \ref{sect:ki0nApp}. Those solutions are necessarily expressed in terms of recurrence relations but the expressions in \ref{sect:ki0nApp} are significantly simpler than the recurrence relations that have appeared elsewhere, such as in \cite{Bena:2015bea,Bena:2016ypk}.

\section{Localizing momentum in the $(1,0,n)$  superstrata}
\label{sect: 10nExample}

In this section we give examples of $(1,0,n)$ multimode superstrata in which the momentum wave  is ``localized''  on the $y$-circle at infinity. This is a notable departure from a single mode superstrata in which the momentum charge is  evenly smeared around the $y$ circle. We will focus on the details of the three-dimensional geometry, producing several contour plots highlighting the behavior of the $y$-circle for various choices of mode function $F(\xi)$. 

\subsection{Metric as an $S^{3}$ fibration}
The general $(1,0,n)$ multimode metric (\ref{sixmet}), with $\beta$, $\cP,\,\omega,\,\cF$ given by \eqref{betadefn}, (\ref{cP10n}) and (\ref{Fom10n}), can be written as an $S^3$ fibered over a three-dimensional base manifold $\cK$ with metric $d\hat{s}_{\cK}^{2}$: 
\begin{equation}
ds_{(1,0,n)}^{2} = d\hat{s}_{\cK}^{2}+d\hat{s}_{S^{3}}^{2}~.
\end{equation}
The sphere part $d\hat{s}_{S^{3}}^{2}$ is schematically of the form:
\begin{align*}
d\hat{s}_{S^{3}}^{2} = g_{\theta \theta} \, d\theta^{2} + g_{11}(d\varphi_{1}+A_t^{(1)}\, dt)^{2} +g_{22}(d\varphi_{2}+A_t^{(2)}\,dt+A_y^{(2)}\,dy)^{2}~,
\end{align*}
and the three-dimensional part takes the form:
\begin{equation}
d\hat{s}_{\cK}^{2} = \hat{g}_{tt}\, dt^{2}+\hat{g}_{yy}(dy+\hat{A}^{(y)}_{t}\,dt)^{2}+\hat{g}_{rr}\,dr^{2}~.
\end{equation}

The geometry of the three-dimensional metric is driven by the behavior of the $y$-circle. For the single-mode superstrata this circle has three distinct phases.  As $r\to \infty$, the circle grows as in (\ref{STmet}) to produce asymptotically AdS$_{3}$.  As $r\to 0$, the circle pinches off smoothly giving a cap with bounded curvature.  In the  intermediate regime, the radius of the $y$-circle stabilizes to approximate a long BTZ throat. We will demonstrate this behavior in the next section and see how multiple excited modes modify this behavior. 

To fully analyze the $y$-circle we rewrite $\hat g_{yy}$ as: 
\begin{align}
\hat{g}_{yy}= \left(\frac{r^{2}\Sigma\sqrt{1- \frac{\abs{F}^{2}\sin^{2}\theta}{2(a^{2}+r^{2})}}}{a^{3}R_{y}\left( c \cos^{2}\theta +2 \Sigma\right)\left(1 - \frac{\abs{F}^{2}}{2(a^{2}+r^{2})} \right)}\right) \, \tilde{g}_{yy} \qquad \text{where} \qquad \tilde{g}_{yy}= 2a^{2}+c-\abs{F}^{2}~, \label{ghatyy}
\end{align}
and $c$ is given in \eqref{cval10n}. Note that whereas the factor in parentheses, $\hat{g}_{yy}/\tilde g_{yy}$, is always non-negative (see, for example, Section \ref{sect:10nreg}), $\tilde{g}_{yy}$ (and thus  $\hat{g}_{yy}$) may be negative in certain regions. However, as discussed in Section \ref{sect:10ni}, the regions where $\tilde{g}_{yy}<0$ will not lead to CTCs. The factor $\tilde{g}_{yy}$ is more sensitive than $\hat g_{yy}$ to the behavior of $F(\xi)$ and will be used to illustrate the multimode superstata of the examples in Sections \ref{sec:nonlocmomex} and \ref{sec:locmomex} below.

\subsection{Limits of the three-dimensional metric}
The three-dimensional metric $d\hat{s}_{\mathcal{K}}^{2}$ has three distinct scaling limits: $r\to 0$, $r\to \infty$, and an intermediate $a<r<\sqrt{c}$ region which becomes dominant for $a \ll r \ll\sqrt{c}$. 

\begin{itemize}

\item \underline{The metric near the cap}: 

\noindent 
To see that the cap is smooth one simply expands the metric around $r=0$.  To do this, it is convenient to define the dimensionless variable $\rho = r/a$, and then one finds:
\begin{align}
\lim_{\rho\to 0}d\hat{s}_{\mathcal{K}}^{2} &= \sqrt{Q_{1}Q_{5}}\left[(1-\rho^{2})d\rho^{2}- \frac{a^{4}R_{y}^{2}}{(Q_{1}Q_{5})^{2}}(1+\rho^{2})\,dt^{2} + \frac{\rho^{2}}{R_{y}^{2}} \left(dy+ \frac{R_{y}^{2}\, c}{2Q_{1}Q_{5}}\, dt \right)^{2} \right]\notag\\
& \qquad \qquad\qquad\qquad + \frac{b_{1}^{2}R_{y}^{2}\rho^{2}}{4\sqrt{Q_{1}Q_{5}}}\left( \frac{a^{4}R_{y}^{2}(3+\cos 2\theta)}{2(Q_{1}Q_{5})^{2}}\, dt^{2} + \sin^{2}\theta \, d\rho^{2}\right) + O(\rho^{3}) \,.
\label{quadord}
\end{align}
The important point is that at $O(\rho^{2})$ there are no corrections to $dy^2$ and so there is no conical singularity at $\rho =0$ and the metric limits precisely to that of 
flat Minkowski space.

More generally, one finds that, at higher order, the cap metric takes the form:
\begin{equation}
\lim_{\rho\to 0}d\hat{s}_{\mathcal{K}}^{2} = \sqrt{Q_{1}Q_{5}}\left[\frac{d\rho^{2}}{\rho^{2} +1 }- \frac{a^{4}R_{y}^{2}}{(Q_{1}Q_{5})^{2}}(1+\rho^{2})\,dt^{2} + \frac{\rho^{2}}{R_{y}^{2}} \left(dy+ \frac{R_{y}^{2}\, c}{2Q_{1}Q_{5}}\, dt \right)^{2} \right] ~+~ O(|F|^2)~,
\label{genexp}
\end{equation}
where $O(|F|^2)$ denotes terms that vanish at the same rate as $|F|^2$ as $r\to 0$.  Indeed, (\ref{quadord}) shows the $|F|^2$ corrections at $O(\rho^{2})$.  The expansion (\ref{genexp}) shows that the cap metric is, in fact, that of a global AdS$_3$, corrected by $|F|^2$. By defining $n_0=\min\{n\geq 1, b_n \neq0\}$, $|F|^2$ vanishes as $O(\rho^{2n_0})$  at $\rho=0$, then the cap metric is that of global AdS$_3$ to the same order.  This property of the cap metric was an important element in understanding  the scattering from single mode superstrata \cite{Bena:2019azk}.

\item \underline{The metric at infinity}: 

\noindent We set $\rho \equiv r(Q_{1}Q_{5})^{-1/2}$ and take the limit as $\rho\to \infty$. The geometry then becomes:
\begin{align}
\lim_{\rho\to \infty}d\hat{s}_{\mathcal{K}}^{2} &=   \sqrt{Q_{1}Q_{5}}\left[\frac{d\rho^{2}}{\rho^{2}}+\rho^{2}\,dy^{2}-\rho^{2}\,dt^{2}  +\frac{(c-\abs{F_{\infty}(v)}^{2})\rho^{2}}{2 a^{2}}(dy+dt)^{2}   \right]+O(\rho^{0})~,
\end{align}
where we have kept the correct leading term in the $dy^{2}$ coefficient to avoid degeneracy. As discussed in Section \ref{sect:hom-gauge}, we must perform the gauge transformation:
\begin{align}
t' = t+ \frac{1}{\sqrt{2}}\, f \qquad \text{and} \qquad y' = y- \frac{1}{\sqrt{2}}\, f~, \label{coordgauge}
\end{align}
where $f$ is defined in (\ref{gtrf10n}), to obtain $AdS_3$ in its canonical form. Indeed, applying this gauge transformation gives:
\begin{align}
\lim_{\rho\to \infty}d\hat{s}_{\mathcal{K}}^{2} &=   \sqrt{Q_{1}Q_{5}}\left[\frac{d\rho^{2}}{\rho^{2}}+\rho^{2}\,(dy')^{2} -\rho^{2}\,(dt')^{2} \right]+ O(\rho^{0})~. \label{ds3rtoinfgauge}
\end{align}
as expected.

\item \underline{The intermediate BTZ throat}: 

\noindent We consider the long throat regime $a \ll c$ and take $r \gtrsim a $. Then, we scale $\rho = r/\sqrt{Q_{1}Q_{5}}$, and work in the same gauge that (\ref{ds3rtoinfgauge}) is expressed ({\it i.e.} we use the solution as presented in (\ref{F-gauged-10n})-(\ref{om-gauged-10n})). After defining
\begin{equation}
\widetilde{F}(v)= \frac{1}{2}\sum_{m,n}b_{m}b_{n}(m+n)e^{i \frac{\sqrt{2}}{R_{y}}(m-n)v}~,
\end{equation}
the three-dimensional base may be written as:
\begin{align}
\lim_{a\to 0}d\hat{s}_{\mathcal{K}}^{2} &= \sqrt{Q_{1}Q_{5}} \left( \frac{d\rho^{2}}{\rho^{2}}-\rho^{2}\,dt^{2} +\rho^{2}\,dy^{2} \right) \notag \\
& \qquad +\frac{\sqrt{Q_{1}Q_{5}}}{R_{y}^{2}} \left(  -\frac{1}{4}+ \frac{\widetilde{F}(v)}{c}+\frac{\abs{F_{\infty}(v)}^{2}}{2c}-\frac{\abs{F_{\infty}(v)}^{4}}{4c^{2}} \right)\left(dy+dt \right)^{2}  +O(a^{2})\,,
\end{align}
which is the form of extremal-BTZ geometry with oscillatory perturbations coming from the second line. For a single mode solution with $F(\xi)=b\xi^{n}$, this reduces exactly to the BTZ result since 
\begin{equation}
\left[  -\frac{1}{4}+ \frac{\widetilde{F}(v)}{c}+\frac{\abs{F_{\infty}(v)}^{2}}{2c}-\frac{\abs{F_{\infty}(v)}^{4}}{4c^{2}} \right]_{F(\xi)= b\xi^{n}} = n~.
\end{equation}
We see that depending on the distribution of the $\left\lbrace b_{n} \right\rbrace$, the size of the oscillatory perturbation on top of the BTZ throat can be tuned. 
\end{itemize}

\subsection{Non-localized momentum wave examples}\label{sec:nonlocmomex}

Perhaps the simplest way to see the effect of having multiple modes is to compare a single mode solution to a two-mode solution, the general form of which may be written as
\be F(\xi) = b_{n_1}\, \xi^{n_1} \,+\, b_{n_2}\, \xi^{n_2}\,.\end{equation}
The charge, $Q_P$, of such a solution is:
\be Q_P = \frac12 (n_1\, b_{n_1}^2 + {n_2}\, b_{n_2}^2),\end{equation}
and the constraint (\ref{reg10n}) reads $c = b_{n_1}^2 + b_{n_1}^2$. 

To consider the effect of even more modes, we can consider a sum of sine functions\footnote{The cosine is a poor choice since it has a constant piece, in conflict with the definition (\ref{G10n}).}, thus giving an infinite mode superposition. So we consider $F(\xi)$ of the form 
\begin{align*}
F(\xi)=\tilde{b}_{\alpha}\sin \alpha \xi + \tilde{b}_{\beta}\sin \beta\xi~,
\end{align*}
carrying charge 
\begin{align}
Q_{P} = \frac{1}{2}\sum_{n=0}^{\infty} (2n+1) \left(\frac{\tilde{b}_{\alpha}\alpha^{2n+1}+\tilde{b}_{\beta}\beta^{2n+1}}{(2n+1)!} \right)^{2}~,
\end{align}
the constraint (\ref{reg10n}) now reads:
\begin{align}
c=\sum_{n=0}^{\infty}  \left(\frac{\tilde{b}_{\alpha}\alpha^{2n+1}+\tilde{b}_{\beta}\beta^{2n+1}}{(2n+1)!} \right)^{2} ~.
\end{align}

\begin{figure}[H]\centering
\begin{subfigure}{0.49\textwidth}\centering
 \includegraphics[width=\textwidth]{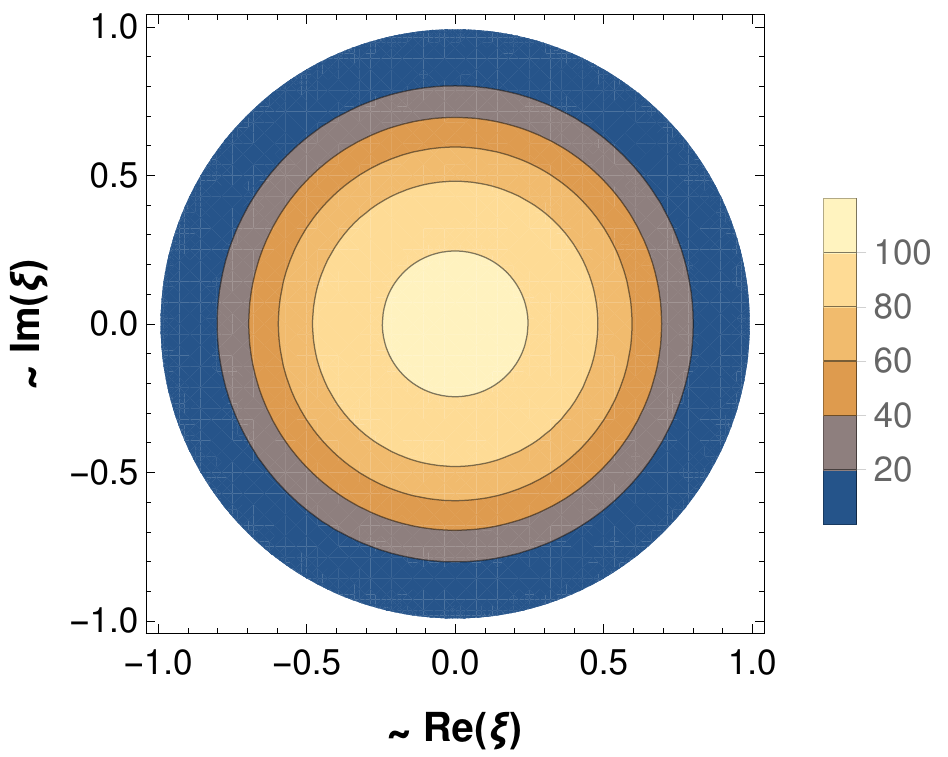}
 \caption{$F(\xi) = 10\xi^2 $}
\end{subfigure}\hspace*{0.04\textwidth}
\begin{subfigure}{0.49\textwidth}\centering
 \includegraphics[width=\textwidth]{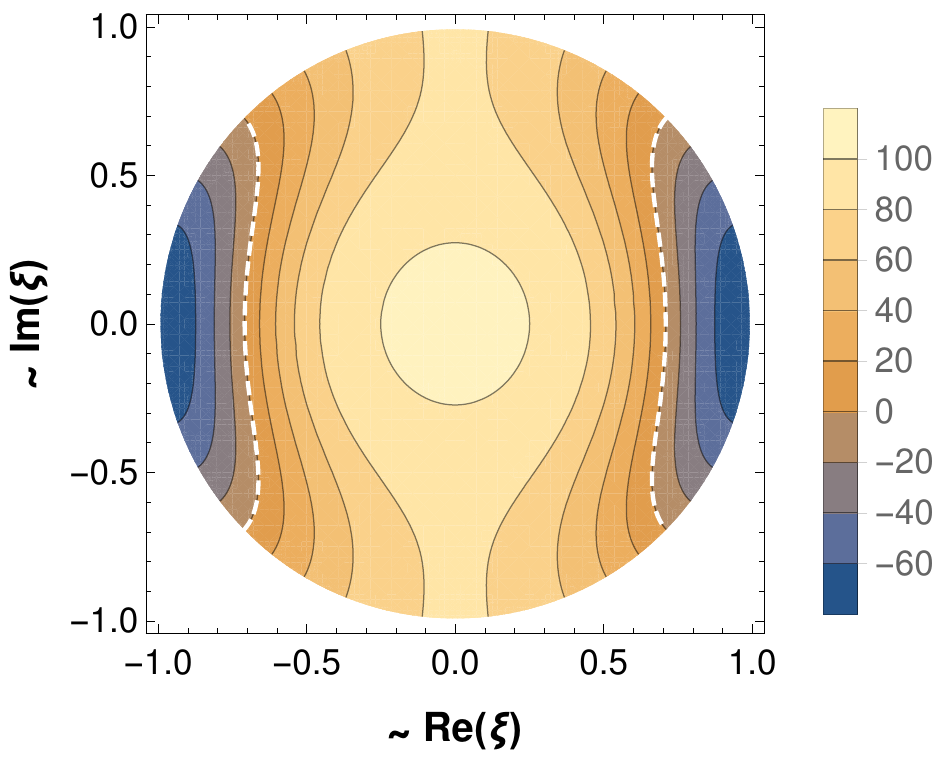}
 \caption{$F(\xi) = \sqrt{20}(2\xi^{2}+\xi^{4})$}
\end{subfigure}
 \caption{Contour plots of $\tilde{g}_{yy}$ with $(a=1,R_{y}=1,\theta=0,c=100)$  for specified $F(\xi)$. Plotted on the unit disc $|\xi|\leq 1$ with radial direction rescaled to $\tan r$. The boundaries between positive and negative domains ($\tilde{g}_{yy}=0$), if present, are separated by a dashed white line.}
 \label{fig:NonLOCplots1}
\end{figure}

\begin{figure}[H]\centering
\begin{subfigure}{0.49\textwidth}\centering
 \includegraphics[width=\textwidth]{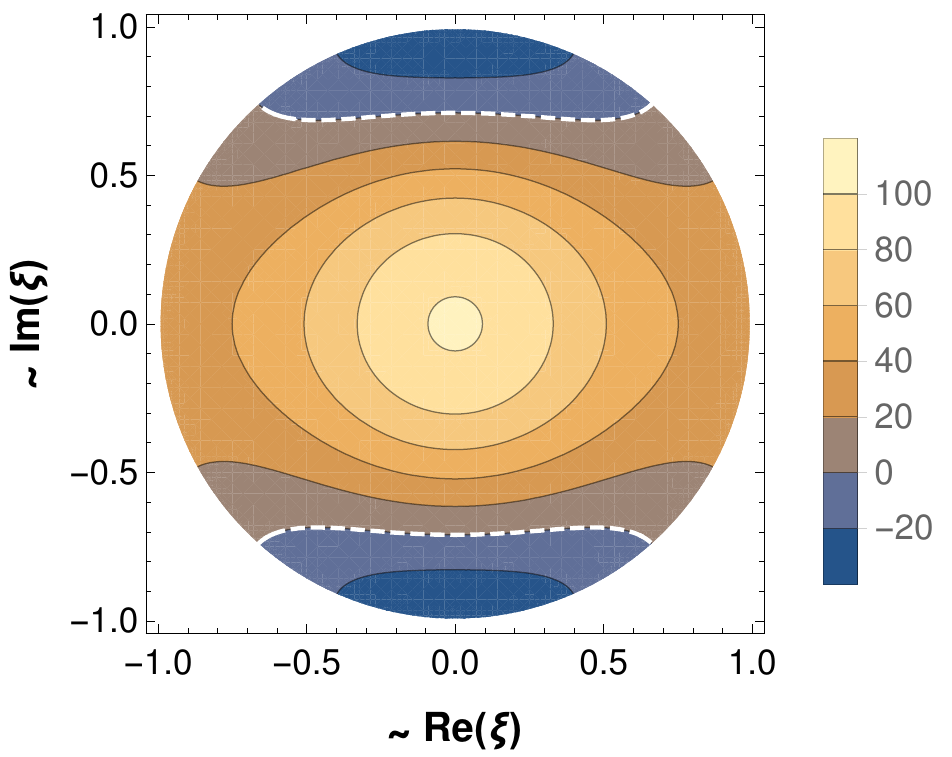}
 \caption{$F(\xi) = 9.86\sin \xi$}
\end{subfigure}\hspace*{0.04\textwidth}
\begin{subfigure}{0.49\textwidth}\centering
 \includegraphics[width=\textwidth]{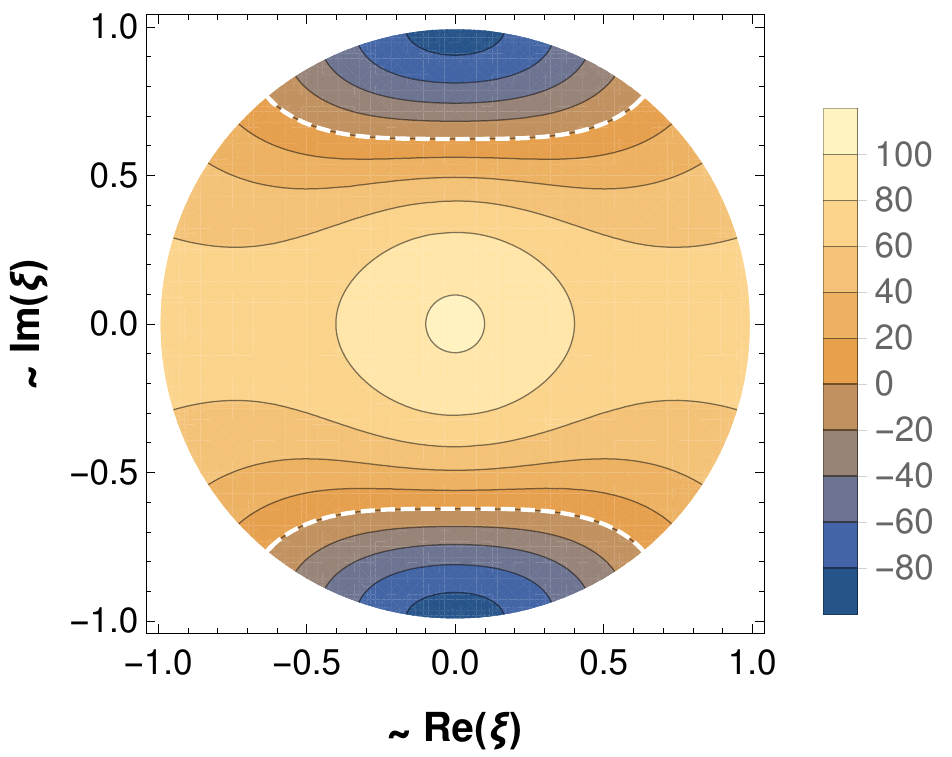}
 \caption{$F(\xi) = 2.30(2\sin \xi+\sin 2\xi)$}
\end{subfigure}
 \caption{Contour plots of $\tilde{g}_{yy}$ with $(a=1,R_{y}=1,\theta=0,c=100)$  for specified $F(\xi)$. Plotted on the unit disc $|\xi|\leq 1$ with radial direction rescaled to $\tan r$. The boundaries between positive and negative domains ($\tilde{g}_{yy}=0$) are separated by a dashed white line.}
 \label{fig:NonLOCplots2}
\end{figure}

To analyze how the geometry is altered by adding more modes, as in the examples above, we have produced contour plots of $\tilde{g}_{yy}$ (as defined in (\ref{ghatyy})) for $F(\xi)$ proportional to the functions
\begin{align*}
\xi^{2}~, \qquad 2\xi^{2}+\xi^{4}~, \qquad \sin \xi ~, \qquad 2\sin \xi +\sin 2\xi~,
\end{align*}
in Fig. \ref{fig:NonLOCplots1} and \ref{fig:NonLOCplots2}. In all such contour plots (Figs. \ref{fig:NonLOCplots1},  \ref{fig:NonLOCplots2}, \ref{fig:LOCplots1}, \ref{fig:LOCplots2}) we set $(a=1,R_{y}=1,\theta=0)$ and scale $F(\xi)$ so that $c=100$. These choices ensure there is a large throat region for $a<r<\sqrt{c}$, as is necessary for microstate geometries. Also note that the plots are made on the complex $\xi$-plane, although the radial direction has been rescaled to $\tan r$, to emphasize the different features.   

It is clear from the plots that the $y$ circle for solutions with more than one mode no longer deforms uniformly as $r$ varies. This is obvious in the plots since the circular symmetry is broken when more than one mode is turned on. We also see that the negative domains, when present, does not wrap the origin. This is a necessary condition to avoid CTC's in the geometry (see Section \ref{10n-y-circle}), and all of the examples we consider clearly exhibit this property.

\subsection{Localized momentum wave examples}\label{sec:locmomex}

For the solutions of the previous section, the momentum charge was smeared  around the $y$-circle, but for multi-mode solutions this is no longer necessary. We can demonstrate this most strongly by considering holomorphic functions $F$ so that $\abs{F}$ has a sharp peak on the $\abs{\xi}=1$ disk boundary. A simple candidate for such an $F$ is built by fixing it such that $|F|^2$ is proportional to a \emph{wrapped Cauchy distribution} on the boundary:
\begin{equation}\label{eq:cauchydistr}
 F(\xi) = c^{1/2} (e^{2\gamma}-1)^{1/2}\frac{\xi}{e^{\gamma} - \xi e^{-i\kappa}}.
 \end{equation}
 Note that the constant $c$ corresponds to L2-norm of $F$ at the boundary previously defined in \eqref{cval10n}. The corresponding  Fourier coefficients are (for $n\geq 1$):
\be b_n =  c^{1/2} (e^{2\gamma}-1)^{1/2} e^{-(n-1)i\kappa} e^{-n\gamma}.\end{equation}
These are real when $\kappa=0$ or $\kappa=\pi$. The momentum charge of this solution is:
\be Q_P = c \,(\coth\gamma + 1) \sim  \frac{c}{\gamma} + O(\gamma^0),\end{equation}
where we give the behaviour for small $\gamma$. Note that (\ref{reg10n}) is satisfied by construction.

\begin{figure}[H]\centering
\begin{subfigure}{0.49\textwidth}\centering
 \includegraphics[width=\textwidth]{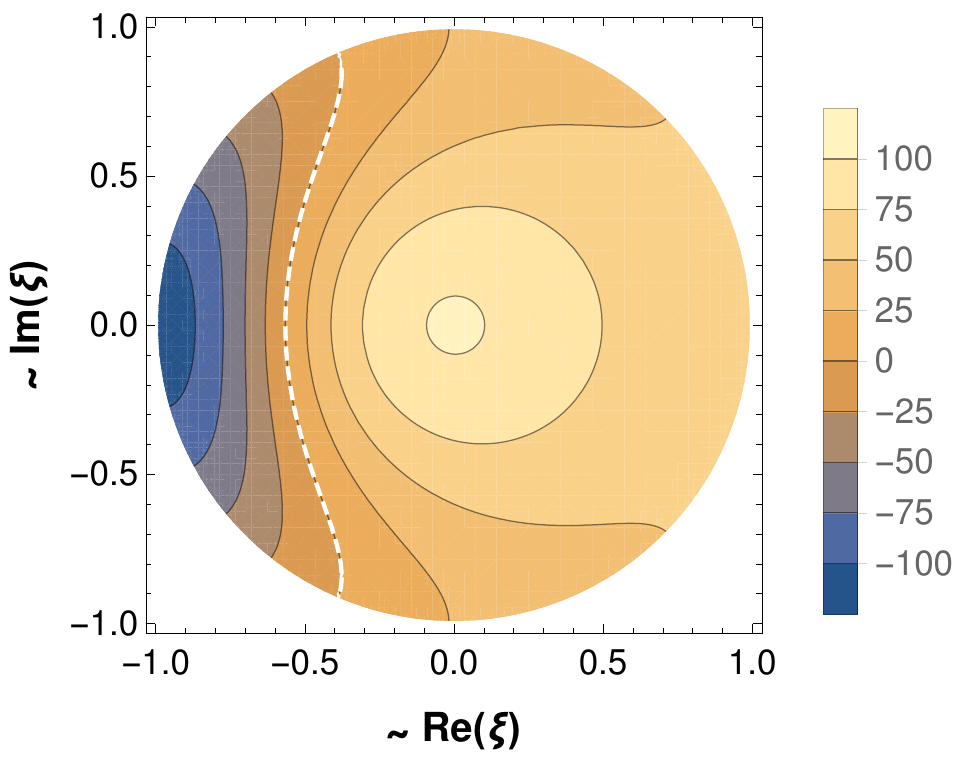}
 \caption{$\kappa=\pi/\sqrt{2}$, $\gamma =1$}
\end{subfigure}\hspace*{0.04\textwidth}
\begin{subfigure}{0.49\textwidth}\centering
 \includegraphics[width=\textwidth]{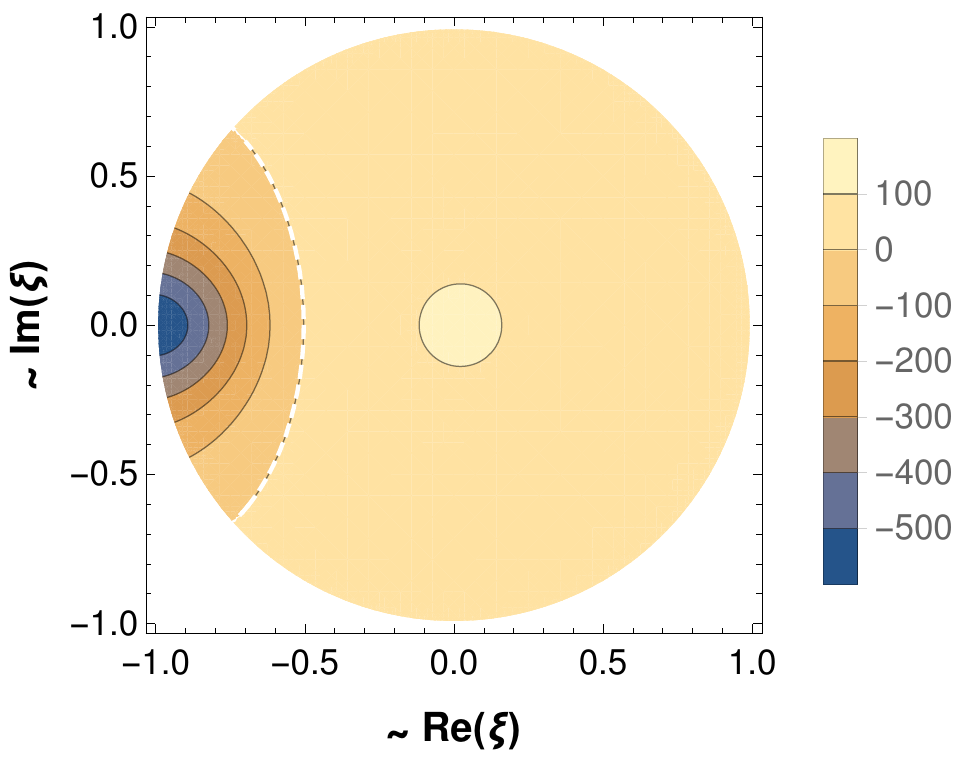}
 \caption{$\kappa=\pi/\sqrt{2}$, $\gamma=0.3$}
\end{subfigure}
 \caption{Contour plots of $\tilde{g}_{yy}$ with $(a=1,R_{y}=1,\theta=0,c=100)$  for specified $F(\xi)$. Plotted on the unit disc $|\xi|\leq 1$ with radial direction rescaled to $\tan r$. The boundaries between positive and negative domains ($\tilde{g}_{yy}=0$) are separated by a dashed white line.}
 \label{fig:LOCplots1}
\end{figure}

\begin{figure}[h]\centering
\begin{subfigure}{0.6\textwidth}\centering
 \includegraphics[width=\textwidth]{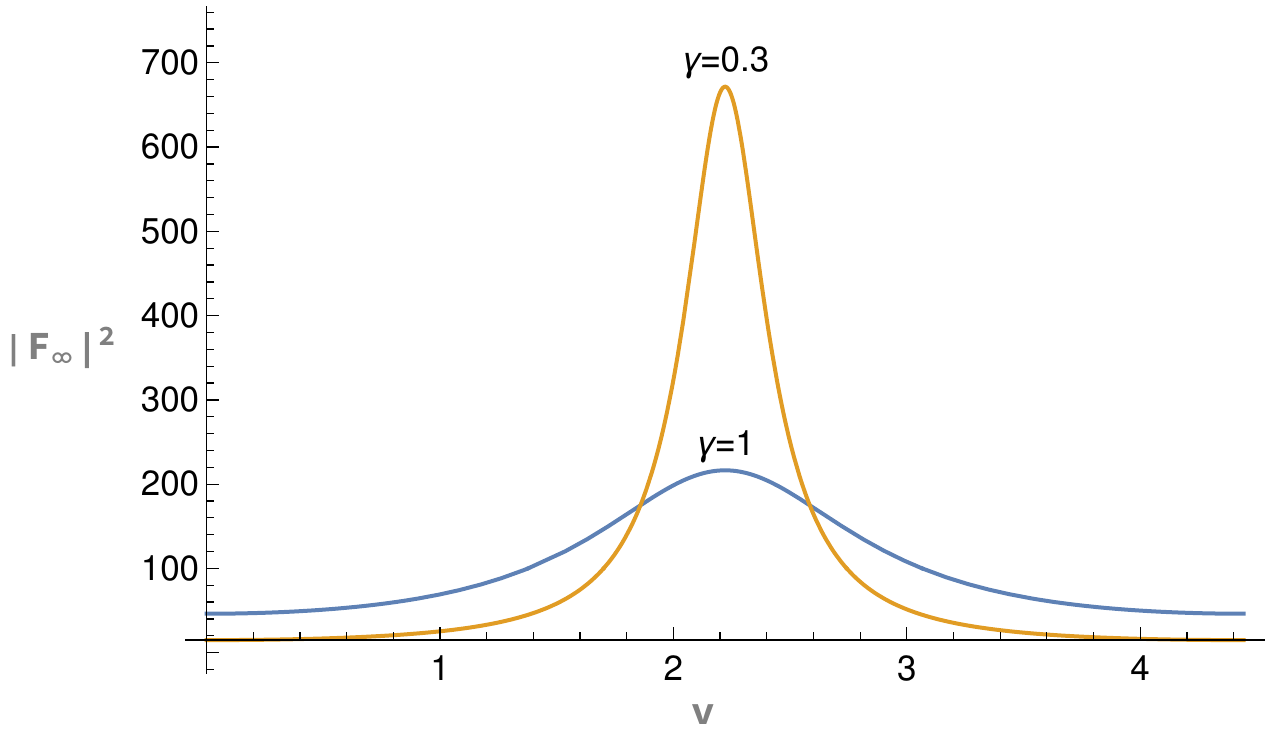}
\end{subfigure}\hspace*{0.04\textwidth}
 \caption{Plot of $\abs{F_{\infty}(v)}^{2}$ for $v\in [0,\sqrt{2}\pi)$ where $F$ is proportional to a Cauchy distribution (\ref{eq:cauchydistr}) with $\gamma$ as labeled and $\kappa=\pi/\sqrt{2}$.} 
 \label{fig:PeaksPlot}
\end{figure}

 There are two free parameters: $\kappa$ determines the location of the peak of the function in the compact direction as $\sqrt{2}v/R_y\sim \kappa$; $\gamma$ determines the strength of this peak. The only constraint on $\gamma$ is $\gamma>0$; taking $\gamma$ smaller makes the peak more pronounced. Quantitatively, if we write:
\be Q_P =  \frac{1}{\sqrt{2}\pi R_y}\int_0^{\sqrt{2}\pi R_y} \delta Q_{P}(v)\,dv\,,\end{equation}
where we define $\delta Q_{P}(v)$ through (\ref{QP10n}), then we see that for small $\gamma$:
\be \delta Q_{P}(v=0) \sim \frac{4c}{\gamma^2}.\end{equation}
Since $Q_P\sim c/\gamma$, we see that taking $\gamma$ small clearly does ``localize'' the momentum charge at $v\sim\kappa$. We illustrate this in Fig. \ref{fig:LOCplots1} by plotting $\tilde{g}_{yy}$ with $\kappa=\pi/\sqrt{2}$ for $\gamma\in (1,0.3)$. The figures show qualitatively, very different behavior to the single mode solution in Fig. \ref{fig:NonLOCplots1}. As the peak strength increases the negative domain becomes more localized to the boundary. To show how peaked the momentum waves are for these examples we also plot $\abs{F_{\infty}(v)}^{2}$ on the boundary in Fig. \ref{fig:PeaksPlot}.

\begin{figure}[h]\centering
\begin{subfigure}{0.49\textwidth}\centering
 \includegraphics[width=\textwidth]{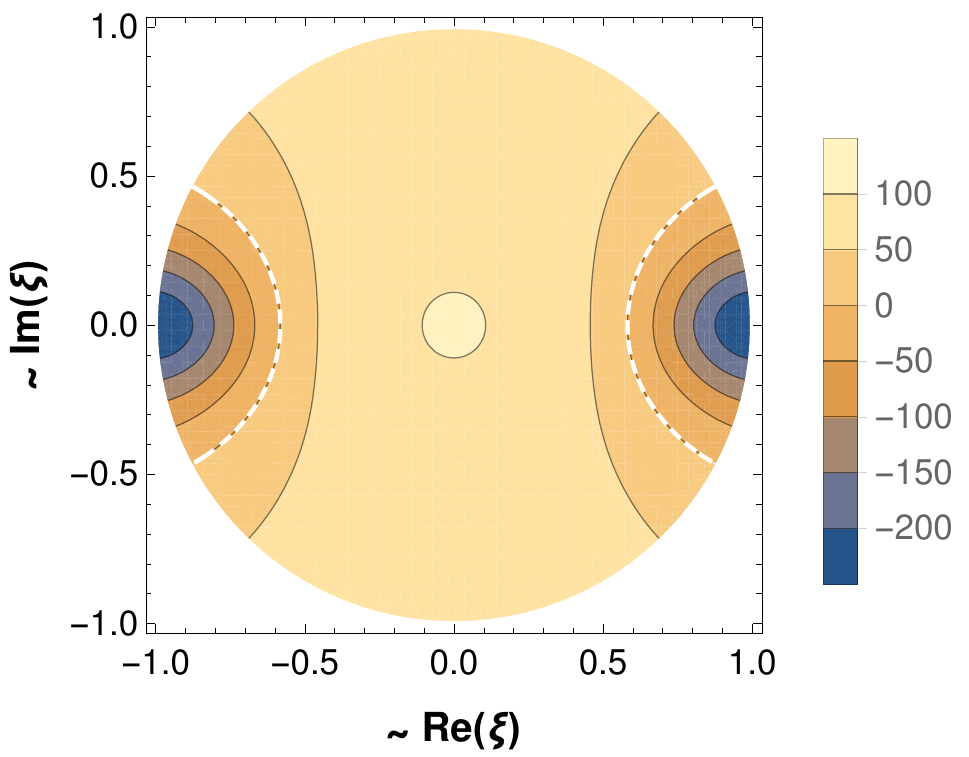}
 \caption{$(\kappa_{1}=\pi/\sqrt{2},\gamma_{1}=0.3)$ and $(\kappa_{2}=0,\gamma_{2}=0.3)$}
\end{subfigure}\hspace*{0.04\textwidth}
\begin{subfigure}{0.49\textwidth}\centering
 \includegraphics[width=\textwidth]{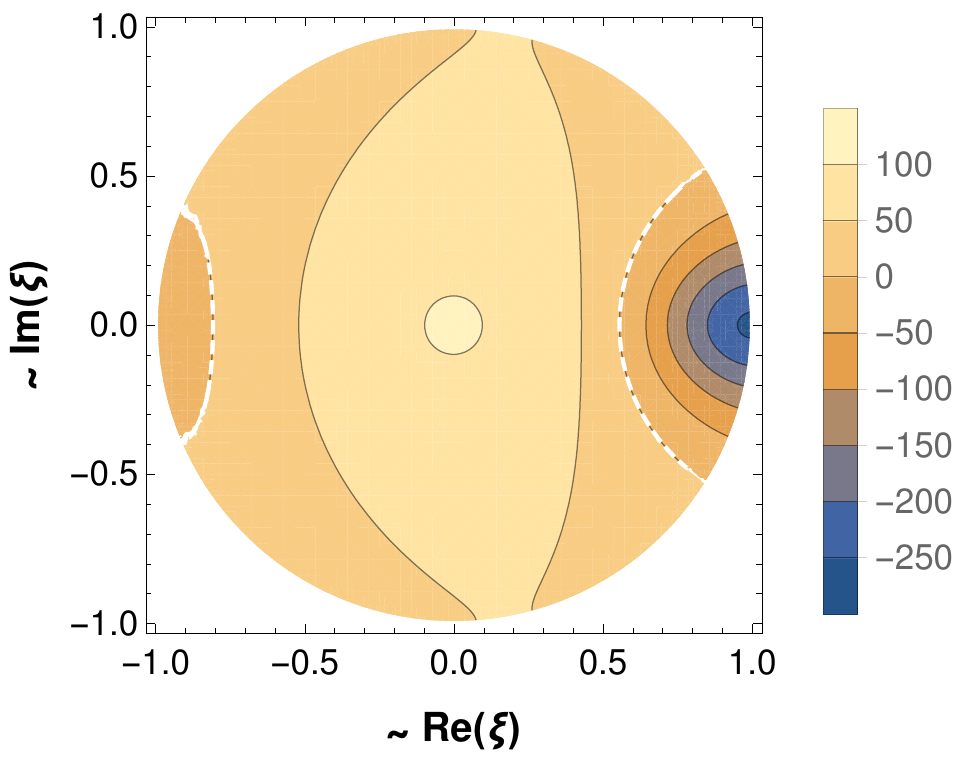}
 \caption{$(\kappa_{1}=\pi/\sqrt{2},\gamma_{1}=1)$ and $(\kappa_{2}=0,\gamma_{2}=0.3)$}
\end{subfigure}
 \caption{Contour plots of $\tilde{g}_{yy}$ with $(a=1,R_{y}=1,\theta=0,c=100)$  for specified $F(\xi)$. Plotted on the unit disc $|\xi|\leq 1$ with radial direction rescaled to $\tan r$. The boundaries between positive and negative domains ($\tilde{g}_{yy}=0$) are separated by a dashed white line.}
 \label{fig:LOCplots2}
\end{figure}

It is also easy to take linear combinations of such distributions, for example:
\begin{equation}\label{eq:twocauchydistr}
 F(\xi) = C \left(  (e^{2\gamma_1}-1)^{1/2}\frac{\xi}{e^{\gamma_1} - \xi e^{-i\kappa_1}} + (e^{2\gamma_2}-1)^{1/2}\frac{\xi}{e^{\gamma_2} - \xi e^{-i\kappa_2}} \right),
 \end{equation}
where $C$ is a constant such that (\ref{reg10n}) is satisfied. Fixing $C$ so that $c=100$ again, we have made plots with $(\kappa_{1}=\pi/\sqrt{2},\kappa_{2}=0)$ in Fig. \ref{fig:LOCplots2}. The figures show the behavior one would expect, as the strength of one of the peaks decreases the distribution interpolates smoothly back to the form of the single peak solutions.

\section{The emergence of holomorphy  in superstrata}
\label{sect:superstrata}

Having provided the new families of metrics, we now return to the beginning and give more complete details of our new holomorphic-wave superstrata.  In particular, we give the expressions for the fields that make up electromagnetic fluxes and show how these lead to the solutions described in Section \ref{sect:Examples}.  We will start with standard approach solving of the BPS equations and highlight how the holomorphic functions naturally arise.

\subsection{The first BPS layer}
\label{sect:Layer1}

The construction of the superstrata starts by requiring one to solve the first layer of BPS equations: a linear system of equations for three pairs of functions and two-forms, $(Z_1,\Theta_2)$,  $(Z_2,\Theta_1)$ and $(Z_4,\Theta_4)$, on the $\IR^4$ base (with coordinates $r,\theta,\varphi_1,\varphi_2$):\footnote{Those equations are linear because the connection one-form $\beta$ is not affected by the superstratum modes and retains the seed value \eqref{betadefn}.}
\begin{eqnarray}
\label{BPSLayer1}
 *_4 \mathcal{D} \dot{Z}_1 &=&  \mathcal{D} \Theta_2\,, \qquad \mathcal{D}*_4\mathcal{D}Z_1 ~=~ -\Theta_2\wedge d\beta\,,\qquad \Theta_2~=~*_4 \Theta_2\,,
\cr
 *_4 \mathcal{D} \dot{Z}_2 &=&  \mathcal{D} \Theta_1\,,\qquad \mathcal{D}*_4\mathcal{D}Z_2 ~=~ -\Theta_1\wedge d\beta\,,\qquad \Theta_1~=~*_4 \Theta_1\,,
\label{BPSlayer1} \\
 *_4 \mathcal{D} \dot{Z}_4 &=&  \mathcal{D}  \Theta_4\,,\qquad \mathcal{D}*_4\mathcal{D}Z_4 ~=~ -\Theta_4\wedge d\beta\,,\qquad \Theta_4~=~*_4 \Theta_4\,.
\nonumber
\end{eqnarray}
where the dot denotes $\frac{\partial}{\partial v}$ and $\mathcal{D}$ is defined by
\begin{equation}
\mathcal{D} ~\equiv~  d_4 ~-~ \beta\wedge \frac{\partial}{\partial v}\,.
\end{equation}
The operator $ d_4$ is simply the exterior derivative on $\IR^4$.

The $Z_I$ and  $\Theta_J$  determine the tensor gauge fields of IIB supergravity.  For details see \cite{Giusto:2013rxa,Bena:2011dd,Bena:2015bea,Bena:2017geu,Bena:2017xbt}. As noted earlier, the metric warp factor,   $\cP$,  is determined by the solution of the first layer of the BPS system and is given by (\ref{Pform}).

\subsubsection{The fundamental solutions}
\label{sect:basesols}

There are two classes of fundamental modes to the first layer of the BPS system.  They were first obtained by solution generating methods \cite{Mathur:2011gz, Lunin:2012gp,Shigemori:2013lta,Giusto:2013bda}. The  ``original'' class is defined by:
\begin{equation}
\begin{aligned}
 \widetilde{z}_{k,m,n} &=\,R_y \,\frac{\Delta_{k,m,n}}{\Sigma}\, \cos{v_{k,m,n}}, \\
 \widetilde{\vartheta}_{k,m,n}&=-\sqrt{2}\,
\Delta_{k,m,n}
\biggl[\left((m+n)\,r\sin\theta +n\left({m\over k}-1\right){\Sigma\over r \sin\theta}  \right)\Omega^{(1)}\sin{v_{k,m,n}} \\
&\hspace{20ex}
 +\left(m\left({n\over k}+1\right)\Omega^{(2)} +\left({m\over k}-1\right)n\, \Omega^{(3)}\right) \cos{v_{k,m,n}} \biggr]
\,,
\end{aligned} 
\label{zthetatilde}
\end{equation}
where
\begin{equation}
\label{selfdualbasis}
\begin{aligned}
\Omega^{(1)} &~\equiv~ \frac{dr\wedge d\theta}{(r^2+a^2)\cos\theta} + \frac{r\sin\theta}{\Sigma} d\varphi_1\wedge d\varphi_2\,,\\
\Omega^{(2)} &~\equiv~  \frac{r}{r^2+a^2} dr\wedge d\varphi_2 + \tan\theta\, d\theta\wedge d\varphi_1\,,\\
 \Omega^{(3)} &~\equiv~ \frac{dr\wedge d\varphi_1}{r} - \cot\theta\, d\theta\wedge d\varphi_2\,.
\end{aligned}
\end{equation}
The ``supercharged'' class is defined by \cite{Ceplak:2018pws}:
\begin{equation}
\label{zthetahat}
 \widehat{z}_{k,m,n} ~=~ 0\,, \qquad \widehat{\vartheta}_{k,m,n} =
\sqrt{2}\, \Delta_{k, m, n}\left[\,\frac{\Sigma}{r\sin\theta}\, \Omega^{(1)}\, \sin{v_{k,m,n}}+ \left(\Omega^{(2)} + \Omega^{(3)}\right)\cos{v_{k,m,n}}\,\right]\,.
\end{equation}

Both these sets functions and fields satisfy:
\begin{equation}
*_4\cD\,\dot{{z}}_{k,m,n} = \cD\,{\vartheta}_{k,m,n}, \qquad \cD *_4 \cD\, {z}_{k,m,n} = - {\vartheta}_{k,m,n} \wedge d\beta,\qquad {\vartheta}_{k,m,n} = *_4 {\vartheta}_{k,m,n},\\
\end{equation}
and so general superpositions can be used to create solutions to (\ref{BPSLayer1}).

The supercharged solutions are particularly simple and it is worth noting that the supercharged two-forms, $\widehat{\vartheta}_{k,m,n}$, satisfy simpler equations: 
\begin{equation}
 \cD\,\widehat{\vartheta}_{k,m,n} = 0, \qquad \widehat{\vartheta}_{k,m,n} \wedge d\beta=0,\qquad \widehat{\vartheta}_{k,m,n} = *_4 \widehat{\vartheta}_{k,m,n}\,.
 \label{schg1stlayer}
\end{equation}
%

\subsubsection{The first layer of the superstratum}
\label{sect:Layer1ss}

From the perspective of the holographic dictionary, and from solution generating techniques,  the most fundamental part of the solution is $(Z_4, \Theta_4)$.  This is because the superstratum states {\it at linear order} only involve excitations of $(Z_4, \Theta_4)$.  The other fields (including the other $Z_I$ and $\Theta_J$) are not excited until second, or even higher orders in the perturbative momentum states. Indeed, solution generating methods also indicate that the superstratum states should not involve excitations of $(Z_2,\Theta_1)$.  Thus the superstratum retains the almost-trivial ``supertube'' Ansatz for $(Z_2,\Theta_1)$:
\begin{equation}
\qquad Z_2 ~=~ \frac{Q_5}{\Sigma} \,, \qquad  \Theta_1 ~=~ 0\,.
\label{Z2Theta1}
\end{equation}
All the other fields can then, in principle,  depend on the details of the superstratum fluctuations.   The limitation of solution generating methods is that they only produce results of global actions of the ``small'' $\Neql{4}$ superconformal algebra.  However, the linearity of the BPS equations means that one can take arbitrary, independent superpositions of such solutions and arrive at the holographic dual of general coherent superspositions of states of the form (\ref{SSCFTstates}).  

From the perspective of supergravity, there appears to be complete democracy between all the pairs $(Z_I,\Theta_J)$ and one can, in principle, use (\ref{zthetatilde}) and (\ref{zthetahat}) to excite them all independently.   However, smoothness of the overall solution requires the imposition of ``coiffuring constraints'' 
\cite{Bena:2013ora,Bena:2014rea,Bena:2015bea,Bena:2016agb,Heidmann:2019zws} between Fourier coefficients of the $(Z_I,\Theta_J)$.  These constraints typically mean that if one chooses excitations of two of the pairs $(Z_I,\Theta_J)$, then the third is fixed\footnote{There can be some restrictions in that the constraints are quadratics with reality requirements.}    \cite{Bena:2015bea}.

It is, however, important to remember the guiding principle underlying the  basic superstrata that describe the holographic dual of the states (\ref{SSCFTstates}): $(Z_4,\Theta_4)$ is fundamental, and $(Z_2,\Theta_1)$ should remain trivial\footnote{As indicated above, there are more general   options  in supergravity \cite{Bena:2015bea} but these are not the holographic duals of the states (\ref{SSCFTstates}).}. 

Thus we start from (\ref{Z2Theta1}) and take: 
\begin{equation}
\qquad Z_4 ~=~   \sum_{k,m,n} \,b_{k,m,n} \, \widetilde{z}_{k,m,n} \,, \qquad  \Theta_4 ~=~    \sum_{k,m,n} \, \Big( \,b_{k,m,n} \, \widetilde{\vartheta}_{k,m,n} ~+~ c_{k,m,n} \, \widehat{\vartheta}_{k,m,n} \Big)\,,
\label{Z4expansion}
\end{equation}
These fields lead one to introduce the two fundamental holomorphic functions:
\begin{equation}
G_1(\xi,\chi,\eta) ~\equiv~   \sum_{k,m,n} \,b_{k,m,n} \, \chi^k \,  \mu^{m} \, \xi^n \,, \qquad G_2(\xi,\chi,\eta)  ~\equiv~   \sum_{k,m,n} \,c_{k,m,n} \, \chi^k \,  \mu^{m} \, \xi^n \,,
\label{sec:emergence:holfns}
\end{equation}
then one has
\begin{equation}
 Z_4 ~=~ \frac{R_y }{2\,\Sigma}\,\left(G_1 + \bar{G}_1 \right) \,, \qquad   \Theta_4 ~=~  \widetilde \Theta_4 ~+~   \frac{1}{\sqrt{2}}\, \bigg[\,\frac{i\Sigma}{r\sin\theta}\, \Omega^{(1)}\, \left( \bar{G}_2-G_2 \right)   + \Big(\Omega^{(2)} + \Omega^{(3)}\Big) \, \left(G_2 + \bar{G}_2 \right) \bigg] \,,
\end{equation}
 where $\widetilde \Theta_4$ is linear in real and imaginary combinations of $G_1$, its derivatives and primatives that will be explicitly derived in Section \ref{sect:superstrata:complex}. (That is, one needs to integrate or differentiate $G_1$ so as to get the  factors of $k,m$ and $n$ that appear in  $\widetilde{\vartheta}_{k,m,n}$ given in (\ref{zthetatilde}).)  The important point is that $Z_4$ and $\Theta_4$ have elementary expressions that are linear in terms of $G_1$ and $G_2$.

We are still left with the apparent freedom of choosing another set of Fourier series, or two more holomorphic functions, $H_1$ and $H_2$
\begin{equation}
\begin{aligned}
\qquad Z_1 &~=~   \frac{Q_1}{\Sigma} ~+~\frac{R_y}{2\, Q_5} \,  \sum_{k,m,n} \,b'_{k,m,n} \, \widetilde{z}_{k,m,n}  ~=~ \frac{Q_1}{\Sigma} ~+~  \frac{R_y^2}{4\, Q_5\,\Sigma}\,\left(H_1 + \bar{H}_1 \right)  \,, \\
  \Theta_2 &~=~   \frac{R_y}{2\, Q_5} \,  \sum_{k,m,n} \, \Big( \,b'_{k,m,n} \, \widetilde{\vartheta}_{k,m,n} ~+~ c'_{k,m,n} \, \widehat{\vartheta}_{k,m,n} \Big) \\
 & ~=~  \widetilde \Theta_2 ~+~    \frac{R_y}{2\sqrt{2}\,  Q_5} \,  \bigg[\,\frac{i\Sigma}{r\sin\theta}\, \Omega^{(1)}\, \left( \bar{H}_2-H_2 \right)   + \Big(\Omega^{(2)} + \Omega^{(3)}\Big) \, \left(H_2 + \bar{H}_2 \right) \bigg] \,,
\end{aligned}
\label{Z1expansion}
\end{equation}
where, for later convenience, we have included a factor of $\frac{R_y}{2\, Q_5}$.  Again $\widetilde \Theta_2$ is the appropriate linear combination of  real and imaginary parts of $H_1$, its derivatives and primatives.

 {\it A priori}, all the Fourier coefficients, $b, c, b'$ and $c'$ are independent.   However, smoothness requires that $(Z_1,  \Theta_2)$ are completely determined by  the coiffuring constraints \cite{Bena:2013ora,Bena:2014rea,Bena:2015bea,Bena:2016agb,Heidmann:2019zws}:
\begin{align}
b'_{k, m,n} & ~=\sum_{k_1 + k_2 =k \atop {m_1 + m_2 =m \atop n_1 + n_2 =n}} \, b_{k_1,m_1,n_1} \, b_{k_2,m_2,n_2}   \,,   \label{coiff1}\\
 \qquad   c'_{k, m,n}  &~=\sum_{k_1 + k_2 =k \atop {m_1 + m_2 =m \atop n_1 + n_2 =n}} \,  \bigg( 2\,b_{k_1,m_1,n_1} \, c_{k_2,m_2,n_2} - \frac{\,(k_1 n_2 - k_2 n_1)(k_1 m_2 - k_2 m_1) }{k_1 k_2 (k_1 +k_2)} \,b_{ k_1,m_1,n_1} \, b_{k_2,m_2,n_2}\bigg)   \,, 
 \label{coiff2} 
\end{align}
While these constraints originally appeared somewhat mysterious, they are simple when written in terms of the holomorphic functions.  In particular, 
\begin{equation}
H_1  ~=~G_1^2 \,, \qquad   H_2  ~=~2\, G_1 \, G_2 ~+~  \cQ(G_1)   \,,
 \label{coiff3} 
\end{equation}
where $\cQ(G_1)$ is a quadratic operator on $G_1$, its derivatives and primatives, arranged so as to create the second, more complicated term in (\ref{coiff2}).

Observe that, as an immediate consequence of  (\ref{Z4expansion}), (\ref{Z1expansion}) and (\ref{coiff3}) that we have
\begin{equation}
\cP   ~=~     Z_1 \, Z_2  -  Z_4^2 ~=~ \frac{1}{\Sigma^2}\,\big(Q_1 Q_5 ~-~  \coeff{1}{2}\,R_y^2\, |G_1|^2\,\big)   \,,
\end{equation}
Coiffuring actually means the cancellation of the purely holomorphic and purely anti-holomorphic parts and it sets the pattern for the second BPS layer.

To summarize, the solution to the first layer of BPS equations is defined by: (\ref{Z2Theta1}),  (\ref{Z4expansion}), (\ref{Z1expansion}), (\ref{sec:emergence:holfns})  and  (\ref{coiff2}) 
and it should be viewed as a momentum wave excitation, defined through $G_1$ and $G_2$, on the ``harmonic'' Ansatz for the supertube:
\begin{equation}
Z_1 ~=~ \frac{Q_1}{\Sigma}  \,, \qquad Z_2 ~=~ \frac{Q_5}{\Sigma} \,, \qquad   Z_4 ~=~ 0  \,, \qquad \Theta_I ~=~ 0  \,.
\label{STelectric}
 \end{equation}
%

\subsection{The second BPS layer}
\label{sect:Layer2}

The first BPS layer determines the three-form fluxes in the six-dimensional solution and some of the warp factors in the metric. 
The remaining components of the solution is determined through the second layer of BPS equations.  These are linear equations  for $\omega$ and $\mathcal{F}$:
\begin{eqnarray}
\label{BPSlayer2}
\mathcal{D} \omega + *_4  \mathcal{D}\omega +  \mathcal{F}\, d\beta &=&  Z_1 \Theta_1+ Z_2 \Theta_2  -2\,Z_4 \Theta_4 \,, \\
 *_4\mathcal{D} *_4\!\big(\dot{\omega} - \coeff{1}{2}\,\mathcal{D} \mathcal{F} \big) 
 &=& \partial_v^2 (Z_1 Z_2 - {Z}_4^2)  -(\dot{Z}_1\dot{Z}_2  -(\dot{Z}_4)^2 )-\coeff{1}{2} *_4\!\big(\Theta_1\wedge \Theta_2 - \Theta_4 \wedge \Theta_4\big)\,.
\nonumber
\end{eqnarray}
The sources on the right-hand sides of these equations are determined entirely by the solution to the first layer of BPS equations (\ref{BPSlayer1}).  One should note that the signs of the last equation depend upon duality conventions in $\IR^4$, and these equations are written with the somewhat unusual  duality conventions of \cite{Gutowski:2003rg,Bena:2011dd}.

The basic observation that started the coiffuring of solutions was that, when Fourier modes combine in the sources to (\ref{BPSlayer2}), the Fourier modes that add to make higher wave numbers generally produced singular solutions, whereas the combinations in which the wave numbers subtract from one another generally resulted in smooth solutions.   The imperative thus became to cancel (coiffure away) all the ``high-frequency sources in the second BPS layer.''  In  \cite{Heidmann:2019zws} it was finally shown that this could be achieved for the fully-general multi-mode solution, and the constraints (\ref{coiff1}) and  (\ref{coiff2}) were derived.  In the holomorphic formulation of the solutions, this cancellation simply means that the purely holomorphic and purely anti-holomorphic functions  cancel on the right-hand sides of (\ref{BPSlayer2}).

Coiffuring therefore means that the sources in (\ref{BPSlayer2}), while quadratic in $G_1, G_2, \overline{G}_1$ and $\overline{G}_2$ (and their derivatives and primitives), are actually sesquilinear, that is, they are linear in  $G_1, G_2$ and in $\overline{G}_1, \overline{G}_2$ (and all their derivatives and primitives).  This can be used to  simplify the BPS equations, (\ref{BPSlayer2}), but we will defer this discussion to the next section.

To find the physically correct, smooth solutions we will also need some of the homogeneous solutions to (\ref{BPSlayer2}). As we will see in practice, the holomorphic structure makes the finding of such solutions remarkable simple. 

\section{BPS equations in holomorphic form}
\label{sect:superstrata:complex}

The advantages of expressing the superstrata solutions in terms of holomorphic variables is already fairly evident.  It also gives a much simpler description of the ``coiffuring'' of Fourier modes: the sources of the last layer of BPS equations are sesquilinear in the solutions of the earlier layers.   As one might expect,  recasting all the BPS equations in terms of the complex variables significantly simplifies their form and greatly streamlines their solution. In this section, we  transform the BPS system to complex variables.  This is a necessarily technical exercise but we include it because it provides the superstratum aficionado  with a more accessible system of equations.  

There are two non-trivial parts to solving the second layer of BPS  equations.  First, one must find the particular solutions, and these are more easily extracted from holomorphic power series expansions.  The second step is to find homogeneous solutions that remove singularities and give the  correct asymptotics in the particular solutions.  These homogeneous solutions are also more easily constructed because they typically involve arbitrary holomorphic functions of one variable. In some instances, we find that we can generate the required homogeneous solutions by simply replacement rules (see, for example, (\ref{homsolk01}) and (\ref{SFtransf})).

\subsection{The holomorphic formalism}
\label{sect:complex expressions}
To start with, it is useful to collect various expressions that facilitate a full translation of the BPS fields into holomorphic coordinates (\ref{cplxcoords}).   It is, however, more convenient to introduce:
\begin{equation}
\mu ~\equiv~ \frac{\eta}{\chi}   ~=~    \cot  \theta \, e^{i \big (\frac{\sqrt{2} v}{R_y} - \varphi_1 - \varphi_2\big)}  \,.
 \end{equation}
The goal is to use this and (\ref{cplxcoords}) to transform the coordinates according to:
\begin{align}
(u,v,r,\theta,\varphi_{1},\varphi_{2})\to (u,\xi, \bar{\xi},\chi,\bar{\chi},\mu,\bar{\mu})~, \label{CoordTrans}
\end{align}
where one must keep in mind the $S^{5}$ constraint (\ref{constraint}). The combinations of the self dual forms (\ref{selfdualbasis}) that turn up in the $\Theta_{I}$ are:
\begin{align}
\Omega_{y} ~=~  \frac{1}{\sqrt{2}} \left(-\Omega^{(2)}+i r\sin\theta \, \Omega^{(1)} \right)\, ,  \qquad
\Omega_{z} ~=~  \frac{1}{\sqrt{2}}\left(\Omega^{(3)}+i \left(r\sin\theta - \frac{\Sigma}{r\sin\theta} \right)  \Omega^{(1)} \right)\, . \label{SDyz}
\end{align}
which can be written in complex coordinates as:
\begin{align}
\Omega_{y} ~=~  \frac{1}{\sqrt{2}} \left(-\Omega^{(2)}+ i\left(\frac{a\abs{\xi}\abs{\chi}}{1-\abs{\xi}^{2}}\right) \, \Omega^{(1)} \right)\, ,  \qquad
\Omega_{z} ~=~  \frac{1}{\sqrt{2}}\left(\Omega^{(3)}-i \left(\frac{a\abs{\eta}^{2}}{(1-\abs{\xi}^{2})\abs{\xi}\abs{\chi}} \right)  \Omega^{(1)} \right)\, . \label{SDyz2}
\end{align}
The  differentials can be related as:
\begin{equation}
  \begin{split}  
d\theta&= -\frac{\abs{\chi}^{2}\abs{\mu}}{2(1-\abs{\xi}^{2})}\left(\frac{d \mu}{\mu} + \frac{d\bar{\mu}}{\bar{\mu}} \right)\, , \\
d\varphi_{1} &= \frac{i}{2}\left(\frac{d\bar{\chi}}{\bar{\chi}}-\frac{d\chi}{\chi} \right)\, ,\\
d\varphi_{2}&= \frac{i}{2}\left( \frac{d\mu}{\mu}-\frac{d\bar{\mu}}{\bar{\mu}} -\frac{d\xi}{\xi}+\frac{d\bar{\xi}}{\bar{\xi}} +\frac{d\chi}{\chi}-\frac{d\bar{\chi}}{\bar{\chi}} \right) \, ,
  \end{split}
  ~~~~
  \begin{split}
  dv&= i\frac{R_y}{2\sqrt{2}} \left(\frac{d\bar{\xi}}{\bar{\xi}} - \frac{d\xi}{\xi} \right)\, ,\\
dr&=  \frac{a\abs{\xi}}{2(1-\abs{\xi}^{2})^{3/2}} \left(\frac{d\xi}{\xi} +  \frac{d\bar{\xi}}{\bar{\xi}} \right)\, .
  \end{split}
\end{equation}

The one form $\omega$ appearing in (\ref{sixmet}) and fixed by the second layer BPS equations (\ref{BPSlayer2}) can then be expanded as 
\begin{align}
\omega=\omega_{\xi} \left(\frac{d\xi}{\xi}+\frac{d\bar{\xi}}{\bar{\xi}} \right)-\omega_{\mu}\left(\frac{d\mu}{\mu}+ \frac{d\bar{\mu}}{\bar{\mu}} \right)+i\omega_{\chi} \left(\frac{d\bar{\chi}}{\bar{\chi}}-\frac{d\chi}{\chi}  \right) +2 \omega_{\delta} \, d\varphi_{2} \, ,\label{omegaAnsatz}
\end{align}
where the functions $(\omega_{\xi},\omega_{\mu},\omega_{\chi},\omega_{\delta})$ are necessarily real. The relationship between the components here and the ansatz used in (\ref{om21n}) can be easily worked out from
\begin{align}
\omega=\frac{2a^{2}}{r(a^{2}+r^{2})}\omega_{\xi} \, dr +\frac{4}{\sin 2\theta}\omega_{\mu} \, d\theta +2\omega_{\chi}\, d\varphi_{1} +2\omega_{\delta} \, d\varphi_{2}~. \label{omegaAnsatz2}
\end{align}

It will be extremely useful to introduce a (dependent) set of first order operators:
\begin{equation*}
\begin{split}
\mathcal{L}_{F}&= -i\frac{2\sqrt{2}}{R_y}\left(\xi \partial_{\xi}+\mu \partial_{\mu}\right)~,\\
\mathcal{L}_{\xi}&= \frac{1}{2}\left[\left( \frac{\abs{\xi}^{2}}{1-\abs{\xi}^{2}}\right)\chi \partial_{\chi} -\xi\partial_{\xi} \right]~, \\
\mathcal{L}_{\chi} &= \frac{i}{2}\left[ \left( \frac{1}{1-\abs{\chi}^{2}}\right) \mu \partial_{\mu}-\chi\partial_{\chi}+\left( \frac{\abs{\chi}^{2}}{1-\abs{\chi}^{2}} \right)\xi \partial_{\xi}  \right] ~,
\end{split} ~~~~~~
\begin{split}
\mathcal{L}_{\mu} &= \frac{1}{2}\left[ \mu \partial_{\mu} - \left(\frac{\abs{\mu}^{2}\abs{\chi}^{2}}{1-\abs{\xi}^{2}}\right)\chi\partial_{\chi}  \right] ~,\\
\mathcal{L}_{\delta}&= \frac{i}{2(1-\abs{\chi}^{2})} \left[ \abs{\xi}^{2}\mu \partial_{\mu} - \abs{\mu}^{2}\abs{\chi}^{2} \xi\partial_{\xi} \right] ~.
\end{split}
\end{equation*}
These operators are considered independent of the constraint (\ref{constraint}), that is, the result is the same whether the constraint is used to shuffle the complex variables before or after applying the operator. For example:
\begin{align*}
\mathcal{L}_{\xi}\xi = \mathcal{L}_{\xi}\left( \frac{1-\abs{\chi}^{2}-\abs{\mu}^{2}\abs{\chi}^{2}}{\bar{\xi}}\right)=-\frac{\xi}{2}~.
\end{align*}
However, to be consistent, one has to express everything in the coordinates $(\chi,\bar{\chi},\mu,\bar{\mu},\xi,\bar{\xi})$ without using $(\eta,\bar{\eta})$. 

These complex $\mathcal{L}_{i}$ operators can be combined with their complex conjugate $\bar{\mathcal{L}}_{i}$ to produce the real combinations:
\begin{align*}
\widetilde{\mathcal{L}}_{i} = \mathcal{L}_{i}+\bar{\mathcal{L}}_{i}\, . 
\end{align*}
These operators arise naturally from the gauge transformation (\ref{gaugetrf}), which now read:
\begin{align*}
u&\to u+U~, \qquad\qquad  ~ \, F \to F+ \widetilde{\mathcal{L}}_{F}U~,\\
\omega_{\xi} &\to \omega_{\xi} + \widetilde{\mathcal{L}}_{\xi}U ~, \qquad
\omega_{\chi} \to \omega_{\chi} + \widetilde{\mathcal{L}}_{\chi}U~, \\
\omega_{\mu} &\to \omega_{\mu} + \widetilde{\mathcal{L}}_{\mu}U ~,\qquad 
\omega_{\delta} \to \omega_{\delta} + \widetilde{\mathcal{L}}_{\delta}U~.
\end{align*}
It may also be of interest to convert the expressions of this  section into coordinates that make the $S^{5}$ of (\ref{constraint}) manifest. This is most easily done by rewriting the $\mathcal{L}$ operators in spherical coordinates (see appendix \ref{app:SphericalCoordinates}). 

\subsection{First layer of the superstratum}
\label{sect:basesolsholomorphic1}

Using the expressions of this section we can rewrite some of the first layer fields. To begin with the basic $(\widetilde{\vartheta}_{k,m,n},\widehat{\vartheta}_{k,m,n})$ become
\begin{align*}
\widetilde{\vartheta}_{k,m,n} &=\left[ m\left(1 + \frac{n}{k} \right)   \Omega_{y} +n\left(1-\frac{m}{k} \right)  \Omega_{z} \right]\chi^{k}\mu^{m}\xi^{n} \,+\, c.c. \, , \\
\widehat{\vartheta}_{k,m,n} &= \left(\Omega_{z}-\Omega_{y} \right)\chi^{k}\mu^{m}\xi^{n} \,+\, c.c. \, .
\end{align*}
Upon introducing the integrated pre-potential:
\begin{align*}
G_{1}^{(1)} = \sum_{k,m,n}\frac{b_{k,m,n}}{k} \chi^{k}\mu^{m}\xi^{n} \, ,
\end{align*}
as well as the combinations 
\begin{align}
A=\left[ \left(\chi \partial_{\chi}+\xi\partial_{\xi} \right)\mu\partial_{\mu}G_{1}^{(1)}-G_{2} \right] \qquad \text{and} \qquad B= \left[\left( \chi\partial_{\chi}-\mu\partial_{\mu}\right)\xi\partial_{\xi}G_{1}^{(1)}+G_{2} \right] \, , \label{ABdef}
\end{align}
the first layer fields discussed in Section \ref{sect:Layer1} then read:
\begin{align}
\begin{split}
Z_{1} &= \frac{Q_{1}}{\Sigma} + \frac{R_y^{2}}{4Q_{5}\Sigma} \left( G_{1}^{2}+\bar{G}_{1}^{2}\right)~, \\
Z_{2}&= \frac{Q_{5}}{\Sigma}~, \\
Z_{4} &= \frac{R_y}{2\Sigma} \left(G_{1}+\bar{G}_{1} \right)~, 
\end{split}
~~
\begin{split}
\Theta_1 &= 0 ~,\\
\Theta_2 &=   \frac{R_y}{Q_{5}}\,G_1  \left(A\, \Omega_{y} + B\, \Omega_{z}  \right)\,+\,c.c. ~, \\
\Theta_4 &= A\, \Omega_{y} + B\, \Omega_{z}\,+\,c.c. ~.
\end{split}
\label{1stLayerGenHolo}
\end{align}

\subsection{Second layer of the superstratum}
\label{sect:basesolsholomorphic2}

The second layer fields, $\cF$ and $\omega$, do not have closed-form solutions according to the two generic arbitrary functions $G_1$ and $G_2$. In this section, we rewrite the second layer of BPS equations (\ref{BPSlayer2}) in the holomorphic formalism; those equations were used to construct the families of explicit solutions in Section \ref{sect:Examples}. They can be decomposed as $\mathcal{O}_i = \mathcal{S}_i$, where $\mathcal{S}_i$ are the source terms on the right-hand side of \eqref{BPSlayer2}, induced by the first layer fields, and $\mathcal{O}_i$ are the differential operators acting on $\cF$ and $\omega$ on the left-hand side of \eqref{BPSlayer2}.

\subsubsection{The sources}
\label{sect:basesolsholomorphic2sources}

The sources of the second layer BPS equations (\ref{BPSlayer2}) read: 
\begin{align}
\mathcal{S}_{1} &= Z_{1}\Theta_{1} +Z_{2}\Theta_{2}-2Z_{4}\Theta_{4}~, \\
\mathcal{S}_{2} &= \ddot{\cP}- \left(\dot{Z}_{1}\dot{Z}_{2}-\dot{Z}^{2}_{4} \right) - \frac{1}{2}*_{4} \left(\Theta_{1} \wedge \Theta_{2} - \Theta_{4} \wedge \Theta_{4} \right)~,
\end{align}
where a dot represents a $\partial_{v}$ derivative. Using the first layer fields of the previous section these sources become:
\begin{align*}
\mathcal{S}_{1}&= -\frac{R_y}{\Sigma} \bar{G}_{1} \left[A \Omega_{y}+B\Omega_{z} \right]\,+\,c.c. \,,\\
\mathcal{S}_{2}&=  \frac{1}{2\Sigma^{2}} \left\lbrace 2\bar{G}_{1} \left(\xi\partial_{\xi}+\mu\partial_{\mu}  \right)^{2}G_{1}-\abs{\xi \partial_{\xi}G_{1}+\mu \partial_{\mu}G_{1}}^{2} -  \left[\left( \xi \partial_{\xi}+\mu \partial_{\mu} \right)G_{1} \right]^{2}   \right. \\
& \qquad \qquad\qquad \qquad  + \left. (A+B)^{2}+\abs{A-B}^{2} +2 \left( \abs{A}^{2}\frac{\abs{\xi}^{2}}{\abs{\mu}^{2}} +\abs{B}^{2}\frac{\abs{\mu}^{2}}{\abs{\xi}^{2}}\right) \right\rbrace\,+\,c.c.\,.
\end{align*}
It will be useful in Section \ref{sect:SecondLayerOperatosCC}, to have the first source expanded in the $(\Omega^{(1)},\Omega^{(2)},\Omega^{(3)})$ basis; defining 
\begin{align}
\mathcal{S}_{1} = \mathcal{S}^{(1)}_{1}\Omega^{(1)} + \mathcal{S}^{(2)}_{1}\Omega^{(2)} + \mathcal{S}^{(3)}_{1}\Omega^{(3)} \, ,
\end{align}
we have 
\begin{align*}
\mathcal{S}_{1}^{(1)}&=- \frac{i\,R_y\, \abs{\chi}}{\sqrt{2}\,a\abs{\xi}(1-\abs{\chi}^{2})}\,\left({ \abs{\mu}^{2}\left(G_{1}\bar{B}-\bar{G}_{1}B\right)-\abs{\xi}^{2}\left(G_{1}\bar{A}-\bar{G}_{1}A\right)}\right)~, \\
\mathcal{S}_{1}^{(2)}&= \frac{R_y\,(1-\abs{\xi}^{2})}{\sqrt{2}\,a^{2}(1-\abs{\chi}^{2})} \left( \bar{G}_{1}A \,+\, G_{1} \bar{A} \right)~, \\
\mathcal{S}_{1}^{(3)}&=  -\frac{R_y\,(1-\abs{\xi}^{2})}{\sqrt{2}\,a^{2}(1-\abs{\chi}^{2})} \left( \bar{G}_{1}B \,+\, G_{1} \bar{B} \right)~.
\end{align*}

For certain families of multi-mode solutions these sources become relatively simple. For example, the sources for the $(1,0,n)$ family reduce to
\begin{align*}
\mathcal{S}^{(1,0,n)}_{1} &= - \frac{R_y \abs{\chi}^{2}}{\Sigma}F\, \xi \partial_\xi F  \, \Omega_{z}\,+\,c.c.\,,\\
\mathcal{S}^{(1,0,n)}_{2} &= \frac{\abs{\chi}^{2}}{\Sigma^{2}} \left( \abs{\mu}^{2}\partial_\xi F\,\partial_\xi F\,+\,F(\xi\partial_{\xi})^{2}F \right)\,+\,c.c. \,.
\end{align*}
where $F=F(\xi)$, as defined in Section \ref{sect:10ni}.

\subsubsection{The operators}
\label{sect:SecondLayerOperatosCC}

The second layer BPS operators in (\ref{BPSlayer2}) read:
\begin{align*}
\mathcal{O}_{1}&= \mathcal{D} \omega + *_4  \mathcal{D}\omega +  \mathcal{F}\, d\beta \, , \\
\mathcal{O}_{2} &=  *_4\mathcal{D} *_4\!\big(\dot{\omega} - \coeff{1}{2}\,\mathcal{D} \mathcal{F} \big) \, .
\end{align*}
The first of these operators is explicitly self dual, so we can decompose it into $(\Omega^{(1)},\Omega^{(2)},\Omega^{(3)})$ components as 
\begin{align}
\mathcal{O}_{1} = \mathcal{O}^{(1)}_{1}\Omega^{(1)} + \mathcal{O}^{(2)}_{1}\Omega^{(2)} + \mathcal{O}^{(3)}_{1}\Omega^{(3)} \, . 
\end{align}
These operators when written in terms of the $\mathcal{L}$ operators of Section \ref{sect:complex expressions} become: 
\begin{align}
\mathcal{O}_{1}^{(1)} &=\frac{4a}{\abs{\xi}\abs{\chi}} \left[(1-\abs{\xi}^{2})\left(\widetilde{\mathcal{L}}_{\mu}\omega_{\xi}-\widetilde{\mathcal{L}}_{\xi}\omega_{\mu} \right) +(1-\abs{\chi}^{2})\left(\widetilde{\mathcal{L}}_{\delta}\omega_{\chi}-\widetilde{\mathcal{L}}_{\chi}\omega_{2} \right)  \right]~, \\
\mathcal{O}_{1}^{(2)} &= \frac{4}{\abs{\chi}^{2}}\left( \frac{1-\abs{\xi}^{2}}{\abs{\xi}^{2}}\right)\left[\abs{\chi}^{2}\left(\widetilde{\mathcal{L}}_{\delta}\omega_{\xi} - \widetilde{\mathcal{L}}_{\xi}\omega_{2} \right) +\abs{\xi}^{2}\left(\widetilde{\mathcal{L}}_{\chi}\omega_{\mu} - \widetilde{\mathcal{L}}_{\mu}\omega_{\chi} \right) \right] + \frac{\sqrt{2}R_y\abs{\chi}^{2}\abs{\mu}^{2}\mathcal{F}}{(1-\abs{\chi}^{2})^{2}}~,  \\
\mathcal{O}_{1}^{(3)} &=\frac{4 (1-\abs{\xi}^{2})}{\abs{\mu}^{2}\abs{\chi}^{2}} \left[ \abs{\mu}^{2}\abs{\chi}^{2} \left(\widetilde{\mathcal{L}}_{\chi}\omega_{\xi}-\widetilde{\mathcal{L}}_{\xi}\omega_{\chi} \right)+\left(\widetilde{\mathcal{L}}_{\mu}\omega_{2}-\widetilde{\mathcal{L}}_{\delta}\omega_{\mu}\right) \right]- \frac{\sqrt{2}R_y\abs{\chi}^{2}\abs{\xi}^{2}\mathcal{F}}{(1-\abs{\chi}^{2})^{2}} ~,
\end{align}
and 
\begin{align}
\mathcal{O}_{2}&= \frac{2}{a^{2}}\widetilde{\mathcal{L}}_{F} \left[ \frac{(1-\abs{\xi}^{2})^{3}}{1-\abs{\chi}^{2}} \left( \frac{\widetilde{\mathcal{L}}_{\xi}\omega_{\xi}}{\abs{\xi}^{2}} +\frac{\widetilde{\mathcal{L}}_{\mu}\omega_{\mu}}{\abs{\chi}^{4}\abs{\mu}^{2}} \right)+ \left(\frac{1-\abs{\xi}^{2}}{\abs{\chi}} \right)^{2}\left( \widetilde{\mathcal{L}}_{\chi}\omega_{\chi} + \frac{\widetilde{\mathcal{L}}_{\delta}\omega_{2}}{\abs{\xi}^{2}\abs{\mu}^{2}} \right) \right] \\
&  \qquad \qquad \qquad  \qquad \qquad \qquad \qquad \qquad -\frac{8(1-\abs{\xi}^{2})^{3}}{a^{2}(1-\abs{\chi}^{2})} \left(\frac{1}{\abs{\xi}^{2}} \mathcal{L}_{\xi}\bar{\mathcal{L}}_{\xi}+ \frac{1}{\abs{\mu}^{2}\abs{\chi}^{4}} \bar{\mathcal{L}}_{\mu}\mathcal{L}_{\mu}  \right)\mathcal{F}~. \notag 
\end{align}

In this form it is straightforward to demonstrate gauge invariance, one just needs to use commutators of the $\mathcal{L}$ operators. Using the insight from rewriting the $\mathcal{L}$ operators in spherical coordinates (Appendix \ref{app:SphericalCoordinates}), it is clear that once a Fourier decomposition is made in the periodic directions of the $S^{5}$, the technical difficulties in solving for general multi-mode superstrata arises due to the $(\mathcal{L}_{\xi},\mathcal{L}_{\mu})$ operators, since the other operators become algebraic in the Fourier modes.

\section{Conclusions}
\label{sect:conclusions}

There are many aspects of the microstate geometry programme, but one of the most important is that microstate geometries provide a gravitational mechanism that is able to support horizon-scale structure.  Indeed, there is good evidence  \cite{Gibbons:2013tqa,deLange:2015gca} that microstate geometries embody the only such mechanism. 
This means that any discussion of horizon-scale microstructure in which there is a semi-classical gravitational limit must involve a microstate geometry.  There is therefore a very broad interest in finding and exhibiting families of  microstate geometries.    This perspective has played a major role in the way that we organized this paper: we have catalogued a broad range of new families so that the results are accessible without needing to struggle through the details of their construction. 

One of the major themes of this paper is the holomorphic structure.  As we described in the introduction, the complete class of superstrata will be governed by two unconstrained holomorphic functions of three variables.  Realizing the most general solution is far beyond our current computational powers, but what we have succeeded in doing is to create superstrata that are governed by various unconstrained holomorphic functions of one variable.  This is done by restricting the functions of three variables by fixing particular powers in two of the variables while not constraining the dependence on the third variable.  The result is by far the largest and most extensive families of superstrata that have been constructed to date.  Indeed, much of the study of superstrata has focussed on the ``single mode'' examples that have very few (or no) free parameters.  Now we have explicit solutions based on arbitrary functions.

One of the obvious new things one can do with such an extensive set of degrees of freedom is to localize the momentum-carrying excitations at infinity and examine how they spread and diffuse into the core of the solution, or in the infra-red limit.  We have investigated this in Section \ref{sect: 10nExample} and in future work, we hope to use this to examine ``almost BPS'' superstrata. 

One of the very satisfying aspects of our analysis is that, for the simplest families, we could show that smoothness and the absence of CTC's actually placed no restrictions on the holomorphic functions.  This is a far-from-trivial result.  Indeed, smoothness and the absence of CTC's places  stringent bounds on combinations of  metric coefficients in the supergravity solution, and we naively expected that we would have to place some conditions on the ``Fourier coefficients'' appearing in the power series of our holomorphic functions.  On the contrary, for the $(1,0,n)$ and $(1,1,n)$ families we were able to use the special properties afforded by holomorphy to show that all the crucial bounds were satisfied without any constraints  on the holomorphic function.  This is significant because the holographic dictionary strongly suggests that there should be no constraints on the linear combinations of the corresponding coherent states, apart from requiring the state to have finite norm.  (In supergravity, the finiteness of the norm amounts to the convergence of the power series for the holomorphic functions.)  Based on this observation, we therefore expect that there will be no restriction on  the holomorphic functions  that underlie the most general superstrata.

The holomorphic structure also gives a natural mathematical interpretation to coiffuring the Fourier coefficients: The second layer of BPS equations, which determines the metric structure, must be sesquilinear in the underlying holomorphic function.  Indeed, our analysis has revealed what we suspect is a much deeper, and, as yet, not understood aspect of the holomorphic structure and the characterization of the entire solution in terms of a ``fundamental holomorphic pre-potential\footnote{We have not carefully defined this notion but, loosely, we  mean that all other holomorphic functions are obtained as derivatives of $F$. See, for example,  (\ref{G1k01}), (\ref{G21nApp}), (\ref{G1Fk0n}) and (\ref{PandFk0nGenApp}).}.''

One of the important technical aspects of solving the BPS equations has always been the careful choice of homogeneous solutions so as to cancel singularities and obtain  smooth geometries.  In this paper we have empirically discovered a way to generate the requisite homogeneous solutions by making transformations like:
\begin{equation} 
|F(\chi)|^2 ~\to~  |F(\chi)|^2 ~-~ c(\chi) ~-~c(\bar{\chi}) \,, 
\label{Genhomtrf}
\end{equation}
where $F$ is the  fundamental holomorphic pre-potential of the system. See for example,  (\ref{homsolk01}), (\ref{SFtransf}) and  (\ref{HomSolApp}).   The important point is that the function, $c$, in transformation (\ref{Genhomtrf}) produces homogeneous solutions of the second layer of the system of BPS equations and that these solutions are not just the obvious gauge transformations  (\ref{gaugetrf}).  Moreover, these solutions are precisely the ones  needed for smoothing out singularities in the solution.   

Thus, when expressed in terms of this ``holomorphic pre-potential,''  the solutions of the second layer of the BPS system have an invariance of the form (\ref{Genhomtrf}).  This is very reminiscent of the role, and behaviour, of  K\"ahler potentials and suggests that there should be a way to understand the entire BPS system in terms of such pre-potentials.  

Some important first steps along these lines were made in \cite{Tyukov:2018ypq}, in which it was shown that, in five dimensions, the first layers of the BPS system could be recast in terms of harmonic pre-potentials and solved by differentiating these pre-potentials with respect to moduli.  However, the geometric interpretation of the last layer of the BPS system remained elusive.  In this paper we are revealing yet more evidence that the {\it entire} BPS system, in six-dimensions, must have a much more fundamental formulation in terms of holomorphic pre-potentials.   

Finally, we note that our work contains results that can be tested against the holographic dual CFT. The most basic coiffuring constraints (\ref{coiff1}) were first inspired by the CFT and the effect of global R-symmetry rotations. These basic coiffuring constraints have been tested using precision holography \cite{Giusto:2015dfa,Giusto:2019qig}. Conversely, the newer and more complicated relation, (\ref{coiff2}), first found in \cite{Heidmann:2019zws}, and particularly the relation between  $c'_{k, m,n}$ and $b_{ k_1,m_1,n_1} b_{k_2,m_2,n_2}$, came from supergravity and have yet to be tested and compared with the results from correlators in the CFT.   Moreover,  in Section \ref{sect:ki01} we found that momentum waves necessarily induce changes in the fundamental charge densities of the basic \nBPS{4} D1-D5 system. This seems to reflect the development of vevs in the dual CFT, as has been described in \cite{Giusto:2015dfa}.

More broadly, the quantum mechanics and coherent states of the CFT have obvious complex structures, it would be very interesting to see whether the complex structure and pre-potentials  we are discovering in the supergravity have direct analogues within the CFT.


\section*{Acknowledgments}
\vspace{-2mm}
We  thank  Iosif  Bena, Stefano Giusto, Anton Lukyanenko, Masaki Shigemori for useful discussions.
The work of NW and RW is supported in part by the ERC Grant 787320 - QBH Structure and by the DOE grant DE-SC0011687. DRM is supported by the ERC Starting Grant 679278 Emergent-BH. The work of PH was supported by an ENS Lyon grant, by the ANR grant Black-dS-String  ANR-16-CE31-0004-01 and by NSF grant PHY-1820784. RW is very grateful to the IPhT of CEA-Saclay for hospitality during this project, his research was also supported by a Chateaubriand Fellowship of the Office for Science Technology of the Embassy of France in the United States.

\appendix

\bigskip
\bigskip
\leftline{\bf \Large Appendices}

\section{Compendium of solutions}
\label{app:compendium}

In this appendix, we give a brief but complete overview of the new solutions we constructed in this paper. The six-dimensional metric, using coordinates $u,v,r,\theta,\varphi_1,\varphi_2$, is given by:
\begin{equation}
d s^2_{6} ~=~-\frac{2}{\sqrt{\cP}}\,(d v+\beta)\,\Big[d u+\omega + \frac{\mathcal{F}}{2}(d v+\beta)\Big]+\sqrt{\cP}\,d s^2_4\,,
\end{equation}
where:
\begin{align}
  d s_4^2 &~=~ \Sigma \, \left(\frac{d r^2}{r^2+a^2}+ d\theta^2\right)+(r^2+a^2)\sin^2\theta\,d\varphi_1^2+r^2 \cos^2\theta\,d\varphi_2^2\,,\qquad \qquad \Sigma ~\equiv~  (r^2+a^2 \cos^2\theta)     \,,\nonumber\\
  \beta &~\equiv~  \frac{R_y\, a^2}{\sqrt{2}\,\Sigma}\,(\sin^2\theta\, d\varphi_1 - \cos^2\theta\,d \varphi_2) \,,\qquad\qquad  \cP   ~=~     Z_1 \, Z_2  -  Z_4^2 \,. 
\end{align}
The six-dimensional solution is completely determined by specifying three scalar functions $Z_1, Z_2,Z_4$, three two-forms $\Theta_1, \Theta_2,\Theta_4$ (which appear in the 6D three-forms), as well as the metric one-form $\omega$ and the metric scalar function $\mathcal{F}$. (See also Section \ref{sect:BPS6D}.) We will also need the one-form:
\begin{equation}
 \omega_0 ~\equiv~  \frac{a^2 \, R_y \, }{ \sqrt{2}\,\Sigma}\,  (\sin^2 \theta  d \varphi_1 + \cos^2 \theta \,  d \varphi_2 ) \,.
\end{equation}

We use the complex coordinates defined by:
 \begin{equation}
\xi ~\equiv~\frac{r}{\sqrt{r^2+ a^2}} \, e^{i \frac{\sqrt{2} v}{R_y} } \,, \quad \chi ~\equiv~\frac{a}{\sqrt{r^2+ a^2}} \, \sin \theta \, e^{i \varphi_1} \,, \quad \eta ~\equiv~\frac{a}{\sqrt{r^2+ a^2}} \, \cos \theta \, e^{i \big (\frac{\sqrt{2} v}{R_y} - \varphi_2\big)} \,,
\end{equation}
which satisfy:
 \begin{equation}
|\xi|^2 ~+~ |\chi|^2  ~+~ |\eta|^2 ~=~   1  \,,
\end{equation}
%
Our solutions will depend on two holomorphic functions of these three variables, of which $G_1(\xi,\chi,\eta)$ carries the $q=0$ superstrata mode information and $G_2(\xi,\chi,\eta)$ carries the $q=1$ (supercharged) superstrata modes \eqref{SSCFTstates}. 

\subsection{Solutions with functions of $\xi$}

For the three families in this section, the scalar functions $Z_1,Z_2,Z_4$ and the two-forms $\Theta_1,\Theta_2,\Theta_4$ are completely determined through (\ref{1stLayerGenHolo}) with $G_2\equiv 0$ and $G_1$ given in terms of a holomorphic function $F(\xi)$ as detailed below. For each of the families, we will give the expressions below for $\cP$ (which is actually determined by the $Z_I$), for the scalar function $\cF$, and for the one-form $\omega$. (See also Section \ref{sect:BPS6D} for a brief discussion of how these quantities fully determine the full 6D solution.)

\subsubsection{The $(1,0,n)$ superstrata}
For these, we have $G_2\equiv 0$ (no supercharged modes) and $G_1(\xi,\chi,\eta) = \chi F(\xi)$, where $F$ is an arbitrary holomorphic function of $\xi$ on the complex unit disc satisfying $F(0)=0$:
\be F(\xi) = \sum_{n=1}^{\infty} b_n \xi^n .\ee
We introduce $F_\infty(v)$, the asymptotic limit of $F$:
\begin{equation}
F_\infty (v)  ~\equiv~    \lim_{|\xi| \to 1}  \, F(\xi) ~=~    \lim_{r  \to \infty}  \, F(\xi) ~=~ F(e^{i\frac{\sqrt{2}\,v}{R_y}})   \,.
\label{FinftyApp}
\end{equation}
Note that the metric warp factor simplifies to:
\be \cP   ~=~   \frac{Q_1 Q_5}{\Sigma^2  } ~-~  \frac{a^2 R_y^2}{2\,(r^2+a^2) \,\Sigma^2 }\, |F|^2\, \sin^2 \theta   \,. \ee
The solution is further completely determined by: 
\begin{align}
  \mathcal{F} &= \frac{1}{a^2} (|F|^2 - |F_\infty|^2),\\
  \omega &= \left(1 - \frac{1}{2a^2}(|F_\infty|^2-c)\right)\omega_0 + \frac{R_y}{\sqrt{2}\Sigma} (|F_\infty|^2 - |F|^2)\sin^2\theta d\varphi_1 .
\end{align}
The function $F$ and constant $c$ must satisfy the constraint:
\be \label{app:constraint10n}  c = 2 \left(\frac{Q_1 Q_5}{R_y^2} - a^2\right) = \frac{1}{\sqrt{2}\pi R_y}\int_0^{\sqrt{2}\pi R_y} dv' |F_\infty|^2 = \sum_{n=1}^{\infty} b_n^2.\ee
These solutions are completely regular for any choice of $F$ (see Section \ref{sect:10nreg}); in particular it can be shown that $\mathcal{P}>0$ and that there are no CTC's (see Section \ref{10n-y-circle}). The angular momenta of the solution are:
\be J_L = J_R =  \frac{R_y}{2}a^2,\ee
and the momentum charge $Q_P$ is:
\be \label{app:QP10n} Q_P  = \frac{1}{4\sqrt{2}\pi R_y} \int_0^{\sqrt{2}\pi R_y} dv ( \xi_\infty F'_\infty \Fb_\infty +  \xib_\infty F_\infty \Fb'_\infty) = \frac12 \sum_{n=1}^{\infty} n b_n^2.
\ee

\subsubsection{The $(1,1,n)$ superstrata}
\label{app:11n}
For these solutions, we have $G_2\equiv 0$ and now $G_1(\xi,\chi,\eta) = \eta F(\xi)$, where $F$ is again an arbitrary holomorphic function of $\xi$ on the complex unit disc satisfying $F(0)=0$:
\be F(\xi) = \sum_{n=1}^{\infty} b_n \xi^n .\ee
Here, the metric warp factor simplifies to:
\be \cP   ~=~    \frac{Q_1 Q_5}{\Sigma^2} ~-~  \frac{a^2 R_y^2}{2\,(r^2+a^2) \,\Sigma^2}\, |F|^2\, \cos^2 \theta   \,.\ee
The solution is further completely determined by: 
\begin{align}
  \mathcal{F} &= \frac{1}{a^2} \,\left(|\xi|^2| F|^2 - |F_\infty|^2\right)\,,\\
   \omega &= \left(1 - \frac{1}{2a^2}(|F_\infty|^2-c)\right)\omega_0 + \frac{R_y}{\sqrt{2}\Sigma}\,\left(|F_\infty|^2\sin^2\theta\, d\varphi_1+ |\xi|^2|F|^2\cos^2\theta \,d\varphi_2\right) \,,
\end{align}
where $F_\infty(v)$ is the asymptotic limit of $F$ \eqref{FinftyApp}. The function $F$ and constant $c$ must also satisfy (\ref{app:constraint10n}), just as for $(1,0,n)$ superstrata.
Similarly to the $(1,0,n)$ superstrata, these solutions are completely regular for any choice of $F$ and have no CTC's (see Section \ref{sect:11ni}). The angular momenta of the solution are:
\begin{equation}
 J_L\,=\, \frac{R_y}{2} \,a^2 \, , \qquad J_R\,=\, \frac{R_y}{2} \left(a^2\,+ \,c \right) \,.
\end{equation}
and the momentum charge is:
\be Q_P \,=\frac{1}{2}\sum_{n=1}^{\infty}(n+1)b_{n}^{2}~. \ee

\subsubsection{The $(2,1,n)$ superstrata (without supercharged modes)}

We again set $G_2\equiv 0$ (see Section \ref{sect:21niApp} for the generalization to $(2,1,n)$ solutions including supercharged modes), and we have $G_1(\xi,\chi,\eta) = \chi\, \eta  \,F(\xi)$, where $F$ is an arbitrary holomorphic function of $\xi$ on the complex unit disc satisfying $F(0)=0$:
\be F(\xi) = \sum_{n=1}^{\infty} b_n \xi^n .\ee
The metric warp factor simplifies to:
\be \cP ~=~  \frac{Q_1 Q_5}{\Sigma^2} ~-~  \frac{a^4 R_y^2}{2\,(r^2+a^2)^2\,\Sigma^2   }\, |F|^2\, \sin^2 \theta\, \cos^2 \theta   \,. \ee
The rest of the solution is given by:
\begin{equation}
\begin{split}
\mathcal{F} &\,=\, \frac{1}{4\,a^2}(1 - |\xi|^2)(1 - |\chi|^2)\,\left| F \right|^2 \,+\, \frac{1}{a^2}\left[ K(\xi,\xib) \,-\, \left(c^{(1)} + \bar{c}^{(1)}\right)\right]\,,\\
\omega &\,=\, \left(1 + \frac{c^{(0)}}{2\,a^2}\right) \omega_0 + \frac{i\,R_y}{2\sqrt{2}\, r\, (r^2 + a^2)}\,\left( c^{(1)}-\bar{c}^{(1)}\right) \,dr \\
&\phantom{\,=\,}- \frac{R_y}{2\sqrt{2}\Sigma}\left( |\eta|^2\, \left| F \right|^2 + \left[ K(\xi,\xib) \,-\, \left(c^{(1)} + \bar{c}^{(1)}\right)\right] \right)     \sin^2\theta\, d\varphi_1\\
&\phantom{\,=\,} + \frac{R_y}{2\sqrt{2}\Sigma}\left( |\xi|^2 |\chi|^2 \,\left| F \right|^2 + \left[ K(\xi,\xib) \,-\, \left(c^{(1)} + \bar{c}^{(1)}\right)\right] \right)     \cos^2\theta \,d\varphi_2\,,
\end{split}
\end{equation}
where we have defined the functions:
\begin{align}
K(\xi, \xib) &= \int_0^{|\xi|} x \left| F\Big(x\,e^{i\frac{\sqrt{2}}{R_y}v}\Big) \right|^2 dx,\\
 c^{(1)}(\xi) &= \frac14\sum_{n=1}^{\infty} \frac{b_n^2}{n+1} + \sum_{\ell=1}^\infty\left(\sum_{n=1}^{\infty} \frac{b_n b_{n+\ell}}{2(n+1)+l}\right)\xi^l,
\end{align}
and the constant $c^{(0)}$ satisfies:
\be c^{(0)} = 2 \left(\frac{Q_1 Q_5}{R_y^2} - a^2\right) = \frac{1}{{\sqrt{2}\,\pi R_y} }\, \int_0^{\sqrt{2}\,\pi R_y}  \, K_\infty(v)\, dv = \frac12\sum_{n=1}^{\infty} \frac{b_n^2}{n+1}.\ee
These solutions are completely regular for any choice of $F$ (see Section \ref{sec:21nreg}); we also do not expect the existence of CTC's for any $F$. The angular momenta of the solution are:
\be J_L =  \frac{R_y}{2}a^2, \qquad J_R = \frac{R_y}{2}\left(a^2 + \frac12 c^{(0)}\right) =  \frac{R_y}{2}\left(a^2 + \frac14 \sum_{n=1}^{\infty} \frac{b_n^2}{n+1} \right) ,\ee
and the momentum charge $Q_P$ is:
\be Q_P  = \frac18 \sum_{n=1}^{\infty}  b_n^2.
\ee

\subsection{Solutions with functions of $\chi$}

\subsubsection{The $(k,0,1)$ superstrata}

For superposition of  $(k,0,1)$ superstrata, we have $G_2\equiv 0$ and we set $G_1(\xi,\chi,\eta) = \xi \chi \, \partial_\chi F(\chi)$, where $F$ is an arbitrary holomorphic function of $\chi$ on the complex unit disc satisfying $F(0)=0$:
\be F(\chi) = \sum_{k=1}^{\infty} b_k \chi^k .\ee
The electric gauge fields $Z_I$ and the magnetic two-form fluxes $\Theta_I$ are altered from \eqref{1stLayerGenHolo} by the introduction of a new holomorphic function $\lambda(\chi)$ to
\be
\begin{split}
Z_1 &\,=\,  \frac{Q_1}{\Sigma} + \frac{R_y^2}{4\,Q_5\,\Sigma}\,\left((\xi \chi \, \partial_\chi F)^2+(\bar{\xi} \bar{\chi} \, \partial_{\bar{\chi}} \bar{F})^2  \,+\,2\left( \lambda+\bar{\lambda}\right)\right)\,, \hspace{4.3cm} \Theta_1 \,=\,0\,,\\
Z_2 &\,=\,\frac{Q_5}{\Sigma}\,,\hspace{6.5cm}  \Theta_2 \,=\, \frac{R_y}{Q_5}\,\left((\xi \chi \, \partial_\chi F)^2\,\Omega_z \,+\,(\bar{\xi} \bar{\chi} \, \partial_{\bar{\chi}} \bar{F})^2 \,\bar{\Omega}_z \right)\,,\\
Z_4 &\,=\, \frac{R_y}{2\,\Sigma} \,\left(\xi \chi \, \partial_\chi F + \bar{\xi} \bar{\chi} \, \partial_{\bar{\chi}} \bar{F}  \right)\,, \hspace{5.05cm}  \Theta_4 \,=\, \xi \chi \, \partial_\chi F\,\Omega_z \,+\, \bar{\xi} \bar{\chi} \, \partial_{\bar{\chi}} \bar{F} \,\bar{\Omega}_z\,,
\end{split}
\label{app:k01final}
\ee
where $\lambda(\chi)$ is given by
\be 
\lambda(\chi) ~=~  \sum_{k,\ell =1}^{\infty}  k\,b_k\,b_{k+\ell}\, \chi^\ell \,,
\ee
and $\Omega_z$ is one of the two complex self-dual two-forms of $\mathbb{R}^4$ \eqref{SDyz}.

\noindent As for the six-dimensional metric \eqref{sixmet}, the metric warp factor is:
\be \cP\, =\, \frac{Q_1 Q_5}{\Sigma^2} + \frac{R_y^2}{2\,\Sigma^2}\,\left(\lambda+\bar{\lambda} - |\xi|^2|\chi|^2\,|\partial_\chi F|^2\right)  \, =\, \frac{Q_1 Q_5}{\Sigma^2} + \frac{R_y^2}{2\,\Sigma^2}\,\left(\lambda+\bar{\lambda} - \frac{a^2 r^2}{(a^2+r^2)^2} |\partial_\chi F|^2\sin^2\theta\right)\,,
\ee
and the other metric terms are given by:
\be
\begin{split}
 \mathcal{F} &\,=\, \frac{|F|^2 - c - \bar{c}}{\Sigma}\,,\\
 \omega &\,=\, \left(1 + \frac{|\xi|^2}{2\,a^2}\,\left(\chi \partial_\chi + \bar{\chi}\partial_{\bar{\chi}}\right)\left( \frac{|F|^2 - c - \bar{c}}{1-|\chi|^2}\right)\right)\omega_0    \,-\, \frac{R_y}{\sqrt{2}\,\Sigma}   \frac{1-|\xi|^2}{1-|\chi|^2}\,\left(|F|^2 - c - \bar{c}\right) \sin^2\theta\, d\varphi_1\\  
 &\phantom{\,=\,} \,+\, \frac{i\, R_y\, }{2\sqrt{2}}\frac{|\xi|^2}{r \Sigma}  \left[ \chi \,\partial_\chi \left( |F|^2-c\right) \,-\, \bar\chi\, \partial_{\bar{\chi}}\left( |F|^2- \bar{c}\right)\right] \,dr\,.
\end{split}
\ee
We have introduced the holomorphic function $c(\chi)$, defined by: 
\begin{equation}
c(\chi) \,=\, \frac{1}{2}\,\sum_{k=1}^{\infty} \, b_k^2 \,+\, \sum_{k,\ell=1}^{\infty}  b_k\,b_{k+\ell} \, \chi^\ell\,.
\end{equation}
Finally, there is a constraint to satisfy:
\be \frac{1}{2\pi i}\oint_{|\chi|=1} \bar{F}\partial_\chi F \,d\chi\,=\, \sum_{k=1}^{\infty} \, k\, b_k^2 \,=\, 2 \left(\frac{Q_1 Q_5}{R_y^2} - a^2\right)  \,.\ee
These solutions are completely regular for any choice of $F$ (see Section \ref{sec:k01reg}); we also do not expect the existence of CTC's for any $F$. The angular momenta of the solution are:
\be J_L = J_R =  \frac{R_y}{2}a^2 ,\ee
and the momentum charge $Q_P$ is:
\be Q_P  \,=\, \frac12 \sum_{k=1}^{\infty}  b_k^2\,=\,  \frac{1}{4\pi i}\oint_{|\chi|=1} \frac{|F|^2}{\chi} \,d\chi.
\ee 

\subsection{Solutions with functions of $\eta$}

\subsubsection{The $(k,k,0)$ superstrata}

For superposition of  $(k,k,0)$ superstrata, we have $G_2\equiv 0$ and we set $G_1(\xi,\chi,\eta) = F(\eta)$, where $F$ is an arbitrary holomorphic function of $\eta$ on the complex unit disc satisfying $F(0)=0$:
\be F(\chi) = \sum_{k=1}^{\infty} b_k \eta^k .\ee
The electric gauge fields $Z_I$ and the magnetic two-form fluxes $\Theta_I$ are also altered from \eqref{1stLayerGenHolo} by the introduction of a new holomorphic function $\lambda(\chi)$ to
\begin{equation}
\begin{split}
Z_{1} &= \frac{Q_{1}}{\Sigma} + \frac{R_y^{2}}{4 Q_{5}\,\Sigma} \left(\,F^{2}+ \bar{F}^{2} \,+\, 2\left(\lambda \,+\, \bar{\lambda} \right)\,\right)~,\qquad \hspace{5.16cm} \Theta^{(1)}  \,=\, 0\,, \\
Z_{2} &= \frac{Q_{5}}{\Sigma}~,\qquad\hspace{3cm}  \Theta^{(2)} \,=\, \frac{R_y}{Q_5}\,\left[\eta \left(F\,\partial_\eta F+\partial_\eta \lambda \right)\,\Omega_{y} \,+\,\bar{\eta}\left( \bar{F}\,\partial_{\bar{\eta}} \bar{F} + \partial_{{\bar{\eta}}}\bar{\lambda}\right)\,  \bar{\Omega}_{y}\right] , \\
Z_{4} &= \frac{R_y}{2\Sigma}\left(\,F \,+\,\bar{F} \,\right) ~, \qquad\qquad\qquad\quad \hspace{3.9cm} \,  \Theta^{(4)}  \,=\,\eta\, \partial_\eta F\,\Omega_{y} \,+\,\bar{\eta}\,\partial_{\bar{\eta}} \bar{F}\,  \bar{\Omega}_{y} \,,
\label{app:kk0final}
\end{split}
\end{equation}
where $\lambda(\chi)$ is given by
\be 
\lambda(\chi) ~=~   \sum_{k,\ell =1}^{\infty}  b_k\,b_{k+\ell}\, \eta^\ell \,,
\ee
and $\Omega_y$ is one of the two complex self-dual two-forms of $\mathbb{R}^4$ \eqref{SDyz}.

\noindent As for the six-dimensional metric \eqref{sixmet}, the metric warp factor is:
\be \cP\, =\, \frac{Q_1 Q_5}{\Sigma^2} + \frac{R_y^2}{2\,\Sigma^2}\,\left(\lambda+\bar{\lambda} - |F|^2\right)\,,
\ee
and the other metric terms are given by:
\begin{equation}
\begin{split}
\cF & ~=~   \frac{|F|^2 - c-\lambda-\bar{\lambda}}{\Lambda} \,, \\
\omega & ~=~ \omega_0 ~+~ \frac{R_y}{\sqrt{2} \,\Lambda \Sigma}\,   \left(c+\lambda+\bar{\lambda}-|F|^2 \right)\, \left[ (a^2+r^2) \sin^2 \theta\,d\varphi_1 \,+\, r^2 \cos^2 \theta \,d\varphi_2 \right] \,.
\end{split}
\end{equation}
We have introduced the constant $c$, defined by: 
\be 
\begin{split}
c\,=\, \sum_{k=1}^{\infty} {b_k}^2\,=\, \frac{1}{2\pi i}\,\oint_{|\eta|=1} \frac{|F(\eta)|^2}{\eta} d\eta \,.
\end{split}
\ee
Finally, there is a constraint to satisfy:
\begin{equation} 
\frac{1}{2\pi i}\,\oint_{|\eta|=1} \frac{|F(\eta)|^2}{\eta} d\eta \,=\, \sum_{k=1}^{\infty} \,b_k^2\,=\, 2\left(\frac{Q_1 Q_5}{R_y^2} \,-\, a^2\right)  \,. \\
\end{equation}
These solutions are completely regular for any choice of $F$ (see Section \ref{sec:kk0reg}); we also do not expect the existence of CTC's for any $F$. The angular momenta of the solution are:
\be J_L \,=\,  \frac{R_y}{2}\,a^2\,,\qquad J_R \,=\,  \frac{R_y}{2} \left(a^2\,+\, \frac{1}{2}\,\sum_{k=1}^{\infty} \,b_k^2 \right) \,=\,\frac{R_y}{2} \left(a^2\,+\, \frac{1}{4\pi i}\,\oint_{|\eta|=1} \frac{|F(\eta)|^2}{\eta} d\eta \right) ,\ee
and the momentum charge $Q_P$ is:
\be Q_P  \,=\,  \frac{1}{2}\, \sum_{k=1}^{\infty} \,b_k^2\,=\,\frac{1}{4\pi i}\,\oint_{|\eta|=1} \frac{|F(\eta)|^2}{\eta} d\eta .
\ee 

\subsection{Other solutions}
We discuss the superposition of $(2,1,n)$ superstrata with supercharged modes turned on (i.e. the solution depends on two arbitrary holomorphic functions of $\xi$) in appendix \ref{sect:21niApp}, and the superposition of $(k,0,n)$ superstrata with $n$ fixed (depending on an arbitrary holomorphic function of $\chi$) in appendix \ref{sect:ki0nApp}.

\section{More examples}
\label{app:MoreEx}

\subsection{The $(2,1,n)$ superstrata with supercharged modes}
\label{sect:21niApp}
In Section \ref{sect:21ni}, we have constructed solutions using the original $(2,1,n)$ superstratum modes \cite{Bena:2017upb}. However, for $k=2$, $m=1$, there are supercharged modes in the CFT \cite{Ceplak:2018pws}. In this section, we will extend the family of solutions and then deal with the ``hybrid" $(2,1,n)$ mode that incorporates both the original and the supercharged $(2,1,n)$ modes \cite{Heidmann:2019zws}.  
The underlying holomorphic functions, $G_1$ and $G_2$, are then given by the most general superposition of the $(2,1,n)$ modes determined by arbitrary functions $F$ and $S$ fluctuating with the complex variable $\xi$:
\begin{equation}
\begin{split}
G_1(\xi,\chi,\eta)   &~=~    \chi\, \eta \, F(\xi) ~=~ \frac{a^2 \, F(\xi)}{a^2+r^2} \, \sin \theta\,\cos\theta\, e^{i \big (\frac{\sqrt{2} v}{R_y} + \varphi_1- \varphi_2\big)} \,,  \qquad\quad F(\xi) ~=~ \sum_{n=1}^\infty \, b_n \, \xi^n \,, \\
G_2(\xi,\chi,\eta) &~=~    \chi \, \eta \, \,\xi\, \partial_\xi^3 \left(\xi^2 \, S(\xi) \right) ~=~ \frac{a^2 \, \,\xi\, \partial_\xi^3 \left(\xi^2 \, S(\xi) \right)}{a^2+r^2} \, \sin \theta\,\cos\theta\, e^{i \big (\frac{\sqrt{2} v}{R_y} + \varphi_1- \varphi_2\big)} \,,  \qquad S(\xi) ~=~ \sum_{n=1}^\infty \, c_n \, \xi^n \,, \\
\end{split}
 \label{G21nApp} 
\end{equation}
where $b_n$ and $c_n$ are real Fourier coefficients. 
Defining $G_2$ in this functional form of $S(\xi)$ drastically simplifies the final form of the solutions. The metric warp factor is  given by:
\begin{equation}
\sqrt{\cP}   ~=~    \frac{1}{\Sigma}\,\sqrt{Q_1 Q_5 ~-~  \frac{a^4 R_y^2}{2\,(r^2+a^2)^2 }\, |F|^2\, \sin^2 \theta\, \cos^2 \theta}   \,,
 \label{cP21nApp} 
\end{equation}
We decompose $\omega$ as in \eqref{om21n}. The quantities $\mathcal{F}$ and $ \omegacomp$ are determined by the BPS equations sourced by $G_1$ and $G_2$:
\begin{adjustwidth}{-0.3cm}{0cm}
\be  
\begin{split}
2a^2\mathcal{F} &\,=\, \frac{1}{2}(1 - |\xi|^2)(1 - |\chi|^2)\,\left| F \right|^2 \,+\, 2 K(\xi,\xib) \,+\, \partial_{\xi}^2 \partial_{\bar{\xi}}^2 \left( |\xi|^4\, \left| S \right|^2 \right) \,-\, |\xi|^6 \,\partial_{\xi}^2 \partial_{\bar{\xi}}^2 \left( |\xi|^2\, \left| S \right|^2 \right)  \\
&\phantom{\,=\, } \,+\,  \left(1-|\xi|^2-2|\chi|^2\right)\left(1 - |\xi|^2 \right) \left[ \left(1 - |\xi|^2 \right)  \partial_{\xi}^2 \partial_{\bar{\xi}}^2 \left(\frac{|\xi|^4}{1 - |\xi|^2} \, \left| S \right|^2 \right) \,-\, 4\,\partial_{\xi} \partial_{\bar{\xi}} \left(\frac{|\xi|^4}{\left(1 - |\xi|^2\right)^2} \, \left| S \right|^2 \right) \right] \\
&\phantom{\,=\, } \,+\, \frac{\left(1 - |\xi|^2 \right)\left(1-|\xi|^2-|\chi|^2-|\chi|^2|\xi|^2 \right)}{|\xi|^4} \,\left[ \xi^2 \, \partial_{\xi}^2\left( \frac{|\xi|^4}{1 - |\xi|^2}\, \bar{F} S \right)\,+\, \bar{\xi}^2 \, \partial_{\bar{\xi}}^2\left( \frac{|\xi|^4}{1 - |\xi|^2}\,F \bar{S} \right) \right]\,, \\
 \varpi_r  &\,=\, - \frac{3\, i R_y}{\sqrt{2}\,a^3}\,\sqrt{\frac{|\xi|^{6}}{1-|\xi|^2}}\left[ \xi\, \partial_\xi \left| S \right|^2 - \bar{\xi}\, \partial_{\bar{\xi}} \left| S \right|^2 \,-\, \frac{\left(1 - |\xi|^2 \right)^4}{6\,|\xi|^4} \,  \partial_\xi \partial_{\bar{\xi}} \left(  \frac{|\xi|^4}{(1 - |\xi|^2)^2} \left( \xi\, \partial_\xi \left| S \right|^2 - \bar{\xi}\, \partial_{\bar{\xi}} \left| S \right|^2 \right)\right) \right]\, ,\\
 \varpi_\theta &\,=\,  \frac{i\,R_y\,|\chi|\,\left( 1- |\xi|^2\right)^2\,\sqrt{1-|\xi|^2-|\chi|^2}}{2 \sqrt{2}\,a^2\,|\xi|^4}\,\left[ \xi^2 \, \partial_{\xi}^2\left( \frac{|\xi|^4}{1 - |\xi|^2}\, \bar{F} S \right)\,-\,\bar{\xi}^2 \, \partial_{\bar{\xi}}^2\left( \frac{|\xi|^4}{1 - |\xi|^2}\, F\bar{S} \right) \right]\,, \\
\varpi_1 &\,=\, - \frac{R_y\, |\chi|^2}{2 \sqrt{2}\,a^2\, (1-|\chi|^2)} \,\Biggl[\left( 1-|\xi|^2-|\chi|^2\right)\,\left| F \right|^2 \,+\,  K(\xi,\xib)  \\
&\hspace{3.8cm} \,+\,\frac{(1 - |\xi|^2)\left( 1-|\xi|^2-|\chi|^2\right)}{|\xi|^4} \,\left( \xi^2 \, \partial_{\xi}^2\left( \frac{|\xi|^4}{1 - |\xi|^2}\, \bar{F} S \right)\,+\, \bar{\xi}^2 \, \partial_{\bar{\xi}}^2\left( \frac{|\xi|^4}{1 - |\xi|^2}\,F \bar{S}  \right)\right) \\
&\hspace{2.0cm} \,-\, \left((1 - |\xi|^2)^2\left(\xi \partial_\xi^2 \partial_{\bar{\xi}}+\bar{\xi} \partial_{\bar{\xi}}^2 \partial_{\xi} \right) -4 |\xi|^2(1 - |\xi|^2) \partial_\xi  \partial_{\bar{\xi}}-6\left(\xi \partial_\xi +\bar{\xi} \partial_{\bar{\xi}} \right) \right)\left( \frac{|\xi|^4}{(1 - |\xi|^2)^2} \, |S|^2 \right) \Biggr],\\
\varpi_{2} &\,= \, \frac{R_y\left( 1-|\xi|^2-|\chi|^2\right)}{2 \sqrt{2}\,a^2\, (1-|\chi|^2)} \,\Biggl[|\xi|^2 |\chi|^2 \,\left| F \right|^2 \,+\,  K(\xi,\xib)   \\
&\hspace{3.8cm} \,-\,\frac{|\xi|^2 |\chi|^2 (1 - |\xi|^2)}{|\xi|^4} \,\left(  \xi^2 \, \partial_{\xi}^2\left( \frac{|\xi|^4}{1 - |\xi|^2}\, \bar{F} S \right)\,+\,\bar{\xi}^2 \, \partial_{\bar{\xi}}^2\left( \frac{|\xi|^4}{1 - |\xi|^2}\, F\bar{S}  \right)\right) \\
&\hspace{2.0cm} \,-\, \left((1 - |\xi|^2)^2\left(\xi \partial_\xi^2 \partial_{\bar{\xi}}+\bar{\xi} \partial_{\bar{\xi}}^2 \partial_{\xi} \right) -4 |\xi|^2(1 - |\xi|^2) \partial_\xi  \partial_{\bar{\xi}}-6\left(\xi \partial_\xi +\bar{\xi} \partial_{\bar{\xi}} \right) \right)\left( \frac{|\xi|^4}{(1 - |\xi|^2)^2} \, |S|^2 \right) \Biggr], 
\end{split}
\label{21nMultiGenApp}
\end{equation}
\end{adjustwidth}
where we have once again defined
\be 
K(\xi,\xib) \,\equiv\,  \int_0^{|\xi|} x \left| F(x\,e^{i\frac{\sqrt{2}}{R_y}v}) \right|^2 dx\,.
\ee

The fields are then generated by the arbitrary function $S$, corresponding to the supercharged mode, and the arbitrary function $F$, corresponding to the original mode. In other words, the terms fluctuating as $|F|^2$ arise only from the original $(2,1,n)$ superstrata, the terms in $|S|^2$ arise from the supercharged equivalents, whereas the terms in $\bar{F}S$ and its complex conjugate come from the interaction between the two different modes. 

One can also add non-trivial solutions of the homogeneous second-layer of BPS equations that do not change the characteristics of the solutions. 
For instance we can make the following replacements in \eqref{21nMultiGenApp}:
\begin{equation}
|S|^2 ~\to~ |S|^2 ~-~ c^{(1)}(\xi) ~-~  c^{(1)}(\bar{\xi}) \,, \qquad \bar{F}S ~\to~ \bar{F}S ~-~  c^{(2)}(\bar{\xi})  \qquad F\bar{S}  ~\to~ F\bar{S}~-~ c^{(2)}(\xi) \,,
\label{SFtransf}
\end{equation}
where $c^{(1)}$ and $c^{(2)}$ are arbitrary holomorphic functions.  This, once again, provides a remarkably simple class of homogeneous solutions  that will ultimately be fixed by regularity. In addition, we will need to add three other homogeneous solutions for regularity, which are related to gauge transformations of the solutions \eqref{gaugetrf}. This consists in adding the following homogeneous solutions to $\mathcal{F}$ and $\omega$:
\begin{equation}
\begin{split}
\mathcal{F} &\,\longrightarrow\, \mathcal{F}\,-\, \mathcal{G} \,-\, \bar{\mathcal{G}}\,,\\
\omega &\,\longrightarrow\, \omega \,+\, \frac{\mathcal{G}+\bar{\mathcal{G}}}{2} \,\beta\,+\, \frac{c^{(0)}}{2\,a^2}\,\omega_0\,-\, \frac{ i\sqrt{2} \,R_y}{a^2}\,\frac{|\chi| \sqrt{1-|\xi|^2-|\chi|^2}}{1-|\xi|^2}\,\left(c^{(4)}(\xi)-c^{(4)}(\bar{\xi})\right)\,d\theta \\
&\phantom{\,\longrightarrow\,} \,+\, \frac{i R_y}{2\sqrt{2}\,a} \sqrt{\frac{(1-|\xi|^2)^3}{|\xi|^2}}\,\left(\mathcal{G}-\bar{\mathcal{G}}\right)\,dr,\\
&\,\longrightarrow\, \omega \,+\, \frac{\mathcal{G}}{2} \,\beta \,+\, \frac{c^{(0)}}{2\,a^2}\,\omega_0 \,-\, \frac{ i\sqrt{2} \,R_y}{a^2}\,\cos\theta \,\sin\theta \,\left(c^{(4)}(\xi)-c^{(4)}(\bar{\xi})\right)\,d\theta  \,+\, \frac{i a^2 R_y}{2\sqrt{2}\,r(r^2+a^2)} \,\left(\mathcal{G}-\bar{\mathcal{G}}\right)\,dr,
\end{split}
\label{lastHomSol21nApp}
\end{equation}
where $\beta$ and $\omega_0$ are defined in \eqref{betadefn} and \eqref{angmom0}, $c^{(0)}$ is an arbitrary constant,
\begin{equation}
\mathcal{G} \,\equiv\,\frac{1}{a^2}\left( c^{(3)}(\xi) \,+\, \frac{1-|\xi|^2-2|\chi|^2}{1-|\xi|^2}\, \xi \partial_\xi c^{(4)}(\xi) \right),
\end{equation}
and $c^{(3)}$ and $c^{(4)}$ are \emph{a priori} arbitrary holomorphic functions that will be fixed in the next section.

\subsubsection{Regularity}
First, we require $\cF$ and $\omega$ to decay as $r^{-2}$ in order to have solutions asymptotically AdS$_3\times$S$^3$. For generic $S$, they are strongly divergent and grow as $r^{2}$. The first step is to fix the homogeneous solutions $c^{(1)}$ and $c^{(2)}$ to cancel those divergences. This is done by taking
\be \begin{split}
c^{(1)}(\xi) &\,\equiv\, \frac{1}{2}\,\sum_{n=1}^{\infty} \, (n+1)\,c_n^2 \,+\, \sum_{n,\ell=1}^{\infty} \frac{2(1+n)+\ell}{2+\ell}\, c_n\,c_{n+\ell} \,\, \xi^\ell\,,\\
c^{(2)}(\xi)  &\,\equiv\, \sum_{n=1}^{\infty} b_n \,c_n \,+\, \sum_{n,\ell=1}^{\infty} \left( b_n\,c_{n+\ell}+c_n\,b_{n+\ell}\,\right) \xi^\ell
\end{split}\end{equation}
where $b_n$ and $c_n$ are the Fourier coefficients of $F$ and $S$  respectively \eqref{G21nApp}. Then, $\cF$ goes to a constant and a $v$-dependent oscillating term at the boundary. To make $\cF$ vanish we fix $c^{(3)}$ and $c^{(4)}$ \eqref{lastHomSol21nApp}. It is similar to the gauge transformation performed for the $(1,0,n)$ solutions in Section \ref{sect:hom-gauge}. We find that 
\be 
\begin{split}
c^{(3)}(\xi) &\,\equiv \, \sum^{\infty}_{n=1}\left( n(1+n)(2+n)\,c_n^2 \,+\, \frac{b_n^2}{4(n+1)}\right)\\
&\phantom{\,\equiv \,}  \,+\, \sum_{n,\ell=1}^{\infty} \left( n (2 + \ell + n) (2(n+1) + \ell) \, c_n\,c_{n+\ell}\,+\, \frac{b_n b_{n+\ell}}{2(1+n)+\ell} \right) \, \xi^\ell \,,\\
c^{(4)}(\xi) &\,\equiv \, \sum_{n,\ell=1}^{\infty} b_n c_{n+l}\,\,\xi^\ell\,.
\end{split}
\end{equation}
Once those functions fixed, $\mathcal{F}$ and $\omega$ decay at least as $r^{-2}$ and the solutions are asymptotically AdS$_3\times$S$^3$.

On the plane $\theta = 0$ (resp. $\theta =\frac{\pi}{2}$), the $\varphi_1$-circle shrinks (resp. $\varphi_2$-circle). Thus, all scalars must have no dependence on $\varphi_1$ (resp. $\varphi_2$) and the forms must have vanishing $\varphi_1$-components (resp. $\varphi_2$-components). A trivial check shows that this is automatically satisfied and does not require any regularization. 

At the center of $\mR^4$, $r=0,\,\theta = 0$ ($\xi = 0, \, \chi =0$), the whole $S^{\theta\varphi_1\varphi_2}$ shrinks. One has to switch to the polar coordinates ($\widetilde{r},\,\widetilde{\theta}$) as introduced in \eqref{polarcoordtheta=0}  and take the limit $\widetilde{r}\rightarrow0$ . Straightforwardly, all scalars are independent of $\widetilde{\theta},\, \varphi_1$ and $\varphi_2$ at this limit. As for the angular-momentum one-form, the $\theta$-component, the $\varphi_1$-component indeed vanish when $\widetilde{r}\rightarrow 0$. However, $\omegacomp_2$ goes to a constant which can induce CTC's. We then cancel this constant by fixing the constant $c^{(0)}$ introduced in \eqref{lastHomSol21nApp}:
\be 
c^{(0)} \,\equiv \, \sum^{\infty}_{n=1}\left( 2 n(1+n)(2+n)\,c_n^2 \,+\, \frac{b_n^2}{2(n+1)}\right) \,\equiv\,  \frac{1}{{\sqrt{2}\,\pi R_y} }\, \int_0^{\sqrt{2}\,\pi R_y}  \,\left(  \hat{S}_\infty(v) \,+\, K_\infty(v) \right) dv  \,,
\end{equation}
where we have defined
\begin{align}\nonumber
  \hat{S}_\infty(v) &\,\equiv \,  \lim_{|\xi| \to 1}  \,\xi\, \partial_\xi^3 \left(\xi^2 \, \left|S(\xi)\right|^2\right)+ \bar{\xi}\, \partial_{\bar{\xi}}^3 \left(\bar{\xi}^2 \, \left|S(\bar{\xi})\right|^2\right)~=~    \lim_{r  \to \infty}  \, \,\xi\, \partial_\xi^3 \left(\xi^2 \, \left|S(\xi)\right|^2\right)+ \bar{\xi}\, \partial_{\bar{\xi}}^3 \left(\bar{\xi}^2 \, \left|S(\bar{\xi})\right|^2\right)  \,,\\
 K_\infty(v) &\,\equiv \,  \lim_{|\xi|\rightarrow 1} K(\xi,\xib) =  \lim_{r\rightarrow \infty} K(\xi,\xib)  \,.
\end{align}

At the supertube locus, $r=0,\,\theta = \frac{\pi}{2}$, there are divergences in the scalars and forms of the solutions. As explained in Section \ref{sect:basesols}, the only dangerous term comes from the metric component in front of $d\varphi_1^2$ given by \eqref{danger1}. To remove the singularity, one must impose
\be 
\frac{Q_1\,Q_5}{R_y^2}\,=\, a^2\,+\, \frac{1}{2}\,c^{(0)}\,.
\end{equation}

Finally, as for some other examples before, we will not complete the daunting analysis of CTC's in this geometry for generic $F$ and $S$. However, we strongly believe that this will not lead to any constraints on the choice of holomorphic function and that solutions will be automatically CTC-free. Indeed, holography suggests we should not expect any choice of holomorphic functions to lead to a non-regular geometry.

We have finally constructed a family of smooth asymptotically AdS$_3\times$S$^3$ D1-D5-P solutions parametrized by two holomorphic arbitrary functions of one variable. The holomorphic functions give a very specific shape to the momentum wave of the solution and contribute to the conserved charges.

\subsubsection{Conserved charges}

The angular momenta $J_R$ and $J_L$ and the momentum charge $Q_P$ can be read off from the asymptotic form of $\omega$, $\beta$ and $\cF$ through the expressions \eqref{asmpmoms} and \eqref{Fexp}. For the solutions constructed above, we obtain the supergravity charges:
\be 
\begin{split}
Q_P \,= \, \frac{1}{2} \,\sum^{\infty}_{n=1}\left( n(1+n)^2(2+n)\,c_n^2 \,+\, \frac{b_n^2}{4}\right) \,,\qquad  J_L\,=\, \frac{R_y}{2} \,a^2 \, , \qquad J_R\,=\, \frac{R_y}{2} \left(a^2\,+\, \frac{1}{2}\,c^{(0)} \right)\,,
\end{split}
\end{equation}
from which the quantized charges are obtained by using  (\ref{quantch}).     

\subsection{The $(k_i,0,n)$ superstrata}
\label{sect:ki0nApp}

In Section \ref{sect:ki01}, we have constructed the family of solutions made of a superposition of $(k,0,1)$ superstratum modes. We will extend to superposition of $(k_i,0,n)$ modes where $n$ is a fixed arbitrary integer. This extension is made possible because superposition of $(k_i,0,n)$ follows the same philosophy as superposition of $(k,0,1)$ in a sense that $\cF$ is determined by a Laplacian equation with sources: $*_4 d *_4 d \cF = \textit{sources}$. More precisely, $\cF$ is actually constructed from generating functions $\mathscr{F}_{2k,2,2n}$ satisfying the following equation
\be 
-*_4 d *_4 d \mathscr{F}_{2k,2,2n} \,=\, \frac{\Delta_{2k,2,2n}}{(r^2+a^2)\,\cos^2\theta\, \Sigma}\,=\, \frac{(1-|\xi|^2)^3 \, |\xi|^{2n}\, |\chi|^{2(k-1)}}{a^4\, (1-|\chi|^2)}\,.
\label{GenFuncApp}
\ee
In previous works \cite{Bena:2015bea,Bena:2016agb,Bena:2017geu,Bena:2017xbt,Ceplak:2018pws,Heidmann:2019zws}, this equation has been solved giving rise to a rather complicated generic solution with a double sum. Using the complex-variable formalism, we have been able to drastically simplifies the form of the solutions by removing many terms with vanishing Laplacian. This leads to a more compact form for the solutions of \eqref{GenFuncApp}:
\be 
\mathscr{F}_{2k,2,2n}\,=\, -\,\left(\frac{n! \,(k-1)!}{(k+n)!} \right)^2 \,\frac{1 - |\xi|^2}{a^2} \, \sum_{j=0}^{n} \,(-1)^{n+1-j}  \,\frac{ |\xi|^{2 j}}{(j!)^2}\, \partial_\chi^j \partial_{\bar{\chi}}^j\left[\frac{ |\chi|^{2(k+n)}}{\left(1- |\chi|^2 \right)^{n+1-j}}\right] \,.
\ee
We will use those generating functions to construct superposition of $(k_i,0,n)$ modes with $n$ fixed. 

For $m=0$, there are no supercharged modes \cite{Ceplak:2018pws,Heidmann:2019zws}. The corresponding holomorphic function is then taken to be zero $G_2\equiv 0$. Thus, we consider only the holomorphic function $G_1$ given by the most general superposition of the $(k_i,0,n)$ original modes determined by an arbitrary function $F$ fluctuating with the complex variable $\chi$:
\be 
G_1(\xi,\chi,\mu) \,=\, \frac{1}{n!}\,\xi^n \,\chi \, \partial^n_\chi \left( \chi^{n-1} \, F(\chi)\right) \,.
\label{G1Fk0n}
\ee
We define the real Fourier modes of $F$ as before:
\be 
F(\chi) \,=\, \sum_{k=1}^{\infty} \, b_k \, \chi^k\,.
\label{Fourierk0nApp}
\ee
We could have also defined $G_1(\xi,\chi,\eta)=    \xi^n\,F(\chi) $ but this drastically complicates the final form of the solutions. Inspired by the expression of the generating functions \eqref{GenFuncApp}, we define
\be 
\begin{split}
\alpha_{j,n} &\, \equiv \, \frac{(-1)^{n-j}}{(j!)^2}\,\partial_\chi^j \partial_{\bar{\chi}}^j \left[ \frac{ |\chi|^{2n}}{\left(1- |\chi|^2 \right)^{n+1-j}} \,|F|^2\right]\,, \qquad d\alpha_{j,n}  \, \equiv \, \frac{1}{2}\left(\chi \, \partial_\chi \alpha_{j,n} \,+\, \bar{\chi} \, \partial_{\bar{\chi}} \alpha_{j,n} \right)\,, \\
 \beta_{j,n} &\, \equiv \,\frac{(-1)^{n-j}}{(j!)^2}\,\partial_\chi^j \partial_{\bar{\chi}}^j \left[ \frac{ |\chi|^{2n}}{\left(1- |\chi|^2 \right)^{n+1-j}}\left(\,\chi \partial_\chi \,|F|^2 \,-\,  \bar{\chi} \, \partial_{\bar{\chi}}\,|F|^2 \,\right) \right]\,,  \qquad d\beta_{j,n}  \, \equiv \, \frac{1}{2}\left(\chi \partial_\chi \beta_{j,n} \,-\, \bar{\chi} \, \partial_{\bar{\chi}} \beta_{j,n} \right)\,, 
\end{split}
\label{alphabetaDefApp}
\ee
The scalars $\cP$ and $\cF$ in the six-dimensional metric \eqref{sixmet} are then given by
\be 
\cP \,=\, \frac{1}{\Sigma^2}\,\left[ Q_1 Q_5\,-\, \frac{R_y^2}{2}\,\frac{|\chi|^2\,|\xi|^{2n}}{(n!)^2}\, \partial_\chi^n \partial_{\bar{\chi}}^n \left(|\chi|^{2(n-1)}\,|F|^2 \right) \right]\,,\qquad \cF \,=\, \,\frac{1 - |\xi|^2}{a^2} \, \sum_{j=0}^{n-1} \,|\xi|^{2 j} \,\alpha_{j,n-1}\,.
\label{PandFk0nGenApp}
\ee
As for the angular-momentum one-form $\omega$, we have, using the decomposition \eqref{om21n},
\be 
\begin{split}
 \varpi_r &\,=\, \frac{i\, R_y\,\sqrt{(1 - |\xi|^2)^3}}{2 \sqrt{2}\,a^3\,|\xi|}\, \sum_{j=1}^{n}\, \frac{|\xi|^{2 j}}{j}\, \beta_{j-1,n-1}\,, \qquad \varpi_\theta\,=\, 0\,,\\
 \varpi_1 +\varpi_2 &\,=\, \frac{R_y}{2\sqrt{2} a^2} \left(1-|\xi|^2 \right) \Biggl[  \frac{|\xi|^2+2|\chi|^2}{1-|\chi|^2} \,\left( |\xi|^{2 n}\, \alpha_{n,n-1} - \alpha_{0,n-1} \right)\,+\, \sum_{j=1}^{n} \, |\xi|^{2 j} \, \left( \, \alpha_{j,n} \,-\,  \frac{|\xi|^2+|\chi|^2}{1-|\chi|^2}\, \alpha_{j,n-1} \right)\Biggr] \, , \\
\varpi_1 -\varpi_2 &\,=\, \frac{R_y}{2\sqrt{2} a^2}  \, \Biggl[ \,\, \sum_{j=1}^{n-1}\, |\xi|^{2 j} \, \left[ \, \frac{(1- |\xi|^2)\,(|\xi|^2 - |\chi|^2)}{1-|\chi|^2}\, \alpha_{j,n-1} \,+\, (1 + |\xi|^2)\,\alpha_{j,n} \right] \,+\, \frac{|\xi|^2 \left(1-|\xi|^2 \right) -2 |\chi|^2}{1-|\chi|^2}\, \alpha_{0,n-1}\\
& \hspace{2cm} -2\sum_{j=1}^{n}\, \frac{|\xi|^{2j}}{j^2}\left( j \,d\alpha_{j-1,n-1}+ d\beta_{j-1,n-1} \right)   \,+\,  \frac{1+ |\xi|^2}{1-|\chi|^2}\,|\xi|^{2n} \left(\,|\chi|^2 \,\alpha_{n,n-1}  +(1-|\chi|^2) \,  \alpha_{n,n}\, \right)  \\
& \hspace{2cm}  \,+\, 2 \sum_{j=1}^{n} \frac{(-1)^j}{j} \,|\xi|^{2j} |\chi|^{2-2j}  \,\,\, \left( \alpha_{0,n-1}+ d\alpha_{0,n-1}-d\beta_{0,n-1}\right) \\
& \hspace{2cm}  \,+\, 2 \sum_{j=1}^{n-1} \, \sum_{\ell=j+1}^{n}  \frac{(-1)^{\ell -j} }{\ell} |\xi|^{2\ell} |\chi|^{2(j-\ell)} \, \biggl[\, \left(j \left( 1 +|\chi|^2 \right)+ |\chi|^2 \right) \, \alpha_{j,n} \,-\, \left(1 - |\chi|^2 \right) \,d\alpha_{j,n} \\
& \hspace{7.7cm}  \,-\, \frac{|\chi|^2}{j+1} \, d\beta_{j,n-1} \biggr]\,.
\end{split}
\label{k0nomegaGenApp}
\ee
As for the previous examples, we will need to add non-trivial solutions of the homogeneous BPS equations that do not change the characteristics of the solutions. For the present solutions, we can freely replace all $|F|^2$ in $\alpha_{j,n}$ and $\beta_{j,n}$ \eqref{alphabetaDefApp} by 
\be 
|F|^2 \,\longrightarrow\, |F|^2 -\sum_{\ell=0}^{n-j}\frac{(-1)^\ell\,\left(1- |\chi|^2 \right)^{\ell}}{ |\chi|^{2 \ell}} \,(c^{(\ell)}(\chi) + c^{(\ell)}(\bar{\chi}))\,.
\label{HomSolApp}
\ee
This introduces in total $n$ arbitrary functions, $c^{(\ell)}(\chi)$, corresponding to homogeneous solutions of the BPS equations. Those functions will be necessary to regularize the solutions and must be fixed according to $F$.

The family of solutions above encompasses a closed-form expression for $(k,0,n)$ single-modes\footnote{In previous works, $\varpi_1 -\varpi_2$ is only given by an integral equation.}. More explicitly, for a single mode, $F(\chi) = \chi^k $, we have $\beta_{j,n}=0$ and the sum of $|\xi|^{2j}\,\alpha_{j,n}$ in $\cF$ corresponds to the generating functions $\mathscr{F}_{2k,2,2n}$ given in \eqref{GenFuncApp}. For generic $F(\chi)$, we have the multi-mode solutions with ($k,0,n$) excitations keeping $n$ fixed.

\subsubsection{Regularity}

The regularization of the solutions follows the same ideas as for superposition of $(k,0,1)$ in Section \ref{sec:k01reg}. Indeed, the solutions given by \eqref{PandFk0nGenApp} and \eqref{k0nomegaGenApp} are straightforwardly asymptotically AdS$_3\times$S$^3$ and have potential singularities only at $r=0$ and $\theta=\frac{\pi}{2}$. To deal with those singularities, it is convenient to go through the polar coordinates ($\rho,\vartheta$) around the supertube locus and take the limit $\rho\rightarrow 0$:
$$
r \,\sim\, \rho\, \cos \vartheta\,, \quad \sin \theta \,\sim\, 1-\frac{\rho^2}{2a^2}\, \sin^2 \vartheta \quad \Longrightarrow \quad \xi \sim  \frac{\rho}{a} \, \cos \vartheta\, e^{i\frac{\sqrt{2}}{R_y}v}\,, \quad \chi \sim \left( 1-\frac{\rho^2}{2a^2}\, \sin^2 \vartheta\right)  \, e^{i\phi_{1}}\,.
$$
The first thing to care about is the divergence in $\cF$ which must be cancelled as a necessary condition for having the metric component along $d\varphi_1^2$ to be regular \eqref{danger1}. A quick functional analysis shows that $\cF$ is strongly divergent up to order $\rho^{-2n}$. To cancel those singularities, one has to fix the $n$ homogeneous solutions \eqref{HomSolApp} as follows:
\be 
c^{(\ell)}(\chi) \,\equiv\,  \frac{1}{2}\,\sum_{k=1}^{\infty} \, \binom{k+\ell-1}{\ell} \, b_k^2 \,+\, \sum_{k,j=1}^{\infty} \binom{k+\ell-1}{\ell}\, b_k\,b_{k+j} \, \chi^j\,,
\ee
where $b_k$ are the Fourier coefficients of $F$. This specific choice nicely cancels the high-order divergent terms in $\omega$ too up to the $\rho^{-2}$ order. 

Next, one needs to cancel the $\rho^{-2}$ divergence in the metric components along $d\varphi_1^2$.   As $\rho \rightarrow 0$, one has
\begin{equation} 
\frac{\sqrt{Q_1 Q_5}}{R_y^2}\,\,g_{\varphi_1 \varphi_1}\, \sim \, \frac{1}{\rho^2}\, \left[\frac{Q_1 Q_5}{R_y^2} \,-\, a^2 \,-\, \frac{1}{2} \sum_{k=1}^{\infty} \, \binom{k+n-1}{n} \, b_k^2 \,\,-\,\,  \sum_{\ell ,k=1}^{\infty} \binom{k+n-1}{n} \, b_k\,b_{k+\ell}\, \cos(\ell \,\varphi_1) \right]\,.
\label{k01divApp}
\end{equation}
As in Section \ref{sect:ki01}, the expression (\ref{k01divApp}) represents a charge density fluctuation.  We therefore have to address this by allowing such a variation in the original supertube charge density.  This means that $Z_1$ needs to be allowed to vary according to 
\begin{equation} 
Z_1 ~=~  \frac{Q_{1}}{\Sigma} ~+~  \frac{R_y^2}{2\,  Q_5\, \Sigma} \, \big(\lambda (\chi) ~+~ \lambda (\bar{\chi})\big) ~+~ ....  \,,
\end{equation}
where  $\lambda$ is an arbitrary holomorphic function and the $...$ denotes the fluctuating modes associated with the momentum wave.   (See Section \ref{sect:basesols} for more details of the momentum wave terms.). This modifies $\cP$ in (\ref{PandFk0nGenApp}):
\begin{equation} 
\sqrt{\cP} ~=~  \frac{1}{\Sigma} \,\sqrt{Q_1 Q_5 ~+~  \frac{R_y^2}{2}\,  \left[\lambda (\chi) ~+~ \lambda (\bar{\chi}) -  \frac{|\chi|^2\,|\xi|^{2n}}{(n!)^2}\, \partial_\chi^n \partial_{\bar{\chi}}^n \left(|\chi|^{2(n-1)}\,|F|^2 \right)\right]}\,,  
\end{equation}
One can then cancel the divergence in $g_{\varphi_1\varphi_1}$ by taking 
\begin{equation} 
\begin{split}
\frac{Q_1 Q_5}{R_y^2} &~=~ a^2 \,+\, \frac{1}{2} \sum_{k=1}^{\infty} \, \binom{k+n-1}{n} \, b_k^2  ~=~ a^2\,+\,  \frac{1}{4\pi i}\oint_{|\chi|=1} \chi^{n-1}\bar{F}\partial^n_\chi F \,d\chi\,, \\
\lambda(\chi) &~=~   \sum_{k,\ell=1}^{\infty}\binom{k+n-1}{n} \,b_k\,b_{k+\ell}\, \, \chi^\ell \,.
\end{split}
\end{equation}
Thus, the metric is regular in the neighborhood of $r=0$. As for the $(k,0,1)$ solutions, one needs to check that the charge density fluctuations must not cause $\cP$ to be negative somewhere:
\begin{equation} 
\frac{2\,Q_1 Q_5}{R_y^2} ~+~  \,  \lambda (\chi) ~+~ \lambda (\bar{\chi}) -  \frac{|\chi|^2\,|\xi|^{2n}}{(n!)^2}\, \partial_\chi^n \partial_{\bar{\chi}}^n \left(|\chi|^{2(n-1)}\,|F|^2 \right) ~>~ 0 \,.
\end{equation}
Using that $|\xi|^2\leq 1-|\chi|^2$, it is actually sufficient to prove that 
\begin{equation} 
\frac{2\,Q_1 Q_5}{R_y^2} ~+~  \,  \lambda (\chi) ~+~ \lambda (\bar{\chi}) -  \frac{|\chi|^2\,(1-|\chi|^2)^{n}}{(n!)^2}\, \partial_\chi^n \partial_{\bar{\chi}}^n \left(|\chi|^{2(n-1)}\,|F|^2 \right) ~>~ 0 \,,
\end{equation}
The left-hand side gives in terms of modes, omitting the $a^2$,
\begin{equation} 
\sum_{k=1}^{\infty} \, \binom{k+n-1}{n}\, b_k^2 ~+~   2\sum_{k,\ell =1}^{\infty} \binom{k+n-1}{n}\,b_k\,b_{k+\ell}\, |\chi|^\ell \cos(\ell \,\varphi_1) -  \left(1-|\chi|^2 \right)^n\,  \bigg| \sum_{k=1}^{\infty}   \binom{k+n-1}{n}\,b_{k} \, \chi^k \bigg|^2  \,.
\end{equation}
After reordering we have
\begin{equation} 
 \sum_{k=1}^{\infty} \, \left( \binom{k+n-1}{n}- \binom{k+n-2}{n}|\chi|^2\right) \bigg| \sum_{\ell=0}^{\infty}  b_{k+\ell} \,e^{i\, \ell\, \varphi_1} \bigg|^2 - \left(1-|\chi|^2 \right)^n \,\bigg|\sum_{k=1}^{\infty} \binom{k+n-2}{n-1} \chi^k \sum_{\ell=0}^{\infty}  b_{k+\ell} \,\chi^\ell\bigg|^2 \,.
\label{cool-resumApp}
\end{equation}
Using the Cauchy-Schwarz inequality and $|\chi|^\ell\leq1$,
\be 
\left(1-|\chi|^2 \right)^n \,\bigg|\sum_{k=1}^{\infty} \binom{k+n-2}{n-1} \chi^k \sum_{\ell=0}^{\infty}  b_{k+\ell} \,\chi^\ell\bigg|^2 \,\leq\,  \sum_{k=1}^{\infty} \, |\chi|^2 \bigg| \sum_{\ell=0}^{\infty}  b_{k+\ell} \,e^{i\, \ell\, \varphi_1} \bigg|^2
\ee
and noticing that $|\chi|^2 \leq  \binom{k+n-1}{n}- \binom{k+n-2}{n}|\chi|^2 $ for any $k$, we have shown that \eqref{cool-resumApp} is always greater than $a^2$ and thus $\cP$  is manifestly positive. Thus, the required charge density variations (\ref{QQrepl})  in $Z_1$, preserve the smoothness of the metric and place no restrictions on the Fourier coefficients, and hence on $F(\chi)$.

\subsubsection{Conserved charges}

The angular momenta $J_R$ and $J_L$ and the momentum charge $Q_P$ can be read off from the asymptotic form of $\omega$, $\beta$ and $\cF$ through the expressions \eqref{asmpmoms} and \eqref{Fexp}. For the solutions constructed above, we obtain the supergravity charges:
\be 
\begin{split}
Q_P \,= \, \frac{1}{2} \,\sum^{\infty}_{k=1} \binom{k+n-1}{n-1} \, b_k^2\,=\,  \frac{1}{4\pi i}\oint_{|\chi|=1} \chi^{n-2}\,\bar{F}\,\partial^{n-1}_\chi F \,d\chi\,,\qquad  J_L\,=\, J_R\,=\, \frac{R_y}{2} \,a^2 \,.
\end{split}
\end{equation}

\section{Spherical coordinates}
\label{app:SphericalCoordinates}

The complex coordinates $(\chi,\eta,\xi)$ form an $S^{5}$ (see (\ref{constraint})). We can connect them to to the canonical spherical coordinates as follows. First, introduce a rescaled radial coordinate $\psi$ by 
\begin{equation}
\frac{r}{a}=\tan \psi~, \label{rescaled r to psi}
\end{equation}
the complex coordinates then read
\begin{align*}
\chi = \sin\theta \cos \psi\,  e^{i \varphi_{1}}~, \qquad \eta = \cos \theta \cos \psi \, e^{i \left(  \frac{\sqrt{2}v}{R}-\varphi_{2} \right)}~, \qquad
\xi = \sin \psi \,e^{i \frac{\sqrt{2}v}{R}} ~.
\end{align*}
Defining new angles by 
\begin{align*}
\alpha = \frac{\sqrt{2}}{R}v ~, \qquad \beta =  \frac{\sqrt{2}}{R}v-\varphi_{2} ~, \qquad \gamma =\varphi_{1}~,
\end{align*}
we are left with the canonical coordinates on a 5-sphere
\begin{align*}
\chi =  e^{i \gamma} \sin\theta \cos \psi ~, \qquad  
\eta = e^{i \beta} \cos \theta \cos \psi  ~, \qquad
\xi = e^{i \alpha} \sin \psi  ~.
\end{align*}

To transcribe the results of Section \ref{sect:SecondLayerOperatosCC} into spherical coordinates one will also need 
\begin{align*}
\begin{split}
\widetilde{\mathcal{L}}_{F}&= -\frac{2\sqrt{2}}{R} \left(\partial_{\alpha}+\partial_{\beta} \right)~,\\
\widetilde{\mathcal{L}}_{\xi}&= - \frac{1}{2}\tan \psi \, \partial_{\psi} ~,\\
\widetilde{\mathcal{L}}_{\chi}&= \frac{\partial_{\alpha}+ \partial_{\beta} }{2(\csc^{2}\theta \sec^{2}\psi-1)}  -\frac{1}{2}\partial_{\gamma} ~, 
\end{split}
\begin{split}
\widetilde{\mathcal{L}}_{\mu}&=- \frac{1}{4}\sin 2\theta \, \partial_{\theta}~, \\
\widetilde{\mathcal{L}}_{\delta}&= \frac{\partial_{\alpha} -\sec^{2}\theta \tan^{2}\psi \, \partial_{\beta} }{2(\tan^{2}\theta - \sec^{2}\theta \sec^{2}\psi)} ~.
\end{split}
\end{align*}
This formulation makes it clear that once a Fourier decomposition is made on the periodic directions $(\alpha,\beta,\gamma)$, the operators $(\widetilde{\mathcal{L}}_{F},\widetilde{\mathcal{L}}_{\chi},\widetilde{\mathcal{L}}_{\delta})$ become algebraic, while the operators $(\widetilde{\mathcal{L}}_{\xi},\widetilde{\mathcal{L}}_{\mu})$ still need to be dealt with.

\section{Collection of proofs}
\label{app:ProofGen}

\subsection{Proofs of the positivity of $\cP$}
\label{app:ProofP}
For the families of solutions constructed in Section \ref{sect:Examples}, regularity requires that the warp factor, $\cP$ \eqref{cPhol}, remains positive everywhere. In this section, we prove that it is true for all the examples treated and for generic arbitrary holomorphic functions $F$.

\subsubsection{For the $(1,0,n)$ superstrata}
\label{app:ProofP10n}
For the solutions constructed from a superposition of $(1,0,n)$ superstrata, the positivity of $\cP$ requires
 \begin{equation}
Q_1 Q_5  ~-~  \frac{a^2 R_y^2}{2\,(r^2+a^2) }\, |F(\xi)|^2\, \sin^2 \theta \,>\,0 \,,
\end{equation}
for any holomorphic function $F$ of the variable $\xi$ defined in \eqref{cplxcoords}. We can re-write this condition as:
\begin{equation}
 \frac{2\, Q_1 Q_5}{R_y^2} ~>~ \frac{a^2}{a^2 + r^2} |F(\xi)|^2 ~=~  (1-|\xi|^2) |F(\xi)|^2.
\end{equation}
Using the Cauchy-Schwarz inequality\footnote{We thank A. Lukyanenko for suggesting this.} for $|\xi|<1$, we have:
\begin{equation}
 (1-|\xi|^2) |F(\xi)|^2 ~=~  (1-|\xi|^2)\Big|\sum_n b_n \xi^n \Big|^2 ~\leq~ (1-|\xi|^2)\Big(\sum_n b_n^2\Big) \Big(\sum_m |\xi|^{2m}\Big)  ~=~  c ~<~ \frac{2\, Q_1 Q_5}{R_y^2}\,,
\end{equation}
where the last inequality follows from  (\ref{reg10n}).
Equivalently, we can take the Cauchy integral formula at $z=\xi$ for the holomorphic function $g(z) = F(z)^2 (1-\xib z)$ to get:
\begin{equation}
 (1-|\xi|^2) |F(\xi)|^2 ~=~  |g(\xi)|  ~=~  \left| \frac{1}{\sqrt{2}\pi R_y}\oint_{|y|=1} F(y)^2 \frac{1-\xib y}{y-\xi}dy \right| \leq  \frac{1}{\sqrt{2}\pi R_y}\oint_{|y|=1} |F(y)|^2 ~<~  \frac{2\, Q_1 Q_5}{R_y^2}.
 \end{equation}
Either way, we conclude that $\mathcal{P} >0$ is satisfied everywhere.

\subsubsection{For the $(2,1,n)$ superstrata}
\label{app:ProofP21n}

For the solutions constructed from a superposition of $(2,1,n)$ superstrata, the positivity of $\cP$ requires
\begin{equation}
 \frac{2\, Q_1 Q_5}{R_y^2} ~>~ \frac{a^4}{(r^2+a^2)^2 }\, |F|^2\, \sin^2 \theta\, \cos^2 \theta  \,,
\end{equation}
for any holomorphic function $F$ of the variable $\xi$ defined in \eqref{cplxcoords}. First, we take the maximum values for $\sin^2 \theta\, \cos^2 \theta $, that is $\frac{1}{4}$,
\begin{equation}
\frac{a^4}{(r^2+a^2)^2 }\, |F|^2\, \sin^2 \theta\, \cos^2 \theta ~\leq~ \frac{(1-|\xi|^2)^2}{4} \, |F(\xi)|^2 ~=~  \frac{(1-|\xi|^2)^2}{4} \Big|\sum_n b_n \,\xi^n \Big|^2 \,,
\end{equation}
then, using the Cauchy-Schwarz inequality for $|\xi|<1$, we have:
\begin{equation}
 \frac{(1-|\xi|^2)^2}{4} \Big|\sum_n b_n \xi^n \Big|^2 ~\leq~ \frac{(1-|\xi|^2)^2}{4}\Big(\sum_n \frac{b_n^2}{2(n+1)} \Big) \Big(\sum_m 2(m+1) |\xi|^{2m}\Big)  ~=~  \frac{|\xi|^2\,(2-|\xi|^2)}{2} \, c^{(0)} ~<~ \frac{2\, Q_1 Q_5}{R_y^2}\,,
\end{equation}
where the last inequality follows from  (\ref{reg21n}).

\subsubsection{For the $(k,0,1)$ superstrata}
\label{app:ProofPk01}

For the solutions constructed from a superposition of $(k,0,1)$ superstrata, the positivity of $\cP$ requires
\begin{equation} 
\frac{2\,Q_1 Q_5}{R_y^2} ~+~  \,  \lambda (\chi) ~+~ \lambda (\bar{\chi}) -  |\xi|^2|\chi|^2\, \partial_\chi \partial_{\bar{\chi}} |F|^2 ~>~ 0 \,,
\label{boundk01-1}
\end{equation}
for any holomorphic function $F$ of the variable $\chi$ defined in \eqref{cplxcoords}, and where $\lambda$ is given according to the Fourier coefficients of $F$ in \eqref{lambdaDef}. Because $|\xi|^2\leq 1-|\chi|^2$, it is actually sufficient to prove that 
\begin{equation} 
\frac{2\,Q_1 Q_5}{R_y^2} ~+~  \,  \lambda (\chi) ~+~ \lambda (\bar{\chi}) - |\chi|^2\left(1-|\chi|^2 \right)\, \partial_\chi \partial_{\bar{\chi}} |F|^2 ~>~ 0 \,,
\label{boundk01-1bis}
\end{equation}
or, in terms of modes:
\begin{equation} 
2 \,a^2 ~+~   \sum_{k=1}^{\infty} \, k\, b_k^2 ~+~  2 \sum_{k,\,\ell =1}^{\infty}  k\,b_k\,b_{k+\ell}\, |\chi|^\ell\, \cos(\ell \,\varphi_1) - \left(1-|\chi|^2 \right)\,  \bigg| \sum_{k=1}^{\infty}  k\,b_{k} \, \chi^k \bigg|^2 ~>~   0  \,.
\label{boundk01-2}
\end{equation}
One can reorder (\ref{boundk01-2}) so that the left-hand side is:
\begin{equation} 
2a^2 ~+~ \sum_{k=1}^{\infty} \, \left(k-(k-1)|\chi|^2\right) \bigg| \sum_{\ell=0}^{\infty}  b_{k+\ell} \,\chi^\ell \bigg|^2 \,-\, \left(1-|\chi|^2 \right) \,\bigg|\sum_{k=1}^{\infty} \chi^k \sum_{\ell=0}^{\infty}  b_{k+\ell} \,\chi^\ell\bigg|^2 \,.
\label{cool-resum}
\end{equation}
Using the Cauchy-Schwarz inequality,
\be 
\left(1-|\chi|^2 \right) \,\bigg|\sum_{k=1}^{\infty} \chi^k \sum_{\ell=0}^{\infty}  b_{k+\ell} \,\chi^\ell\bigg|^2 \,\leq\,  |\chi|^2\, \sum_{k=1}^{\infty}  \bigg| \sum_{\ell=0}^{\infty}  b_{k+\ell} \,\chi^\ell \bigg|^2
\ee
and noticing that $|\chi|^2 \leq k-(k-1)|\chi|^2$ for any $k$, we have shown that \eqref{cool-resum} is always greater than $a^2$ and thus $\cP$  is manifestly positive. 

\subsubsection{For the $(k,k,0)$ superstrata}
\label{app:ProofPkk0}

For the solutions constructed from a superposition of $(k,k,0)$ superstrata, the positivity of $\cP$ requires
\begin{equation} 
\frac{2 \,Q_1 Q_5}{R_y^2} ~+~  \,  \lambda (\eta) ~+~ \lambda (\bar{\eta})  ~>~ |F(\eta)|^2 \,,
\label{boundkk0-1}
\end{equation}
for any holomorphic function $F$ of the variable $\eta$ defined in \eqref{cplxcoords}, and where $\lambda$ is given according to the Fourier coefficients of $F$ as in \eqref{lambdaDefkk0}. This is trivially shown by reordering appropriately the left hand-side using the regularity condition \eqref{RegCond}. We obtain
\be 
2\,a^2 \,+\, (1-|\eta|^2)\sum_{k=1}^\infty \left|\sum_{\ell=0}^{\infty} b_{k+\ell}\,\eta^\ell \right|^2  \,+\, | F(\eta)|^2\,,
\label{reordering}
\ee
and that proves that \eqref{boundkk0-1} is true for any arbitrary holomorphic function $F$. Thus $\sqrt{\cP}$ is well-defined everywhere. 

\subsection{Proof of the claim in Section \ref{10n-y-circle}}
\label{app:Proof}

\noindent Consider the open unit disk $D \equiv \{\xi \in \IC : 0 \le |\xi | < 1 \}$.  We consider an analytic function $F$ on $\bar{D}$ that is not constant. Define
\begin{equation}
c   ~\equiv~    \frac{1}{2\pi}\, \int_0^{2\pi} \, \big | F(e^{i \theta}) \big|^2 \, d \theta   \,,
 \label{cDefn} 
\end{equation}

\noindent{\bf Claim}:  There is no  simple closed curve,  $\gamma: [0,2\pi] \to   D$, that  winds once around the origin  and on  which one has 
\begin{equation}
\forall \theta\,,\qquad  \big | F(\gamma(\theta)) \big|^2 ~-~  c  ~\ge~ 0 \,,
 \label{CTC} 
\end{equation}
for all $\theta \in [0,2\pi]$.

\noindent{\bf Proof}: We use proof by contradiction and thus start by assuming that such a closed curve exists. Define the Taylor expansion of $F$:
\be 
F(\xi) \,=\, \sum_{n=0}^{\infty} b_n \,\xi^n\,,
\ee
which implies that $c\,=\, \sum_{n=0}^{\infty} |b_n|^2$. We  divide the proof according to the number of zeroes of $F$.
\begin{itemize}
\item $F$ has no zeroes in $D$:
\end{itemize}
Thus, $F$ has no zeroes in the open subset, denoted $U$, inside the closed curve $\gamma$. One can then apply the minimum modulus principle on $U$ and $F$ must take its minimum value at the boundary, that is on $\gamma$. However, $0 \in U$ by definition and\footnote{We have a strict inequality  $|b_0|^2 \,<\, c$ because $F$ is not constant. There exists an other $n>0$ with $b_n \neq 0$ and then $c\,=\, \sum_{n=0}^{\infty} |b_n|^2 > |b_0|^2$.} $$|F(0)|^2  \,=\, |b_0|^2 \,<\, c \,\leq\, |F(\gamma(\theta))|^2\,, $$ 
which is in contradiction with having the minimum at the boundary of $U$.

\begin{itemize}
\item $F$ has zeroes in $D$:
\end{itemize}

We denote the zeroes of $F$ in $D$ as $z_i$, $i={1,\ldots,p}$. For simplicity, we have taking into account the multiplicity of the zeroes by considering that some $z_i$ can be equal. Let us now define the functions: 
\begin{equation}
G(\xi) ~\equiv~   \prod_{i=1}^p \frac{1-\bar{z}_i \,\xi}{\xi - z_i} \, F(\xi)  \,.
 \label{newfns} 
\end{equation}
By construction, $G$, is analytic on $\bar{D}$ and has no zeroes on $D$. We introduce the Taylor expansion of $G$ as follows
\be 
G(\xi)\,=\, \sum_{n=0}^\infty \, a_n\, \xi^n\,.
\ee
Also note that\footnote{We use that $\left|\frac{1-\bar{z}_i \,\xi}{\xi - z_i} \right|=1$ when $|\xi|=1$.}
\begin{equation}
c ~=~   \frac{1}{2\pi}\, \int_0^{2\pi} \, \big | G(e^{i \theta}) \big|^2 \, d \theta\,=\,\sum_{n=0}^{\infty} |a_n|^2.
 \label{inteql1} 
\end{equation}
Moreover, by noticing that 
\be 
\left|\frac{1-\bar{z}_i \,\xi}{\xi-z_i }\right|^2 \,=\, 1\,+\, \frac{(1-|\xi|^2)(1-|z_i|^2)}{|\xi-z_i |^2} \,>\, 1\,,
\label{AppImportantIneq}
\ee
because $\xi,z_i \in D$, 
one can easily prove that $|G(\gamma(\theta))|^2 \,>\, c$, $\forall \theta\in [0,2\pi]$ from \eqref{newfns}. Since $G$ is non-vanishing, we can apply the minimum modulus principle for $G$ on $U$ and evaluate at $0\in U$:
$$|G(0)|^2  \,=\, |a_0|^2 \,\leq\, c \,<\, |G(\gamma(\theta))|^2\,, $$ 
which is in contradiction with having the minimum at the boundary of $U$.

\noindent Thus, such a curve $\gamma$ does not exist and we have proven the claim.

\newpage
\begin{adjustwidth}{-1mm}{-1mm} 
\bibliographystyle{utphys}      
\bibliography{microstates}       

\end{adjustwidth}


\end{document}